\documentclass[12pt,a4paper]{article}
\usepackage{amsmath,amsfonts,amsthm,graphicx}
\allowdisplaybreaks[3]
\numberwithin{equation}{section} 
\newtheorem{theorem}{Theorem}[section]
\newtheorem{lemma}[theorem]{Lemma}

\usepackage[body={15cm,21.5cm}]{geometry}
\newcommand{\Qf}{{\mathbf Q}}
\newcommand{\Qb}{{\mathbb Q}}
\newcommand{\Qbb}{\overline{\mathbb Q}}

\newcommand{\Tf}{{\mathbf T}}
\newcommand{\Tb}{{\mathbb T}}
\newcommand{\Tbb}{\overline{\mathbb T}}
\newcommand{\Ds}{{\mathsf D}}
\newcommand{\Yb}{{\mathbb Y}}
\newcommand{\Def}{{\mathbf \Delta}}
\begin{document}
\title{
Wronskian 
solutions of the $T$, $Q$ and $Y$-systems related to infinite 
dimensional unitarizable modules of  the general linear superalgebra $gl(M|N)$
}
\author{Zengo Tsuboi
\footnote{An additional post member at 
Osaka City University Advanced Mathematical Institute (until 31 March 2012); 
Present address: Department of Theoretical Physics, 
Research School of Physics and Engineering,
%ANU College of Physical and Mathematical Sciences, 
the Australian National University, ACT 0200, Australia; 
E-mail: ztsuboi$\bullet$gmail.com}
\\
Institut f\"{u}r Mathematik und Institut f\"{u}r Physik, \\
Humboldt-Universit\"{a}t zu Berlin, 
\\ 
IRIS Haus, Zum Gro{\ss}en Windkanal 6,  12489 Berlin, Germany
% \&  
\\
} 
\date{}
\maketitle
%%%%%%%%%%%%%%%%%%%%%%%%%%%%%%%%
\begin{abstract}
In \cite{T09} (Z. Tsuboi, Nucl. Phys. B 826 (2010) 399 [arXiv:0906.2039]), we proposed Wronskian-like solutions of the 
T-system for $[M,N]$-hook of the general linear 
superalgebra $gl(M|N)$. 
We have generalized these Wronskian-like solutions to the ones for the general T-hook, 
which is a union of $[M_{1},N_{1}]$-hook and $[M_{2},N_{2}]$-hook 
($M=M_{1}+M_{1}$, $N=N_{1}+N_{2}$). 
These solutions are related to Weyl-type supercharacter formulas of infinite 
dimensional unitarizable modules of $gl(M|N)$. 
Our solutions also include  a 
Wronskian-like solution discussed in \cite{GKLT10} 
(N.\ Gromov, V.\ Kazakov, S.\ Leurent, Z.\ Tsuboi, JHEP 1101 (2011) 155 [arXiv:1010.2720]) 
in relation to the $AdS_{5}/CFT_{4}$ spectral problem. 
\end{abstract}
%%%%%%%
PACS 2008: 
02.20.Uw  02.30.Ik  75.10.Pq  11.25.Hf
\\
\\\\
e-Print: arXiv:1109.5524 [hep-th]
\\
Report number: HU-Mathematik-2011-21,  HU-EP-11/ 45
\\
Journal-ref: Nuclear Physics B 870 [FS] (2013) 92-137\\
DOI: 10.1016/j.nuclphysb.2013.01.007
%%%%%%%%%%%%%%%%%%%%%%%%%%%%%%%
%
\setlength{\baselineskip}{14.6pt}
%\setlength{\abovedisplayskip}{13pt}
%\setlength{\belowdisplayskip}{13pt}
%%%%%%%%%%%%%%%%%%%%%%%%%%%%%%%%%%%%%%%%%%%%%%%%%%%
\newpage 
\tableofcontents 
%%%%%%%%%%%%%%%%%%%%%%%%%%%%%%%%%%%%%%%%%%%%%%%%%%%%%%%%%%%%%%%%%%%%%%%%%%%
\section{Introduction} 
\begin{figure}
  \begin{center}
    \setlength{\unitlength}{1.5pt}
    \begin{picture}(160,55) 
      {\thicklines \put(80,50){\vector(0,-1){50}}}
      {\thicklines \put(40,0){\line(0,1){30}}}
       {\thicklines \put(110,0){\line(0,3){20}}}
       {\thicklines \put(110,20){\line(1,0){50}}}
       {\thicklines \put(0,30){\line(1,0){40}}}
       {\thicklines \put(0,50){\vector(1,0){160}}}
      {\linethickness{0.2pt} \put(80,20){\line(1,0){30}}}
      {\linethickness{0.2pt} \put(110,20){\line(0,1){30}}}
      {\linethickness{0.2pt} \put(40,30){\line(1,0){40}}}
      {\linethickness{0.2pt} \put(40,30){\line(0,1){20}}}
       \put(78,-5){$a$}
       \put(163,48){$s$}
       \put(69,19){$M_{1}$}
       \put(69,29){$M_{2}$}
       \put(108,52){$N_{1}$}
       \put(38,52){$-N_{2}$}
       \put(78,52){$0$}
    \end{picture}
  \end{center}
  \caption{ 
A generalized T-hook (a union of $[M_{1},N_{1}]$-hook and $[M_{2},N_{2}]$-hook) for the $T$-system for $gl(M_{1}+M_{2}|N_{1}+N_{2})$ 
\cite{GKV09,Hegedus09}: 
$T_{a,s}(u)=0$ if $\{a<0\}$ or $\{a>M_{1},s>N_{1}\}$ or $\{a>M_{2},s<-N_{2}\}$.}
  \label{gene-T-hook}
\end{figure}
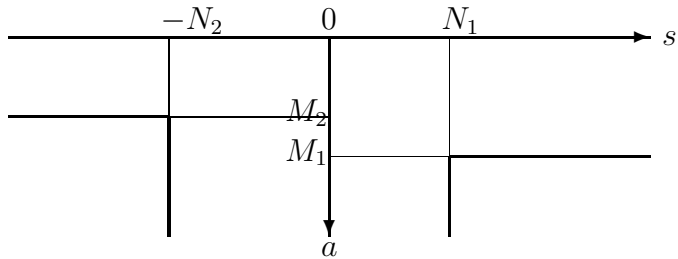

The T-system is a class of functional relations among transfer matrices of quantum integrable systems 
 related to Yangians $Y(\mathfrak{g})$ or quantum affine algebras $U_{q}(\hat{\mathfrak{g}})$ 
associated with simple Lie algebras $\mathfrak{g}$ 
\cite{KNS10}. 
In relation to finite dimensional representations of $\mathfrak{g}=gl(M|N)$ or $sl(M|N)$, 
it was proposed for $(M,N)=(2,0)$ case in \cite{KR87}, for $(M,N)=(M,0)$ in \cite{KNS93} and 
 for any non-negative integers $(M,N)$ in \cite{T97,T98}. 
 It is a certain Hirota bilinear difference equation \cite{Hirota81} of the form:  
\begin{align}
T_{a,s}(u-1)T_{a,s}(u+1)=T_{a,s-1}(u) T_{a,s+1}(u) + T_{a-1,s}(u) T_{a+1,s}(u), \quad 
a,s \in {\mathbb Z}, \quad u \in {\mathbb C} 
\end{align}
with specific boundary conditions on the mutually commuting dependent variables $T_{a,s}(u)$ 
which correspond to transfer matrices (T-operators) or their eigenvalues. 
In particular for finite dimensional representations of $\mathfrak{g}=gl(M|N)$, it is defined on $[M,N]$-hook 
(cf.\ Figure \ref{MN-hook}). 
Recently, T-system defined on a T-hook, which is a union of 2-copies of 
$[2,2]$-hooks (cf.\ Figure \ref{gene-T-hook}) appeared in the study of the $AdS_{5}/CFT_{4}$ duality
\footnote{In this paper, we do not discuss integrable systems related to this duality explicitly. 
For an overview of the AdS/CFT integrability, see \cite{Beisert10}. 
Some explanations on the T-system (or Y-system) in AdS/CFT can be seen for example in review papers 
\cite{Serban10,Volin10-1,GK10,KNS10}.}
 \cite{GKV09} (see \cite{BFT09} for an integral form of it). 
This T-system was further generalized \cite{Hegedus09} to the T-system on generalized T-hook, which is a 
union of $[M_{1},N_{1}]$-hook and $[M_{2},N_{2}]$-hook 
($M_{1}, M_{2},N_{1},N_{2} \in {\mathbb Z}_{\ge 0}$; cf.\ Figure \ref{gene-T-hook}). 
%%%%%%%%%%%%%%%%%%%%%%%
\begin{figure}[b]
  \begin{center}
    \setlength{\unitlength}{1.5pt}
    \begin{picture}(85,55) 
      {\thicklines \put(0,50){\vector(0,-1){50}}}
       {\thicklines \put(20,0){\line(0,1){20}}}
       {\thicklines \put(20,20){\line(1,0){60}}}
     % {\thicklines \put(30,30){\line(0,1){20}}}
     % {\thicklines \put(0,30){\line(1,0){30}}}
     % {\thicklines \put(0,30){\line(0,1){20}}}
     % {\thicklines \put(0,50){\line(1,0){30}}}
      %\put(0,40){\line(1,0){30}}
       {\thicklines \put(-10,50){\vector(1,0){90}}}
      {\linethickness{0.2pt} \put(0,20){\line(1,0){20}}}
      {\linethickness{0.2pt} \put(20,20){\line(0,1){30}}}
       \put(-2,-5){$a$}
       \put(83,48){$s$}
       \put(-10,19){$M$}
       \put(18,52){$N$}
       \put(-2,52){$0$}
       %\put(-30,32){\tiny $\Ts_{a,s}= 0$}
        % \put(35,32){\tiny $\Ts_{a,s} \ne 0$}
        % \put(35,5){\tiny $\Ts_{a,s} = 0$}
        % \put(35,55){\tiny $\Ts_{a,s} = 0$}
       %\put(1.5,10){$\underleftrightarrow{\quad N\quad }$}
       %\put(50,21){\rotatebox{90}{$\underleftrightarrow{\qquad \rotatebox{-90}{M} \qquad }$}}
    \end{picture}
  \end{center}
  \caption{ 
$[M,N]$-hook for the $T$-system related to 
finite dimensional representations of $gl(M|N)$ \cite{T97,T98,KSZ07}: 
$T_{a,s}(u)=0$ if $\{a<0\}$ or $\{a>M,s>N\}$ or $\{a>0,s<0\}$.}
  \label{MN-hook}
\end{figure}
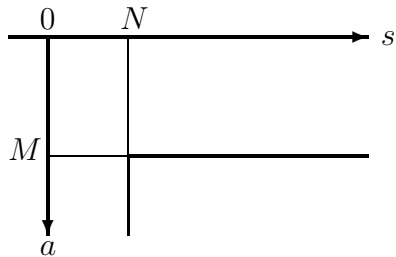
%%%%%%%%%%%%%%%%%%%%%%%%%%%%%%%%%%%%%%%%%%%%%%%%%%%
%%%%%%%%%%%%%%%%%
It was pointed out \cite{GKT10} that these T-systems on T-hooks are related to infinite dimensional 
 unitarizable modules of $gl(M_{1}+M_{2}|N_{1}+N_{2})$ \cite{CLZ03}. 

In our previous paper \cite{T09}, we proposed Wronskian-like
\footnote{To be precise, they are a kind of discrete analogue of Wronskian, 
which is called Casoratian.}
 determinant solutions for the T-system defined on 
$[M,N]$-hook of $gl(M|N)$. 
These non-trivially generalize Wronskian-like solutions for $gl(M)$ case \cite{KLWZ97}. 
In contrast with determinant solutions \cite{T97,T98} based on 
 the quantum supersymmetric Jacobi-Trudi formula 
\footnote{This determinant formula for $gl(M|0)$ case appeared in \cite{BR90}. 
It is often called `Bazhanov-Reshetikhin formula' in literatures of physics. 
However, this formula essentially follows from resolutions of representations of the Yangian by Cherednik. 
In this sense, it may be called `Cherednik-Bazhanov-Reshetikhin formula'.}, 
there is an upper bound for the size of the matrices for these Wronskian-like determinant 
solutions. And thus, they will be suited, for example, for the analysis of transfer matrices with large 
dimensional representations in the auxiliary space. 
In this paper, we will generalize our previous result to the generalized T-hook. 
The size of the matrices of our determinant formulas is {\bf finitely bounded} although they 
are formulas for transfer matrices for  {\bf infinite} dimensional representations in the auxiliary space. 
Some of the determinant expressions for $M_{1}=N_{1}=M_{2}=N_{2}=2$ case have already been 
proposed in \cite{GKLT10} in the study of the $AdS_{5}/CFT_{4}$ spectral problem. 

B\"{a}cklund transformations for the T-system were proposed for $gl(M)$ case in \cite{KLWZ97}, 
for $[M,N]$-hook of $gl(M|N)$ in \cite{KSZ07} and for the generalized T-hook in \cite{Hegedus09}. 
Our Wronskian-like solutions also solve all the functional relations in the intermediate steps of the 
B\"{a}cklund flows. 

The Q-system
\footnote{One should not confuse the Q-system with functional relations among  Baxter Q-operators.} is a degenerated version of the T-system. 
Namely, if one drops the spectral parameter dependence of the T-system, one obtains the Q-system. 
It was proposed as functional relations among characters of Yangians or quantum affine algebras 
(cf.\ \cite{Q-systems}). 
The Q-system for the T-hook was considered  in \cite{Gromov09,GKT10} in relation to 
quasiclassical AdS/CFT, and 
an explicit Wronskian-like determinant 
(first Weyl-type) supercharacter solution was given \cite{GKT10} for case $M_{1}=M_{2}=2,N_{1}=0,N_{2}=4 $. 
In section 2, we first consider a solution of the Q-system for the generalized T-hook in terms of  
supercharacters of infinite dimensional unitarizable modules of $gl(M|N)$ 
and propose a Wronskian-like determinant expression for it. 

The next step is to `Baxterize' 
this supercharacter solution by Baxter Q-functions
%\footnote{To be precise, the `Baxter Q-functions' here are any mutually commuting functions 
%which obey functional relations for Baxter Q-functions called QQ-relation. 
%One needs to impose conditions on analyticity for these functions for applications.}  
so that it satisfies the T-system. 
In this paper, `Baxterize' means to put a spectral parameter into  
the supercharacter. 
This is done in section 3, based on a map from  
 the supercharacters to T-functions. 
% proposed in \cite{T09}. 
%%%%%%
The Baxter Q-operators were introduced by Baxter when he solved the 8-vertex model \cite{Bax72}. 
And the Baxter Q-functions are their eigenvalues. 
However, `Baxter Q-functions' here are any mutually commuting functions (or operators) 
which obey functional relations among them called QQ-relations (Eqs.\ \eqref{QQb}, \eqref{QQf}). 
Thus one needs to impose additional conditions on analyticity for these functions for applications. 
This leads to the Bethe ansatz equations for the corresponding system. 
%%%%%
We also rewrite this determinant solution and obtain three equivalent expressions. 

The Y-system is a system of functional relations related to the thermodynamic Bethe ansatz \cite{Zamolodchikov90}. 
It is related to the T-system by a dependent variable transformation. 
The Y-system is invariant under the gauge transformations for the T-functions (the dependent variables for the T-system). 
In this sense, the Y-system can be viewed as a gauge invariant form of the T-system (cf.\ \cite{KLWZ97}). 
In section 4, we briefly comment on the Y-system for the generalized T-hook. 
Section 5 is devoted to concluding remarks. 
The formulas in the main text contain parameters to define the supercharacters. 
These parameters correspond to the boundary twist or the horizontal field whose  
crucial role in the construction of the Baxter Q-operators was recognized in \cite{BLZ97} first. 
In appendix A (and section 3.8), we present formulas without these parameters. 
We will give a proof of the functional relations in appendix B. 
This supplements some omitted calculations of Ref.\ \cite{GKLT10}. 

Although we did not discuss an explicit operator realization of the Baxter Q-operators, 
the formulas in this paper should also be valid for operators. In particular, 
the fact that our new determinant formulas satisfy the T-system 
is independent of whether $\{ \Qb_{I} \}$ are mutually commuting Baxter Q-operators or 
their eigenvalues.  
%%%%%%%%%%%%%%%%%%%%%%%
\section{Q-system} 
In this section, we introduce supercharacter formulas of some infinite 
dimensional representations of ${\mathfrak g}=gl(M_{1}+M_{2}|N_{1}+N_{2})$, 
and consider functional relations (Q-system) among them. 
The supercharacters here as solutions of the Q-system should be interpreted as the ones for 
the quantum affine superalgebra $ U_{q}(\hat{\mathfrak g})$ (or the super-Yangian 
$Y({\mathfrak g})$). This is possible
\footnote{In general, there is no such evaluation map to the classical algebra for ${\mathfrak g} \ne gl(M|N)$. 
Thus one have to consider the branching of the representations, and to take linear combinations 
of characters of the classical Lie algebra to obtain the solution of the Q-system \cite{Q-systems}.
}
 since there is an evaluation map 
$ U_{q}(\hat{\mathfrak g}) \to U_{q}({\mathfrak g}) $ (or $Y({\mathfrak g}) \to {\mathfrak g} $). 
References relevant to our discussions on the representation theory are for example, 
\cite{CLZ03,CKW09,Kwon06,Conrey05} (and also \cite{GKT10} for $gl(4|4)$ case related to 
the  $AdS_{5}/CFT_{4}$ duality, and \cite{Volin10-2} in relation to the string hypothesis on Bethe roots). 
%%%%%%
\subsection{Supercharacters of unitarizable $gl(M_{1}+M_{2}|N_{1}+N_{2})$ modules} 
Let us introduce sets
\begin{align}
\begin{split}
&{\mathfrak B}_{1}=\{1,2,\dots, M_{1} \}, \qquad 
{\mathfrak B}_{2}=\{M_{1}+1, M_{1}+2,\dots, M_{1}+M_{2} \}, \\[4pt]
& {\mathfrak F}_{1}=\{M_{1}+M_{2}+1,M_{1}+M_{2}+2,\dots, M_{1}+M_{2}+N_{1} \}, \\[4pt]
& 
{\mathfrak F}_{2}=\{M_{1}+M_{2}+N_{1}+1,M_{1}+M_{2}+N_{1}+2,\dots, M_{1}+M_{2}+N_{1}+N_{2} \}, \\[4pt]
&{\mathfrak B}={\mathfrak B}_{1} \sqcup {\mathfrak B}_{2}, 
\quad 
{\mathfrak F}={\mathfrak F}_{1} \sqcup {\mathfrak F}_{2},  
\quad 
{\mathfrak I}={\mathfrak B} \sqcup {\mathfrak F}, 
\quad 
{\mathfrak I}_{1}={\mathfrak B}_{1} \sqcup {\mathfrak F}_{1}, 
\quad 
{\mathfrak I}_{2}={\mathfrak B}_{2} \sqcup {\mathfrak F}_{2}. 
\end{split}
\end{align}
We will use a grading parameter on the set $  {\mathfrak B} \sqcup  {\mathfrak F}$:  
\begin{align} 
p_{a}=1 \quad \text{for} \quad a \in {\mathfrak B}, \qquad 
p_{a}=-1 \quad \text{for} \quad a \in {\mathfrak F}. 
\end{align}
Let us 
take any subsets of the above sets: 
$B_{1} \subset {\mathfrak B}_{1}$, 
$B_{2} \subset {\mathfrak B}_{2}$, 
$F_{1} \subset {\mathfrak F}_{1}$, 
$F_{2}  \subset {\mathfrak F}_{2}$ and 
arrange them as tuples: 
\begin{align}
\begin{split}
& B_{1}=(b_{1}^{(1)},b_{2}^{(1)},\dots, b_{m_{1}}^{(1)}), \quad 
B_{2}=(b_{1}^{(2)},b_{2}^{(2)},\dots, b_{m_{2}}^{(2)}) ,  \\[4pt]
&
F_{1}=(f_{1}^{(1)},f_{2}^{(1)},\dots, f_{n_{1}}^{(1)}), \quad 
F_{2}=(f_{1}^{(2)},f_{2}^{(2)},\dots, f_{n_{2}}^{(2)}). 
\end{split}
\end{align}
For these tuples, we will also use notations: 
\begin{align}
\begin{split}
B&= B_{1} \sqcup B_{2}=(b_{1},b_{2},\dots, b_{m})=
(b_{1}^{(1)},b_{2}^{(1)},\dots, b_{m_{1}}^{(1)}, 
b_{1}^{(2)},b_{2}^{(2)},\dots, b_{m_{2}}^{(2)}), \\[4pt]
F&= F_{1} \sqcup F_{2}=(f_{1},f_{2},\dots, f_{n})=
(f_{1}^{(1)},f_{2}^{(1)},\dots, f_{n_{1}}^{(1)}, 
f_{1}^{(2)},f_{2}^{(2)},\dots, f_{n_{2}}^{(2)}), 
\end{split}
\end{align}
where we write the number of the elements as $m=m_{1}+m_{2}, 
n=n_{1}+n_{2}$. In the same way as above, we will use a symbol $I$ for a tuple 
 given by  any subset of the full set $ {\mathfrak I}$. 
We will distinguish tuples whose elements have different order.  
For example, $(1,3,4)$ and $(1,4,3)$ represent different tuples, while as sets, these are the same
 set $\{1,3,4\}=\{1,4,3\} $. 
In case we need not mind order of the elements of these tuples, we will regard them just as sets. 
For example, $b \in B_{1}$ means $b \in \{ b_{1}^{(1)},b_{2}^{(1)},\dots, b_{m_{1}}^{(1)} \} $. 
%%%%%

A sequence of $a$ integers $\lambda =(\lambda_{1}, \lambda_{2}, \dots, \lambda_{a})$ is called 
a {\em generalized partition of length $a$} if they satisfy $ \lambda_{1} \ge  \lambda_{2} \ge \dots \ge \lambda_{a}$. 
Here we do not assume that these are non-negative integers. If $\lambda_{a} \ge 0 $, then $\lambda$ is 
called a {\em partition}. 
For this generalized partition, we will use the following symbols:
\begin{gather}
\langle x \rangle =\max (x,0) ,
\quad 
\lambda^{+}=(\langle \lambda_{1} \rangle, \langle \lambda_{2} \rangle, \dots, \langle \lambda_{a} \rangle),
\quad 
\lambda^{-}=(\langle -\lambda_{1} \rangle, \langle -\lambda_{2} \rangle, \dots, \langle -\lambda_{a} \rangle). 
\end{gather}
Here $\lambda^{+}$ is a partition and  $\lambda^{-}$ is a reverse ordered partition. 
For the generalized partition $\lambda $, we assign a digram, called a {\em generalized Young diagram} 
(see figure \ref{fig-gyoung}) and 
will use  the same symbol $\lambda $ to denote this diagram. 
%%%%%%%%%%%%%%%
\begin{figure}
  \begin{center}
    \setlength{\unitlength}{1.5pt}
    \begin{picture}(160,55) 
      {\thicklines \put(80,50){\vector(0,-1){50}}}
      {\thicklines \put(40,0){\line(0,1){30}}}
       {\thicklines \put(110,0){\line(0,3){20}}}
       {\thicklines \put(110,20){\line(1,0){50}}}
       {\thicklines \put(0,30){\line(1,0){40}}}
       {\thicklines \put(0,50){\vector(1,0){160}}}
      \put(40,40){\line(1,0){90}}
      \put(60,30){\line(1,0){50}}
      \put(80,20){\line(1,0){20}}
      {\thicklines \put(81,20){\line(0,-1){10}}}
      \put(40,50){\line(0,-1){10}}
      \put(50,50){\line(0,-1){10}}
      \put(60,50){\line(0,-1){20}}
      \put(70,50){\line(0,-1){20}}
      \put(90,50){\line(0,-1){30}}
      \put(100,50){\line(0,-1){30}}
      \put(110,50){\line(0,-1){20}}
      \put(120,50){\line(0,-1){10}}
      \put(130,50){\line(0,-1){10}}
       \put(78,-5){$a$}
       \put(163,48){$s$}
       \put(75,18){$3$}
       \put(75,28){$2$}
       \put(108,52){$3$}
       \put(36,52){$-4$}
       \put(78,52){$0$}
    \end{picture}
  \end{center}
  \caption{ 
A generalized Young diagram in a generalized T-hook: this T-hook is 
a union of $[3,3]$-hook and $[2,4]$-hook for
 $gl(3+2|3+4)$. 
The generalized Young diagram (of length 6) is given by a 
generalized partition 
$\lambda=(5,3,2,0,-2,-4)$, 
$\lambda^{+}=(5,3,2,0,0,0)$,
$\lambda^{-}=(0,0,0,0,2,4)$. 
To denote one `0' in $\lambda$, a length one line is put on the bottom of the diagram.
}
  \label{fig-gyoung}
\end{figure}
We will also use a notation for a partition $\mu^{\prime } =(\mu_{1}^{\prime }, \mu_{2}^{\prime }, \dots ) $ for any 
sequence of non-negative integers 
$\mu =(\mu_{1}, \mu_{2}, \dots ) $, where $\mu^{\prime }_{i} = |\{ k | \mu_{k} \ge i \}|$. 
If $\mu$ is a  partition, then $\mu^{\prime } $ is a transposition of it. 

%%%%%%%%
Let $\{E_{ij}| i,j \in {\mathfrak B} \sqcup {\mathfrak F} \} $ 
be the generators of $gl(M_{1}+M_{2}|N_{1}+N_{2})$, which obey the super-commutation relation
\begin{multline}
[E_{ij},E_{kl} \}= E_{ij} E_{kl} - (-1)^{\frac{(1-p_{i}p_{j})(1-p_{k}p_{l})}{4}} E_{kl} E_{ij} 
 \\ =
\delta_{jk}E_{il} -(-1)^{\frac{(1-p_{i}p_{j})(1-p_{k}p_{l})}{4}}\delta_{li}E_{kj} . 
\end{multline} 
For the fundamental representation, 
$E_{ij}$ is $(M_{1}+M_{2}+N_{1}+N_{2}) \times (M_{1}+M_{2}+N_{1}+N_{2})$ matrix whose $(k,l)$ matrix element is 
$\delta_{ik}\delta_{jl} $. 
The generators $\{E_{ij}| i,j \in B\sqcup F \} $ generate a subalgebra $gl(m_{1}+m_{2}|n_{1}+n_{2})$ of 
$gl(M_{1}+M_{2}|N_{1}+N_{2})$ by a natural embedding. 
A Cartan subalgebra $ {\mathfrak h}(B_{1},B_{2}|F_{1},F_{2}) $ of $gl(m_{1}+m_{2}|n_{1}+n_{2})$ is generated by 
$\{E_{ii} | i \in B_{2} \sqcup F_{2} \sqcup B_{1} \sqcup F_{1} \} $
and a Borel subalgebra $ {\mathfrak b}(B_{1},B_{2}|F_{1},F_{2}) $ is generated by 
$\{ E_{i_{a} i_{b}} | 1 \le a<b \le m+n,  (i_{1},i_{2},\dots, i_{m+n})=B_{2} \sqcup F_{2} \sqcup B_{1} \sqcup F_{1} \} $. 
Let $ {\mathfrak h}^{*}(B_{1},B_{2}|F_{1},F_{2}) $ be a dual space of 
$ {\mathfrak h}(B_{1},B_{2}|F_{1},F_{2})$ with a basis $ \{ \varepsilon_{i} \} $ 
such that $\varepsilon_{i}(E_{jj} )=\delta_{ij} $. 
A bilinear form in ${\mathfrak h}^{*}(B_{1},B_{2}|F_{1},F_{2}) $ is defined as 
$(\varepsilon_{i} |\varepsilon_{j} )=p_{i} \delta_{ij}$. 
%%%%%
Let $\lambda $ be a generalized partition of length $a$ such that 
$\lambda_{m_{1}+1} \le n_{1}  $ and  $-n_{2} \le \lambda_{a-m_{2}} $. 
Let us consider an irreducible highest weight representation of 
$gl(m_{1}+m_{2}|n_{1}+n_{2})$ relative to 
the Borel subalgebra ${\mathfrak b}(B_{1},B_{2}|F_{1},F_{2}) $ 
with the highest weight:
\begin{multline} 
\lambda =
- \sum_{i=1}^{m_{2}} (\langle \lambda^{-}_{a-m_{2}  +i}-n_{2} \rangle +a)\varepsilon_{b^{(2)}_{i}}
- \sum_{i=1}^{n_{2}} ((\lambda^{-})^{\prime }_{n_{2}+1- i} -a)\varepsilon_{f^{(2)}_{i}} 
\\ 
+ \sum_{i=1}^{m_{1}} \lambda^{+}_{i} \varepsilon_{b^{(1)}_{i}} +
\sum_{i=1}^{n_{1}} (\langle (\lambda^{+})^{\prime }_{i}-m_{1} \rangle )\varepsilon_{f^{(1)}_{i}} .
\end{multline}
Here we used the same symbol for the partition and the 
highest weight corresponding to it. 
This representation is infinite dimensional for generic $\lambda $ and is called 
{\it unitarizable module of $gl(m_{1}+m_{2}|n_{1}+n_{2})$} (cf. \cite{CLZ03})
\footnote{It seems that this corresponds to a unitary representation of a real form 
$u(m_{1},m_{2}|n_{1},n_{2})$ of $gl(m_{1}+m_{2}|n_{1}+n_{2})$ (cf. \cite{CLZ03}). 
However, we will concentrate on expressions of (super)characters without 
 discussing the unitarity of the representations in this paper.}.
We denote this representation as $W(B_{1},B_{2}|F_{1},F_{2}; \lambda )$. 
%%%%%
Let us introduce formal exponentials 
\begin{align}
z_{j}=\exp (\varepsilon_{j}). \label{formexp}
\end{align}
$z_{j}$ can be regarded as complex numbers 
under the evaluation $z_{j}=z_{j}(h)=\exp (\varepsilon_{j})(h)=\exp (\varepsilon_{j}(h))$ for any fixed 
$h \in {\mathfrak h}(B_{1},B_{2}|F_{1},F_{2})$. 
The generating function of the supercharacters of the symmetric tensor representations 
of $gl(m_{1}+m_{2}|n_{1}+n_{2})$ is defined as 
\begin{align} 
w(t)=
\frac{\prod_{f \in F_{1}}(1-z_{f}t) \prod_{f \in F_{2}}(1-z_{f}t) }{\prod_{b \in B_{1}}(1-z_{b}t) 
\prod_{b \in B_{2}}(1-z_{b}t) } ,
\end{align}
where $t \in {\mathbb C}$. 
If we expand this with respect to non-negative power of  $t$, we obtain the supercharacters 
$\chi_{(k)}^{B_{1} \sqcup B_{2},\emptyset, F_{1} \sqcup  F_{2},\emptyset} 
=\mathrm{sch} W(B_{1} \sqcup B_{2}, \emptyset |F_{1} \sqcup F_{2},\emptyset; (k))$  of the 
finite dimensional 
symmetric tensor representations 
of $gl(m_{1}+m_{2}|n_{1}+n_{2})$: 
\begin{align}
w(t)=\sum_{k=0}^{\infty} \chi_{(k)}^{B_{1} \sqcup B_{2},\emptyset, F_{1} \sqcup  F_{2},\emptyset}  t^{k}, 
 \label{gen-ch}
\end{align}
where $ \chi_{(k)}^{B_{1} \sqcup B_{2},\emptyset, F_{1} \sqcup  F_{2},\emptyset} =0$ for $k<0$.
On the other hand
\footnote{There is a discussion for quantum case (T-functions) in \cite{BKSZ05}.}
, if we expand $w(t)$ with respect to non-negative power of $t$ 
(for the product indexed by $B_{1}$ and $F_{1}$) and $t^{-1}$ 
(for the product indexed by $B_{2}$ and $F_{2}$), 
we obtain the supercharacters 
$\chi_{(k)}^{B_{1},B_{2},F_{1},F_{2}} =\mathrm{sch} W(B_{1},B_{2}|F_{1},F_{2}; (k) )$
 of  infinite dimensional unitarizable representations of $gl(m_{1}+m_{2}|n_{1}+n_{2})$: 
\begin{align}
w(t)&=
(-1)^{m_{2}}t^{n_{2}-m_{2}}
\left(
\frac{\prod_{ f \in F_{2}} (-z_{f}) }{\prod_{b \in B_{2}} z_{b} }
\right)
\frac{\prod_{f \in F_{1}}(1-z_{f}t) \prod_{f \in F_{2}}(1-z_{f}^{-1}t^{-1}) }
{\prod_{b \in B_{1}}(1-z_{b}t) \prod_{b \in B_{2}}(1-z_{b}^{-1}t^{-1}) }
\nonumber \\
&=
(-1)^{m_{2}} 
\sum_{k \in {\mathbb Z}}\chi_{(k)}^{B_{1},B_{2},F_{1},F_{2}} 
t^{k+n_{2}-m_{2}}. 
 \label{expan-inf}
\end{align} 
Taking the coefficient of $t^{s+n_{2}-m_{2}}$ for $s \in {\mathbb Z} $, one obtains 
\begin{align}
\chi_{s}^{B_{1},B_{2},F_{1},F_{2}} &=
\sum_{k=\max(0,-s)}^{\infty} 
\chi_{(s+k)}^{B_{1},\emptyset ,F_{1},\emptyset }
\chi_{(-k)}^{\emptyset ,B_{2} , \emptyset ,F_{2} }, 
 \label{ch-infsym}
\end{align}
where 
\begin{align}
\frac{\prod_{f \in F_{1}}(1-z_{f}t) }
{\prod_{b \in B_{1}}(1-z_{b}t)  }
&=
\sum_{k=0}^{\infty }\chi_{(k)}^{B_{1}, \emptyset ,F_{1},\emptyset } t^k ,
 \label{gf-right}
\\[6pt]
\left(
\frac{ \prod_{ f \in F_{2}} (-z_{f}) }{ \prod_{b \in B_{2}} z_{b} }
\right)
\frac{\prod_{f \in F_{2}}(1-z_{f}^{-1}t^{-1}) }
{ \prod_{b \in B_{2}}(1-z_{b}^{-1}t^{-1}) }
&=
\sum_{k=0}^{\infty} \chi_{(-k)}^{\emptyset ,B_{2} , \emptyset ,F_{2} }
t^{-k},
\label{gf-left}
\end{align} 
and  $ \chi_{(-k)}^{B_{1}, \emptyset ,F_{1},\emptyset } =\chi_{(k)}^{\emptyset ,B_{2} , \emptyset ,F_{2} }=0$ 
for $k > 0 $.
Substituting \eqref{ch-infsym} into the Jacobi-Trudi type formula \cite{Kwon06}, one can 
obtain the supercharacter 
$\chi_{\lambda }^{B_{1},B_{2},F_{1},F_{2}} =\mathrm{sch} W(B_{1},B_{2}|F_{1},F_{2}; \lambda )$
 labeled by  the generalized partition 
$\lambda=(\lambda_{1},\lambda_{2},\dots, \lambda_{a})$ 
of length $a$: 
\begin{align}
\chi_{\lambda}^{B_{1},B_{2},F_{1},F_{2}} =
\det_{1 \le i,j \le a}(\chi_{(\lambda_{i}-i+j)}^{B_{1},B_{2},F_{1},F_{2}} ). 
 \label{Jacobi-Trudi}
\end{align}
%%%%%%%%%%%%%%%%%%%%
\subsection{Wronskian solution of the Q-system}
From now on, we often use a shorthand notation on the supercharacter
$\chi_{a,s}=\chi_{a,s}^{B_{1},B_{2},F_{1},F_{2}}$. 
In this paper, we restrict our discussions mainly on the supercharacter 
for the rectangular diagram
\footnote{
As for the general Young diagram case, we obtained 
some conjectures on Wronskian-like formulas on supercharacters and 
T-functions. We hope to discuss them in a separate paper. 
} 
$\lambda=(\underbrace{s,s,\dots, s}_{a})=(s^{a})$: 
$\chi_{a,s}:=\chi_{(s^{a})}$ 
for $a \ge 0 $ and $\chi_{a,s}:=0$ for $a<0$. 
The above supercharacter satisfies the so called Q-system 
\begin{align}
(\chi_{a,s})^{2}
=\chi_{a,s-1} \chi_{a,s+1} +
\chi_{a-1,s} \chi_{a+1,s}, 
\quad 
a,s \in {\mathbb Z}
 \label{Q-system}
\end{align}
for the generalized T-hook boundary condition: 
\begin{align}
\chi_{a,s}=0 
\quad  \text{if } 
\quad 
\{a<0\} 
\quad 
\text{or}
\quad 
\{a>m_{1}, s > n_{1}\}  
\quad \text{or} 
\quad \{a>m_{2}, s < -n_{2}\}, 
\label{boundTch}
\end{align}   
and 
\begin{align}
\begin{split}
\chi_{m_{1},n_{1}+a} & =
\left( \frac{\prod_{b \in B}z_{b} }{\prod_{f \in F}(-z_{f}) } \right)^{a} 
\chi_{m_{1}+a,n_{1}}, 
\\[6pt]
\chi_{m_{2},-n_{2}-a}&=
\chi_{m_{2}+a,-n_{2}} 
\quad \text{for} \quad  a \in {\mathbb Z}_{\ge 0}. 
\end{split}
\label{dualT-hook}
\end{align} 
Here $\frac{\prod_{b \in B}z_{b} }{\prod_{f \in F} z_{f} } $ in the above equation is the superdeterminant of a group element of 
$GL(m|n)$ (cf.\ \cite{Zabrodin07}). 
The above equation \eqref{Q-system} is a special case of the Hirota equation. 
It is known that the shape of this equation is invariant under 
the following gauge transformation. 
\begin{align}
%\chi_{a,s} \longrightarrow  
\widetilde{\chi}_{a,s}=
g_{1}g_{2}^{a} g_{3}^{s} g_{4}^{as}
\chi_{a,s}  ,
\label{gaugeqsystem}
\end{align}
where $g_{1},g_{2},g_{3},g_{4}$ are arbitrary non-zero complex parameters. 
%%%%%%%%%%%%%%%%%%%%%%%%%%%

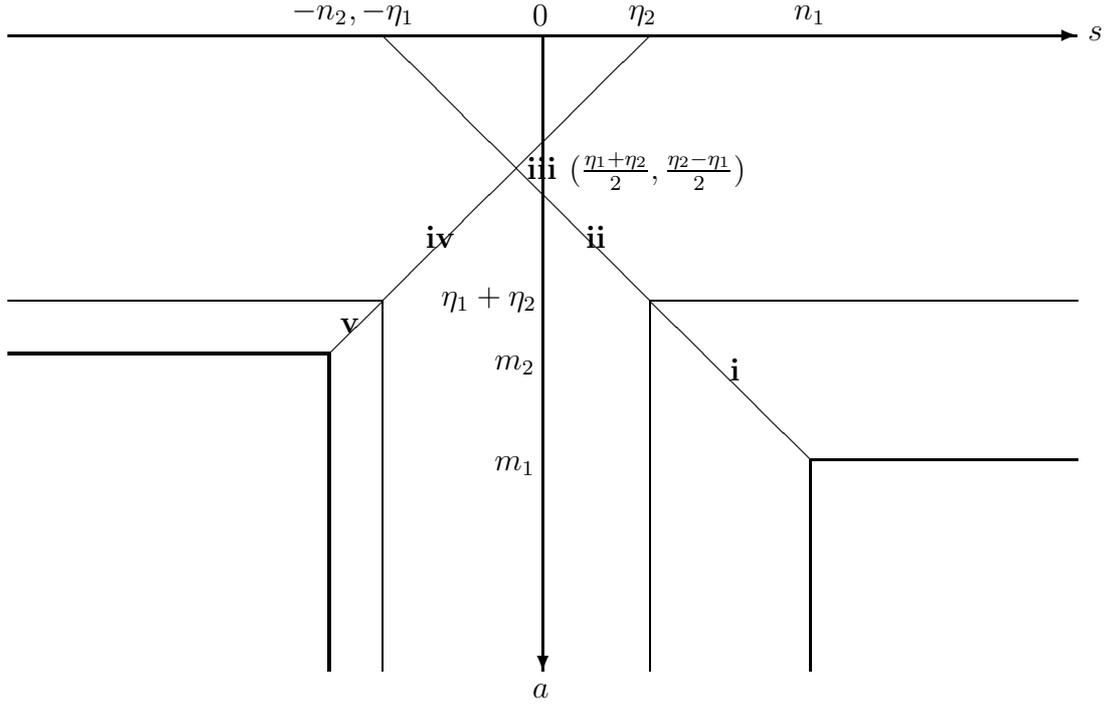
\begin{figure}
  \begin{center}
    \setlength{\unitlength}{2pt}
    \begin{picture}(205,120)     
       {\thicklines \put(0,120){\vector(1,0){200}}}
       {\thicklines \put(100,120){\vector(0,-1){120}}}
       {\thicklines \put(0,60){\line(1,0){60}}}
       {\thicklines \put(60,60){\line(0,-1){60}}}
       {\thicklines \put(150,40){\line(1,0){50}}}
       {\thicklines \put(150,40){\line(0,-1){40}}}
       {\linethickness{0.2pt} \put(0,70){\line(1,0){70}}}
       {\linethickness{0.2pt} \put(70,70){\line(0,-1){70}}}
       {\linethickness{0.2pt} \put(120,70){\line(1,0){80}}}
       {\linethickness{0.2pt} \put(120,70){\line(0,-1){70}}}
       {\linethickness{0.2pt} \put(120,120){\line(-1,-1){60}}}
       {\linethickness{0.2pt} \put(70,120){\line(1,-1){80}}}
       \put(98,-5){$a$}
       \put(202,119){$s$}
       \put(91,57){$m_{2}$}
       \put(91,38){$m_{1}$}
       \put(81,69){$\eta_{1}+\eta_{2}$}
       \put(53,123){$-n_{2},$}
       \put(66,123){$-\eta_{1}$}
        \put(98,122){$0$}
       \put(116,123){$\eta_{2}$}
       \put(147,123){$n_{1}$}
       \put(135,55){\bf i}
       \put(108,80){\bf ii}
       \put(97,93){\bf iii $(\frac{\eta_{1}+\eta_{2}}{2},\frac{\eta_{2}-\eta_{1}}{2})$}
       \put(78,80){\bf iv}
       \put(62,64){{\bf v}}
       %
        %\put(94,94){{\bf $\bullet $}}
    \end{picture}
  \end{center}
  \caption{The domain for the sparse determinant solutions  of T-and Q-systems 
for $\eta_{1}+\eta_{2}>0$ case. The lines corresponding to the boundaries 
 of the domain of the definition on $(a,s)$ are drawn by thine lines. 
The lines $a=s+\eta_{1}$ and $a=-s+\eta_{2}$ intersect at 
$(a,s)=(\frac{\eta_{1}+\eta_{2}}{2},\frac{\eta_{2}-\eta_{1}}{2})$. 
The numbers (i)-(iv) correspond to the ones in Appendix B.}
  \label{T-hook-sparse}
\end{figure}

\begin{figure}
  \begin{center}
    \setlength{\unitlength}{2pt}
    \begin{picture}(205,120)     
       {\thicklines \put(0,120){\vector(1,0){200}}}
       {\thicklines \put(100,120){\vector(0,-1){120}}}
       {\thicklines \put(40,90){\line(-1,0){40}}}
       {\thicklines \put(40,90){\line(0,-1){90}}}
       {\thicklines \put(170,40){\line(1,0){30}}}
       {\thicklines \put(170,40){\line(0,-1){40}}}
       {\linethickness{0.2pt} \put(70,120){\line(0,-1){120}}}
       {\linethickness{0.2pt} \put(90,120){\line(0,-1){120}}}
       {\linethickness{0.2pt} \put(40,90){\line(1,1){30}}}
       {\linethickness{0.2pt} \put(170,40){\line(-1,1){80}}}
       \put(98,-5){$a$}
       \put(202,119){$s$}
       \put(91,87){$m_{2}$}
       \put(91,38){$m_{1}$}
       \put(34,123){$-n_{2}$}
       \put(86,123){$-\eta_{1}$}
        \put(98,122){$0$}
       \put(66,123){$\eta_{2}$}
       \put(167,123){$n_{1}$}
       \put(135,75){\bf i}
       \put(52,104){{\bf v}}
       %
        %\put(94,94){{\bf $\bullet $}}
    \end{picture}
  \end{center}
  \caption{The domain for the sparse determinant solutions  of T-and Q-systems 
for $\eta_{1}+\eta_{2}<0$ case. The lines corresponding to the boundaries 
 of the domain of the definition on $(a,s)$ are drawn by thine lines. 
The numbers (i) and (v) correspond to the ones in Appendix B.}
  \label{T-hook-sparse2}
\end{figure}
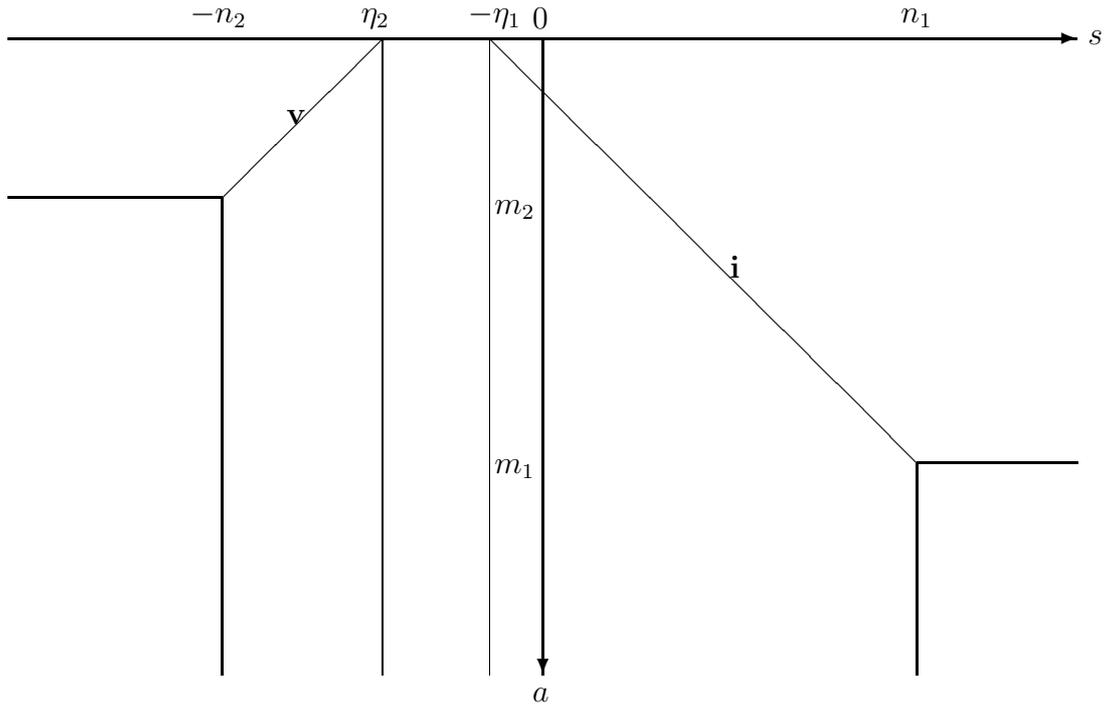

We have a determinant expression of the supercharacter. 
Let us introduce a determinant
\footnote{
In this paper, we use a notation for a matrix whose matrix elements are labeled by tuples 
$J=(j_{1},j_{2},\dots, j_{a})$, 
$K=(k_{1},k_{2},\dots, k_{b})$: 
\begin{align}
\left( A_{j,k} \right)_{j \in J \atop k \in K}
=
\begin{pmatrix} 
 A_{j_{1},k_{1}} & A_{j_{1},k_{2}} & \cdots & A_{j_{1},k_{b}} \\ 
 A_{j_{2},k_{1}} & A_{j_{2},k_{2}} & \cdots & A_{j_{2},k_{b}} \\
 \vdots       & \vdots       & \ddots  & \vdots \\
 A_{j_{a},k_{1}} & A_{j_{a},k_{2}} & \cdots & A_{j_{a},k_{b}} 
\end{pmatrix}. 
\end{align}
We also use a notation $(0)_{a\times b}$ for a $a$ by  $b$ zero matrix. 
}
labeled by tuples 
$B_{1}$, $B_{2}$,  $F$,   
$R=(r_{1},r_{2},\dots, r_{a}) $, $S=(s_{1},s_{2},\dots,s_{b})$, 
$T_{1}=(t^{(1)}_{1},t^{(1)}_{2},\dots, t^{(1)}_{c_{1}}),  
T_{2}=(t^{(2)}_{1},t^{(2)}_{2},\dots, t^{(2)}_{c_{2}}) $, 
$r_{i},s_{i}, t^{(1)}_{j}, t^{(2)}_{j}, \eta \in {\mathbb C}$: 
\begin{align}  
\varDelta^{B_{1},B_{2},R
,(\eta)}_  
{F,S, T_{1}, T_{2}  
}  
=
\begin{vmatrix}  
\left(\frac{1}{z_{b}- z_{f} } 
\right)_{  
\genfrac{}{}{0pt}{}{b\in B_{1}, }{f \in F} } 
&   
\left(
z_{b }^{s-1} 
\right)_{  
\genfrac{}{}{0pt}{}{b\in B_{1}, }{s \in S} }   
&   
\left(
z_{b }^{t-1} 
\right)_{  
\genfrac{}{}{0pt}{}{b\in B_{1}, }{t \in T_{1}} }   
& (0)_{|B_{1}| \times |T_{2}| }   
  \\[6pt]  
\left( 
\frac{ \left(-\frac{z_{f}}{z_{b}} \right)^{\eta } }{z_{b}-z_{f} }
\right)_{ 
\genfrac{}{}{0pt}{}{b \in B_{2}, }{f \in F} }  
&   
\left(
z_{b }^{s-1} 
\right)_{  
\genfrac{}{}{0pt}{}{b\in B_{2}, }{s \in S} }   
& (0)_{|B_{2}|\times |T_{1}|}   
&   
\left(
z_{b }^{t-1} 
\right)_{  
\genfrac{}{}{0pt}{}{b\in B_{2}, }{t \in T_{2}} }  
   \\[6pt]  
\left(
(-z_{f})^{r-1} 
\right)_{  
\genfrac{}{}{0pt}{}{r \in R, }{f \in F} }  
& (0)_{|R| \times |S| }   
& (0)_{|R| \times |T_{1}| }  
& (0)_{|R| \times |T_{2}| }\\  
\end{vmatrix}  
,  \label{sparcedetch}
\end{align}  
where the number of the elements of the sets
\footnote{$|S|=\mathrm{Card}(S)$ is the number of elements of the set $S$.}
 must satisfy 
$|B_{1}|+|B_{2}|+|R|=|F| + |S|+|T_{1}|+|T_{2}| $. 
This is a minor determinant of an infinite size matrix. 
Let us introduce a notation:
\begin{align}
\langle a,b \rangle =
\begin{cases}
(a,a+1,a+2, \dots, b ) & \text{for} \quad b -a \in {\mathbb Z}_{\ge 0}, \\[3pt]
\emptyset & \text{for}  \quad  b -a  \notin {\mathbb Z}_{\ge 0} .
\end{cases}
\end{align}
The denominator formula of the supercharacter of $gl(m|n)$ can be written as
 \cite{MV03}: 
\begin{align}
\Ds(B_{1},B_{2}|F_{1},F_{2})& = \Ds( B|F)= \nonumber \\
&= 
\varDelta^{B_{1},B_{2}, \langle 1, n-m  \rangle ,(0)}_  
{F, \langle  1, m-n \rangle,  \emptyset,\emptyset }
=
\frac{\prod_{ b,b^{\prime } \in B_{1} \sqcup B_{2}, \atop b \prec b^{\prime } }(z_{b} - z_{b^{\prime}})
\prod_{f,f^{\prime } \in F, \atop f \prec f^{\prime } }(z_{f^{\prime} } - z_{f})
}
{\prod_{ (b,f) \in (B_{1} \sqcup B_{2}) \times F } (z_{b} - z_{f})}, 
\end{align}
where we used a notation for the product: 
$\prod_{ b,b^{\prime } \in (b_{1},b_{2},\dots, b_{m}), \atop b \prec b^{\prime } }(z_{b} - z_{b^{\prime}})
=
\prod_{ 1 \le i<j \le m }(z_{b_{i}} - z_{b_{j}})$. 
Then we find the following new determinant expression of the supercharacter 
(for $(a,s)$ in the generalized T-hook. 
cf.\ Figure \ref{T-hook-sparse}, \ref{T-hook-sparse2})
\footnote{As far as discussions on the solution of the Q-system is concerned, the numbers $(n_{1},n_{2})$ 
are not very important if $n=n_{1}+n_{2}$ is fixed. To change $(n_{1},n_{2})$ for a fixed $n$ corresponds to 
change $s=0$ axis on the $(a,s)$ plain. The same remark can be applied to the solutions of the 
T-system in the next section.}.   
\begin{align}  
\chi_{a,s} &=   
(-1)^{(s+\eta_{1})(s+n_{2}) +\theta } 
\frac{
\varDelta^{B_{1},B_{2},\emptyset,( a-s-\eta_{1})}_  
{F,\emptyset,\langle 1, s+\eta_{1} \rangle ,  
\langle s-a+\eta_{1}+1, -a+\eta_{1}+\eta_{2} \rangle }  
}{\Ds(B_{1},B_{2}|F_{1},F_{2})}  
\nonumber  \\  
& \qquad \text{for} \quad a\ge \max\{s+\eta_{1}, -s+\eta_{2}, 0 \},  
\quad   
-\eta_{1} \le s \le \eta_{2},  
\label{sparsech1}  
\\[6pt]
%%%% 
\chi_{a,s} &=   
(-1)^{(s+\eta_{1})(s+n_{2}) +\theta }   
\frac{ 
\varDelta^{B_{1},B_{2}, \langle  a-s-\eta_{1}+1, a-\eta_{1}-\eta_{2} \rangle ,  
(a-s-\eta_{1})}_  
{F,\emptyset, \langle 1, s+\eta_{1} \rangle ,\emptyset}  
}{\Ds(B_{1},B_{2}|F_{1},F_{2})}  
\nonumber  \\   
& \qquad \text{for} \quad a\ge \max\{s+\eta_{1}, -s+\eta_{2}, 0 \},  
\quad   
s \ge \max\{ -\eta_{1}, \eta_{2}\},  
\label{sparsech2}  
\\[6pt]
%%%%%
\chi_{a,s} &=   
(-1)^{(s+\eta_{1})m_{2} +\theta }   
\frac{
\varDelta^{B_{1},B_{2},  
\langle 1, -s-\eta_{1} \rangle ,  
(a-s-\eta_{1})}_  
{F,  
\emptyset, \emptyset, \langle s-a+\eta_{1}+1, -a+\eta_{1}+\eta_{2} \rangle }  
}{\Ds(B_{1},B_{2}|F_{1},F_{2})}  
\nonumber \\ 
 & \qquad \text{for} \quad a\ge \max\{s+\eta_{1}, -s+\eta_{2}, 0 \},  
\quad   
s \le \min\{ -\eta_{1}, \eta_{2}\},  
\label{sparsech3}  
\end{align}  
%%%%  
\begin{align}  
\chi_{a,s} & =   
(-1)^{(s+\eta_{1})m_{2} +\theta }
\frac{
\varDelta^{B_{1},B_{2},  
(1,2,\dots,-s-\eta_{1},a-s-\eta_{1}+1,a-s-\eta_{1}+2,\dots, a-\eta_{1}-\eta_{2} ),  
(a-s-\eta_{1})}_  
{F,  
\emptyset, \emptyset,\emptyset}  
}{\Ds(B_{1},B_{2}|F_{1},F_{2})}  
\nonumber  \\ 
& \qquad \text{for} \quad a\ge \max\{s+\eta_{1}, -s+\eta_{2}, 0 \},  
\quad   
\eta_{2}  \le s \le  -\eta_{1},   
\label{sparsech4}  
\\[6pt]
%%%%%%%
\chi_{a,s} &=   
(-1)^{a m_{2} +\theta }
\frac{
\varDelta^{B_{1},B_{2},\emptyset ,  
(0)}_  
{F,  
\langle 1, -a+\eta_{1}+\eta_{2} \rangle , 
\langle s-a+\eta_{1}+1, s+\eta_{1} \rangle ,\emptyset}  
}{\Ds(B_{1},B_{2}|F_{1},F_{2})}  
\nonumber \\
&
\qquad \text{for} \quad a\le \min \{ s+\eta_{1}, \eta_{1}+\eta_{2} \},   
\label{sparsech5}  
\\[6pt]
%%%% 
\chi_{a,s} &=   
(-1)^{a(\eta_{1}+n_{2}+1) +\theta }  
\frac{
\varDelta^{B_{1},B_{2},\langle 1, a-\eta_{1}-\eta_{2} \rangle ,  
(0)}_  
{F,  
\emptyset, \langle s-a+\eta_{1}+1, s+\eta_{1} \rangle ,\emptyset}  
}{\Ds(B_{1},B_{2}|F_{1},F_{2})}  
\nonumber \\
&
\qquad \text{for} \quad \eta_{1}+\eta_{2} \le a \le s+\eta_{1},   
\label{sparsech6}  
\\[6pt] 
%%%%%%%%%%%%%%%%%%%%%%%%%%%%%%%%%%%%    
\chi_{a,s} &=   
(-1)^{a(1+m_{2}) +\theta }   
\frac{
\varDelta^{B_{1},B_{2},\emptyset ,  
(0)}_  
{F,  
\langle 1, -a+\eta_{1}+\eta_{2}\rangle , 
\emptyset, 
\langle s-a+\eta_{1}+1, s+\eta_{1}\rangle }  
}{\Ds(B_{1},B_{2}|F_{1},F_{2})}  
\nonumber \\
&
\qquad \text{for} \quad a\le \min \{ -s+\eta_{2}, \eta_{1}+\eta_{2} \},   
\label{sparsech7}  
\\[6pt]
%%%%%%%%%%%%%%%%%%%%%%%%%%%%%%%%%%%%  
\chi_{a,s} &=   
(-1)^{a(\eta_{1}+n_{2}) +\theta }
\frac{
\varDelta^{B_{1},B_{2}, \langle  1, a-\eta_{1}-\eta_{2} \rangle  ,  
(0)}_  
{F,  
\emptyset,\emptyset, 
\langle s-a+\eta_{1}+1, s+\eta_{1} \rangle  }  
}{\Ds(B_{1},B_{2}|F_{1},F_{2})}  
\nonumber \\
& 
\qquad \text{for} \quad \eta_{1}+\eta_{2} \le a \le -s+\eta_{2} ,
\label{sparsech8}  
\end{align}  
where we introduced  symbols 
$\eta_{1}=|B_{1}|-|F_{1}|=m_{1}-n_{1},   
\eta_{2}=|B_{2}|-|F_{2}|=m_{2}-n_{2}$ and 
\begin{align}
\theta :=
\frac{(|B_{1}|+|B_{2}|)(|B_{1}|+|B_{2}|-1)}{2}+\frac{|F|(|F|-1)}{2}. \label{sgnsp}
\end{align}
%%%%%%%%%%%%%%%%%%%%%%%%%
%%%%%%%%%%%%
%%%%%%%%%%%%%%%%%%%%%%%%%
%%%%%%%%%%%%%%%%%
If one changes the order of elements of the tuples $B_{1}$, $B_{2}$, $F_{1}$, $F_{2}$, 
 signs of the denominator and the numerator of these formulas change. But 
they cancel one another, and thus the sign of the formulas do not change in total. 
In this sense, we can treat these tuples in $\chi_{a,s}$ just as sets. 
For $F_{2}=B_{2}=\emptyset $, 
these determinants reduce to determinant expressions for 
character formulas of finite dimensional representations of $gl(m_{1}|n_{1})$ \cite{MV03}. 
The above formulas \eqref{sparsech1}-\eqref{sparsech8} 
take the following values at the boundary of the T-hook: 
\begin{align} 
 & \chi_{a,n_{1}} =   
\left( \frac{\prod_{b \in B_{2}}z_{b} }{\prod_{f \in F}(-z_{f}) } \right)^{m_{1}-a} 
\frac{\prod_{(b,f) \in B_{1} \times F } (z_{b}-z_{f}) }
{\prod_{(b^{\prime },b) \in B_{2} \times B_{1}  } (z_{b^{\prime}}-z_{b}) } 
\qquad  \text{for} \quad a \ge  m_{1},  
\label{boundchUR}  
\\[6pt]
& \chi_{a,-n_{2}}  =   
\frac{ \prod_{b \in B_{2}}z_{b}^{\eta_{1}-a-n_{2}} 
\prod_{(b,f) \in B_{2} \times F } (z_{b}-z_{f}) }
{\prod_{(b^{\prime },b) \in B_{2} \times B_{1}  } (z_{b^{\prime}}-z_{b}) } 
 \qquad \text{for} \quad a\ge  m_{2},  
\label{boundchUL}  
\\[6pt] 
& \chi_{m_{1},s} =   
\frac{ \prod_{b \in B_{1}}z_{b}^{s-n_{1}} 
\prod_{(b,f) \in B_{1} \times F } (z_{b}-z_{f}) }
{\prod_{(b^{\prime },b) \in B_{2} \times B_{1}  } (z_{b^{\prime}}-z_{b}) }  
\quad  \text{for} \quad s\ge  n_{1},  
\label{boundchR}  
\\[6pt]  
&   
\chi_{m_{2},s}   =
\frac{ \prod_{b \in B_{2}}z_{b}^{s-m_{2}+\eta_{1}} 
\prod_{(b,f) \in B_{2} \times F } (z_{b}-z_{f}) }
{\prod_{(b^{\prime },b) \in B_{2} \times B_{1}  } (z_{b^{\prime}}-z_{b}) } 
\qquad 
  \text{for} \quad s\le  -n_{2}, \label{boundchL}  
\\[6pt]  
& \chi_{0,s} =   1 
\qquad  \text{for} \quad s \in {\mathbb Z}, 
\label{boundchD}  
\end{align}  
%%%%%%%%%%%%%%%%%%%%%%%%%
Then we arrived at 
\begin{theorem}
Let $\chi_{a,s} $ be defined by \eqref{sparsech1}-\eqref{sparsech8} 
and \eqref{boundTch}. Then, $\chi_{a,s} $ solves the Q-system 
\eqref{Q-system} with the boundary conditions \eqref{dualT-hook} and \eqref{boundchUR}-\eqref{boundchD}. 
\end{theorem}
A proof of this theorem will be given in the case of the T-functions in Appendix A. 
%%%%%%%%%%%%%%%%%%%%

\begin{figure}
  \begin{center}
    \setlength{\unitlength}{1.5pt}
    \begin{picture}(100,100) 
      {\thicklines \put(10,100){\vector(0,-1){100}}}
      {\thicklines \put(10,50){\vector(1,0){90}}}
      {\thicklines \put(40,20){\line(1,0){60}}}
       {\thicklines \put(40,20){\line(0,-1){20}}}
       {\thicklines \put(50,70){\line(1,0){50}}}
       {\thicklines \put(50,70){\line(0,1){30}}}
      {\linethickness{0.2pt} \put(10,20){\line(1,0){30}}}
      {\linethickness{0.2pt} \put(10,70){\line(1,0){40}}}
      {\linethickness{0.2pt} \put(50,50){\line(0,1){20}}}
      {\linethickness{0.2pt} \put(40,50){\line(0,-1){30}}}
       \put(8,-5){$a$}
       \put(102,48){$s$}
       \put(-1,18){$m_{1}$}
       \put(-7,68){$-m_{2}$}
       \put(38,52){$n_{1}$}
       \put(48,45){$n_{2}$}
       \put(4,48){$0$}
    \end{picture}
  \end{center}
  \caption{ 
A rotated generalized T-hook for the supercharacters (section 2.2) and T-functions (section 3.6): 
${\check \chi}_{a,s}^{B_{1},B_{2},F_{1},F_{2}}=
{\check \Tb}_{a,s}^{B_{1},B_{2},F_{1},F_{2}}=\check{\overline{ \Tb}}_{a,s}^{B_{1},B_{2},F_{1},F_{2}}=0$ if $\{s<0\}$ or $\{a>m_{1},s>n_{1}\}$ or $\{a<-m_{2},s>n_{2}\}$.}
  \label{rot-T-hook}
\end{figure}
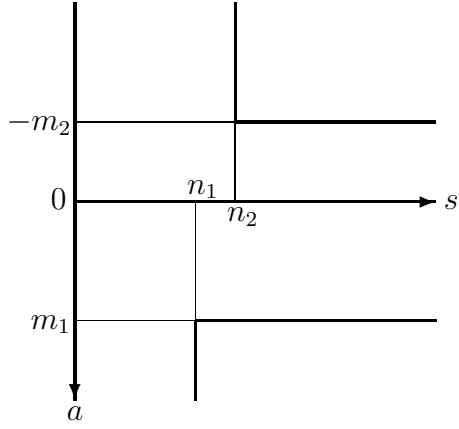

In contrast to  $w(t)$ \eqref{expan-inf}, 
the function $w(-t)^{-1}$ generates the supercharacters of the anti-symmetric tensor representations if 
we expand it with respect to non-negative power of $t$. 
Then we can repeat a similar calculation as 
\eqref{expan-inf}-\eqref{Jacobi-Trudi} for $w(-t)^{-1}$. 
\begin{align}
w(-t)^{-1}&=
(-1)^{n_{2}}t^{m_{2}-n_{2}}
\left(
\frac{\prod_{ b \in B_{2}} z_{b} }{\prod_{f \in F_{2}} (-z_{f}) }
\right)
\frac{\prod_{b \in B_{1}}(1+z_{b}t) \prod_{b \in B_{2}}(1+z_{b}^{-1}t^{-1}) }
{\prod_{f \in F_{1}}(1+z_{f}t) \prod_{f \in F_{2}}(1+z_{f}^{-1}t^{-1}) }
\nonumber \\
&=
(-1)^{n_{2}} 
\sum_{k \in {\mathbb Z}} \check{\chi}_{(1^{k})}^{B_{1},B_{2},F_{1},F_{2}} 
t^{k+m_{2}-n_{2}},
 \label{expan-inf-du}
\end{align} 
where we have expanded each factor as follows
\begin{align}
\frac{\prod_{b \in B_{1}}(1+z_{b}t) }
{\prod_{f \in F_{1}}(1+z_{f}t)  }
&=
\sum_{k=0}^{\infty }\chi_{(1^{k})}^{B_{1}, \emptyset ,F_{1},\emptyset } t^k ,
 \label{gf-right-du}
\\[6pt]
\left(
\frac{ \prod_{ b \in B_{2}} z_{b} }{ \prod_{f \in F_{2}} (-z_{f}) }
\right)
\frac{\prod_{b \in B_{2}}(1+z_{b}^{-1}t^{-1}) }
{ \prod_{f \in F_{2}}(1+z_{f}^{-1}t^{-1}) }
&=
\sum_{k=0}^{\infty} \chi_{(1^{-k})}^{\emptyset ,B_{2} , \emptyset ,F_{2} }
t^{-k},
\label{gf-left-du}
\end{align} 
and  $ \chi_{(1^{-k})}^{B_{1}, \emptyset ,F_{1},\emptyset } =\chi_{(1^{k})}^{\emptyset ,B_{2} , \emptyset ,F_{2} }=0$ 
for $k > 0 $.  
Here we formally extend the Young diagram $(1^k)$ for $k\ge 0$ to 
 negative direction ($k<0$) with respect to $s=0$ axis in $(a,s)$-plane. 
Then the coefficient of $t^{a+m_{2}-n_{2}}$ for $a \in {\mathbb Z} $ in \eqref{expan-inf-du} is given as follows: 
\begin{align}
\check{\chi}_{(1^a)}^{B_{1},B_{2},F_{1},F_{2}} &=
\sum_{k=\max(0,-a)}^{\infty} 
\chi_{(1^{a+k})}^{B_{1},\emptyset ,F_{1},\emptyset }
\chi_{(1^{-k})}^{\emptyset ,B_{2} , \emptyset ,F_{2} }. 
 \label{ch-infsym-du}
\end{align}
We define more general function $\check{\chi}_{a,s}^{B_{1},B_{2},F_{1},F_{2}}$ by 
substituting \eqref{ch-infsym-du} into the Jacobi-Trudi type formula (which is `dual' to \eqref{Jacobi-Trudi}): 
\begin{align}
\check{\chi}_{a,s}^{B_{1},B_{2},F_{1},F_{2}} =
\det_{1 \le i,j \le s}(\check{\chi}_{(1^{a-i+j})}^{B_{1},B_{2},F_{1},F_{2}} ), 
 \label{Jacobi-Trudi-du}
\end{align}
where $s \in {\mathbb Z}_{\ge 0}$, and $\check{\chi}_{a,s}^{B_{1},B_{2},F_{1},F_{2}} :=0$ for $s<0$. 
Note that the function $\check{\chi}_{a,s}^{B_{1},B_{2},F_{1},F_{2}} $ 
satisfies Hirota equation defined on the 90 degree rotated T-hook (see Figure \ref{rot-T-hook}). 
%%%%%%%%%%%%%%%%%%%%%%%%%%%%%%%%%%%%%%%%%%%%%%%%%%%
%%%%%%%%%%%%%%%%%
%%%%%%%%%%%%%%%%%%%%%%%%%%%%%%%%%%%%%%%%%%%%%%%%%%%%%%%%%%%%%%
\subsection{B\"{a}cklund transformations for the Q-system}
There are functional relations among supercharacters for 
subalgebras. 
These are a kind of B\"{a}cklund transformations 
in the soliton theory, and connect  supercharacters of $gl(M|N)$ to the trivial ones for $gl(0|0)$. 
We find the determinant formulas \eqref{sparsech1}-\eqref{sparsech8} satisfy the following 
B\"{a}cklund transformations.  
For $a,s \in {\mathbb Z}$, $b \in B_{1}$, $f \in F_{1}$,  
$B_{1}^{\prime}:=B_{1}\setminus \{b \} $ and  $F_{1}^{\prime}:=F_{1}\setminus \{f \} $, 
the B\"{a}cklund transformations for the `right wing' are   
\begin{align}   
& \chi_{a+1,s}  
\chi^{B_{1}^{\prime}}_{a,s}-  
\chi_{a,s}  
\chi^{B_{1}^{\prime}}_{a+1,s}  
 =  z_{b}
\chi_{a+1,s-1}  
\chi^{B_{1}^{\prime}}_{a,s+1},   
\label{bacch1}   
\\[8pt]   
&  \chi_{a,s+1}  
\chi^{B_{1}^{\prime}}_{a,s}-  
\chi_{a,s}  
\chi^{B_{1}^{\prime}}_{a,s+1}  
  =  z_{b}
\chi_{a+1,s}  
\chi^{B_{1}^{\prime}}_{a-1,s+1},   
\label{bacch2}  
\\[8pt]   
%%%    
&  \chi^{F_{1}^{\prime}}_{a+1,s}  
\chi_{a,s}-  
\chi^{F_{1}^{\prime}}_{a,s}  
\chi_{a+1,s}  
  =    z_{f}
\chi^{F_{1}^{\prime}}_{a+1,s-1}  
\chi_{a,s+1},   
\label{bacch3}  
\\[8pt]   
& \chi^{F_{1}^{\prime}}_{a,s+1}  
\chi_{a,s}-  
\chi^{F_{1}^{\prime}}_{a,s}  
\chi_{a,s+1}  
 =  z_{f}
\chi^{F_{1}^{\prime}}_{a+1,s}  
\chi_{a-1,s+1}  .
\label{bacch4}   
\end{align}  
He we omitted the index sets whose size is unchanged: 
$ \chi^{B_{1}^{\prime}}_{a,s}=\chi^{B_{1}^{\prime},B_{2},F_{1}.F_{2}}_{a,s}   $, etc. 
We will use similar notations from now on. 
%%%%%%%%%%%%%%%%%%%%%%%%%%%  
For $b \in B_{2}$, $f \in F_{2}$,   
$B_{2}^{\prime}:=B_{2}\setminus \{b \} $ and  $F_{2}^{\prime}:=F_{2}\setminus \{f \} $, 
the B\"{a}cklund transformations for the `left wing' are   
\begin{align}    
& z_{b} \chi_{a+1,s}  
\chi^{B_{2}^{\prime}}_{a,s}  
-  
\chi_{a,s}  
\chi^{B_{2}^{\prime}}_{a+1,s}  
  = 
\chi_{a+1,s+1}  
\chi^{B_{2}^{\prime}}_{a,s-1},   
\label{bacchL1}   
\\[8pt]   
&  \chi_{a,s-1}  
\chi^{B_{2}^{\prime}}_{a,s}  
-  
\chi_{a,s}  
\chi^{B_{2}^{\prime}}_{a,s-1}  
  = 
\chi_{a+1,s}  
\chi^{B_{2}^{\prime}}_{a-1,s-1},   
\label{bacchL2}  
\\[8pt]   
%%%  
&  z_{f} \chi^{F_{2}^{\prime}}_{a+1,s}  
\chi_{a,s}  
+  
\chi^{F_{2}^{\prime}}_{a,s}  
\chi_{a+1,s}  
 = 
\chi^{F_{2}^{\prime}}_{a+1,s+1}  
\chi_{a,s-1},   
\label{bacchL3}  
\\[8pt]   
& \chi^{F_{2}^{\prime}}_{a,s-1}  
\chi_{a,s}  
-  
\chi^{F_{2}^{\prime}}_{a,s}  
\chi_{a,s-1}  
  =-
\chi^{F_{2}^{\prime}}_{a+1,s}  
\chi_{a-1,s-1}  
\label{bacchL4}  .
\end{align}  
The B\"{a}cklund transformations for the supercharacters \eqref{bacch1}-\eqref{bacch4} 
for $B_{2}=F_{2}=\emptyset $ were introduced in \cite{Zabrodin07}. 
One can check that the boundary conditions 
\eqref{boundTch} and \eqref{boundchUR}-\eqref{boundchD} 
are compatible with the above 
B\"{a}cklund transformations \eqref{bacch1}-\eqref{bacchL4}.
These equations can be written in a more symmetric form by the gauge 
transformation
\footnote{
This corresponds to consider supercharacters of 
 a central extension of $gl(m_{1}+m_{2}|n_{1}+n_{2})$. 
See Remark 3.2 in \cite{CKW09}.
} 
 \eqref{gaugeqsystem} 
for $g_{2}=\frac{ \prod_{b \in B_{2}} z_{b} }{ \prod_{ f \in F_{2}} (-z_{f}) }$ and $ g_{1}=g_{3}=g_{4}=1$. 
The shape of the equations for the right wing 
\eqref{bacch1}-\eqref{bacch4} is invariant under this gauge transformation, while 
equations for the left wing \eqref{bacchL1}-\eqref{bacchL4} become 
\begin{align}    
&  \widetilde{\chi}_{a+1,s}  
\widetilde{\chi}^{B_{2}^{\prime}}_{a,s}  
-  
\widetilde{\chi}_{a,s}  
\widetilde{\chi}^{B_{2}^{\prime}}_{a+1,s}  
  = z_{b}^{-1}
\widetilde{\chi}_{a+1,s+1}  
\widetilde{\chi}^{B_{2}^{\prime}}_{a,s-1},   
\label{bacchL1-2}   
\\[8pt]   
&  \widetilde{\chi}_{a,s-1}  
\widetilde{\chi}^{B_{2}^{\prime}}_{a,s}  
-  
\widetilde{\chi}_{a,s}  
\widetilde{\chi}^{B_{2}^{\prime}}_{a,s-1}  
  = z_{b}^{-1}
\widetilde{\chi}_{a+1,s}  
\widetilde{\chi}^{B_{2}^{\prime}}_{a-1,s-1},   
\label{bacchL2-2}  
\\[8pt]   
%%%  
&  \widetilde{\chi}^{F_{2}^{\prime}}_{a+1,s}  
\widetilde{\chi}_{a,s}  
-  
\widetilde{\chi}^{F_{2}^{\prime}}_{a,s}  
\widetilde{\chi}_{a+1,s}  
 = z_{f}^{-1}
\widetilde{\chi}^{F_{2}^{\prime}}_{a+1,s+1}  
\widetilde{\chi}_{a,s-1},   
\label{bacchL3-2}  
\\[8pt]   
& \widetilde{\chi}^{F_{2}^{\prime}}_{a,s-1}  
\widetilde{\chi}_{a,s}  
-  
\widetilde{\chi}^{F_{2}^{\prime}}_{a,s}  
\widetilde{\chi}_{a,s-1}  
  =z_{f}^{-1}
\widetilde{\chi}^{F_{2}^{\prime}}_{a+1,s}  
\widetilde{\chi}_{a-1,s-1}  
\label{bacchL4-2}  .
\end{align}  
One can see a `left and right symmetry' 
$(z_{b}, z_{f}, s\pm 1) \leftrightarrow (z_{b}^{-1}, z_{f}^{-1},s \mp 1) $ 
between \eqref{bacch1}-\eqref{bacch4} and \eqref{bacchL1-2}-\eqref{bacchL4-2}. 
%%%%
%%%%%%%%%%%%%%%%%%%
\subsection{Discrete transformations on the solutions}
Let us consider the following maps: 
\begin{align}
\sigma(\varepsilon_{i})& =
\begin{cases} -\varepsilon_{M+1-i} 
\quad \text{for} \quad i \in {\mathfrak B}, \\
-\varepsilon_{2M+N+1-i} 
\quad \text{for} \quad i \in {\mathfrak F}, 
\end{cases} 
\label{sigmadual}
\\[6pt]
\tau(\varepsilon_{i})& =-\varepsilon_{M+N+1-i} 
\quad \text{for} \quad i \in {\mathfrak I}.
\label{taudual}
\end{align}
Here $\tau $ changes the parity in the sense $ (\varepsilon_{\tau(i)} |\varepsilon_{\tau(j)}  )=-p_{i}\delta_{ij}$.
We assume that these maps can be lifted to the ones for character variables $z_{i}=e^{\varepsilon_{i}}$ 
($i \in {\mathfrak I}$): 
\begin{align}
\sigma(z_{i})& =
\begin{cases}
z_{M+1-i}^{-1} 
\quad \text{for} \quad i \in {\mathfrak B}, \\
z_{2M+N+1-i}^{-1} 
\quad \text{for} \quad i \in {\mathfrak F}, 
\end{cases}
\label{sigmach}
\\[6pt]
\tau(z_{i})& =z_{M+N+1-i}^{-1} 
\quad \text{for} \quad i \in {\mathfrak I}.
\label{tauch}
\end{align}
We also define 
$\sigma(t)=t^{-1}$ and $\tau(t)=-t^{-1}$ for $t \in {\mathbb C}$.
$\sigma $ is an automorphism of $gl(M|N)$ and in addition to $\sigma $, 
$\tau $ also becomes an automorphism if $M=N$. 
They correspond to ${\mathbb Z}_{2} $ automorphism of $gl(M|N)$ and 
${\mathbb Z}_{2} \times {\mathbb Z}_{2}$ automorphism of $gl(M|M)$ (see for example, \cite{FSS89}).
% for $M>2$.  

We also introduce a permutation group 
 over any subset $J$ of ${\mathfrak I}$ and denote it as $S(J)$. 
 We suppose $ \rho \in S(J)$ acts on $\varepsilon_{i} $ and $z_{i}$ 
 as $\rho ( \varepsilon_{i})=\varepsilon_{\rho(i)}$ and 
 $\rho (z_{i})=z_{\rho (i)}$ for any $i \in J$. 
 In particular, $S( {\mathfrak B}) \times S( {\mathfrak F}) $ 
 corresponds to the Weyl group symmetry of $gl(M|N)$.  
Our supercharacter  solution 
$\chi_{a,s}^{{\mathfrak B}_{1},{\mathfrak B}_{2}, {\mathfrak F}_{1},{\mathfrak F}_{2}  }$ is invariant under 
 $S( {\mathfrak B_{1}}) \times S( {\mathfrak B_{2}}) \times S( {\mathfrak F_{1}}) \times S( {\mathfrak F_{2}}) $. 
 It does not have full Weyl group symmetry. 
 
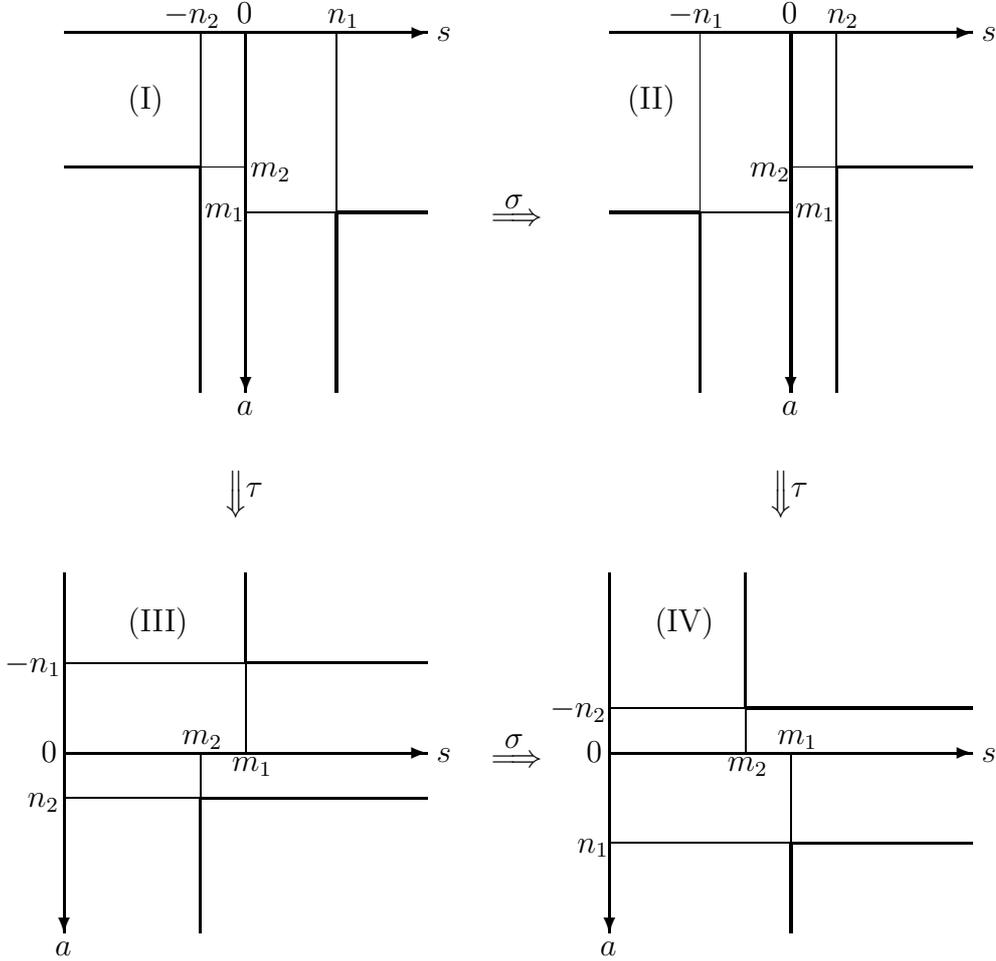
\begin{figure}
  \begin{center}
    \setlength{\unitlength}{1.7pt}
    \begin{picture}(200,200) 
      {\thicklines \put(0,200){\vector(1,0){80}}}
      {\thicklines \put(40,200){\vector(0,-1){80}}}
      {\thicklines \put(120,200){\vector(1,0){80}}}
      {\thicklines \put(160,200){\vector(0,-1){80}}}
      {\thicklines \put(0,40){\vector(1,0){80}}}
      {\thicklines \put(0,80){\vector(0,-1){80}}}
      {\thicklines \put(120,40){\vector(1,0){80}}}
      {\thicklines \put(120,80){\vector(0,-1){80}}}
      {\thicklines \put(30,170){\line(-1,0){30}}}
      {\thicklines \put(30,170){\line(0,-1){50}}}
      {\thicklines \put(60,160){\line(1,0){20}}}
      {\thicklines \put(60,160){\line(0,-1){40}}}
      {\thicklines \put(170,170){\line(1,0){30}}}
      {\thicklines \put(170,170){\line(0,-1){50}}}
      {\thicklines \put(140,160){\line(-1,0){20}}}
      {\thicklines \put(140,160){\line(0,-1){40}}}
      {\thicklines \put(40,60){\line(1,0){40}}}
      {\thicklines \put(40,60){\line(0,1){20}}}
      {\thicklines \put(30,30){\line(1,0){50}}}
      {\thicklines \put(30,30){\line(0,-1){30}}}
      {\thicklines \put(150,50){\line(1,0){50}}}
      {\thicklines \put(150,50){\line(0,1){30}}}
      {\thicklines \put(160,20){\line(1,0){40}}}
      {\thicklines \put(160,20){\line(0,-1){20}}}
      {\linethickness{0.2pt} \put(30,170){\line(1,0){10}}}
      {\linethickness{0.2pt} \put(30,170){\line(0,1){30}}}
      {\linethickness{0.2pt} \put(60,160){\line(-1,0){20}}}
      {\linethickness{0.2pt} \put(60,160){\line(0,1){40}}}
      {\linethickness{0.2pt} \put(170,170){\line(-1,0){10}}}
      {\linethickness{0.2pt} \put(170,170){\line(0,1){30}}}
      {\linethickness{0.2pt} \put(140,160){\line(1,0){20}}}
      {\linethickness{0.2pt} \put(140,160){\line(0,1){40}}}
      {\linethickness{0.2pt} \put(40,60){\line(-1,0){40}}}
      {\linethickness{0.2pt} \put(40,60){\line(0,-1){20}}}
      {\linethickness{0.2pt} \put(30,30){\line(-1,0){30}}}
      {\linethickness{0.2pt} \put(30,30){\line(0,1){10}}}
      {\linethickness{0.2pt} \put(150,50){\line(-1,0){30}}}
      {\linethickness{0.2pt} \put(150,50){\line(0,-1){10}}}
      {\linethickness{0.2pt} \put(160,20){\line(-1,0){40}}}
      {\linethickness{0.2pt} \put(160,20){\line(0,1){20}}}
       \put(97,41){$\sigma $}
       \put(94,37){$\Longrightarrow $}
       \put(97,161){$\sigma $}
       \put(94,157){$\Longrightarrow $}
       \put(36,103){\rotatebox{-90}{$\Longrightarrow $} } 
       \put(40,97){$\tau$}
       \put(156,103){\rotatebox{-90}{$\Longrightarrow $} } 
       \put(160,97){$\tau$}
       \put(-2,-5){$a$}
       \put(118,-5){$a$}
       \put(38,115){$a$}
       \put(158,115){$a$}
       \put(82,38){$s$}
       \put(202,38){$s$}
       \put(82,198){$s$}
       \put(202,198){$s$}
       \put(37,36){$m_{1}$}
       \put(157,42){$m_{1}$}
       \put(31,159){$m_{1}$}
        \put(161,159){$m_{1}$}
       \put(26,42){$m_{2}$}
       \put(146,36){$m_{2}$}
       \put(41,168){$m_{2}$}
       \put(151,168){$m_{2}$}
       \put(-13,58){$-n_{1}$}
       \put(113,18){$n_{1}$}
       \put(58,202){$n_{1}$}
       \put(133,202){$-n_{1}$}
       \put(-8,28){$n_{2}$}
       \put(107,48){$-n_{2}$}
       \put(22,202){$-n_{2}$}
       \put(168,202){$n_{2}$}
       \put(-5,38){$0$}
        \put(115,38){$0$}
        \put(38,202){$0$}
        \put(158,202){$0$}
        \put(14,183){(I)}
        \put(14,67){(III)}
        \put(124,183){(II)}
        \put(130,67){(IV)}
    \end{picture}
  \end{center}
  \caption{
Generalized T-hooks from discrete transformations for supercharacters (section 2.4) 
and T-functions (section 3.6): 
(I): $ \chi_{a,s}^{B_{1} ,B_{2} , F_{1} ,F_{2} }=\Tb_{a,s}^{B_{1} ,B_{2} , F_{1} ,F_{2} }=0$ 
if $\{a<0 \}$, or $\{a>m_{1},s>n_{1} \}$ or $ \{a>m_{2},s<-n_{2} \}$, 
(II): $\chi_{a,s}^{\hat{B}_{1} ,\hat{B}_{2} , \hat{F}_{1} ,\hat{F}_{2} }=
 \overline{\Tb}_{a,s}^{\hat{B}_{1} ,\hat{B}_{2} , \hat{F}_{1} ,\hat{F}_{2} }=0$ 
if $\{a<0 \}$, or $\{a>m_{2},s>n_{2} \}$ or $ \{a>m_{1},s<-n_{1} \}$,
(III): $ \check{\chi}_{a,s}^{\check{B}_{1} ,\check{B}_{2} , \check{F}_{1} ,\check{F}_{2} }=
\check{\Tb}_{a,s}^{\check{B}_{1} ,\check{B}_{2} , \check{F}_{1} ,\check{F}_{2} }=0$ 
if $\{s<0 \}$, or $\{s>m_{2},a>n_{2} \}$ or $ \{s>m_{1},a<-n_{1} \}$,
(IV): $  \check{\chi}_{a,s}^{\hat{\check{B}}_{1} ,\hat{\check{B}}_{2} , \hat{\check{F}}_{1} ,\hat{\check{F}}_{2} }=
\check{\overline{\Tb}}_{a,s}^{\hat{\check{B}}_{1} ,\hat{\check{B}}_{2} , \hat{\check{F}}_{1} ,\hat{\check{F}}_{2} }=0$ 
if $\{s<0 \}$, or $\{s>m_{1},a>n_{1} \}$ or $ \{s>m_{2},a<-n_{2} \}$.}
  \label{4-T-hooks}
\end{figure}

 The map $\sigma $ induces a reflection  of  the T-hook with respect to $s=0$ axis, and 
the map $\tau $ induces $90$ degree rotation and a reflection with respect to $a=0$ axis of the T-hook. 
Let us define $\hat{B}_{1}=\{ M+1-b_{i}^{(2)} \}_{i=1}^{m_{2}}$, 
$\hat{B}_{2}=\{ M+1-b_{i}^{(1)} \}_{i=1}^{m_{1}}$, 
$\hat{F}_{1}=\{ 2M+N+1-f_{i}^{(2)} \}_{i=1}^{n_{2}}$, 
$\hat{F}_{2}=\{ 2M+N+1-f_{i}^{(1)} \}_{i=1}^{n_{1}}$, 
$\hat{B}=\hat{B}_{1} \sqcup \hat{B}_{2} $, 
$\hat{F}=\hat{F}_{1} \sqcup \hat{F}_{2} $. 
Then from \eqref{sigmach}, \eqref{gf-right} and \eqref{gf-left}, 
we obtain
\begin{align}
\sigma (\chi_{(s)}^{B_{1} ,\emptyset , F_{1} ,\emptyset}) &=
\left(
\frac{ \prod_{b \in \hat{B}_{2}} z_{b} }{ \prod_{ f \in \hat{F}_{2}} (-z_{f}) }
\right)
 \chi_{(-s)}^{\emptyset ,\hat{B}_{2} , \emptyset ,\hat{F}_{2} }, \\[6pt]
\sigma (\chi_{(-s)}^{\emptyset, B_{2} ,\emptyset,  F_{2} }) &=
\left(
\frac{ \prod_{b \in \hat{B}_{1}} z_{b} }{ \prod_{ f \in \hat{F}_{1}} (-z_{f}) }
\right)
 \chi_{(s)}^{\hat{B}_{1},\emptyset ,\hat{F}_{1} , \emptyset  },
\end{align}
where $s \in {\mathbb Z}_{\ge 0}$. 
Therefore by \eqref{ch-infsym} and \eqref{Jacobi-Trudi}, we have 
\begin{align}
\sigma (\chi_{a,s}^{B_{1} ,B_{2} , F_{1} ,F_{2} }) =
 \left(
\frac{ \prod_{b \in \hat{B}} z_{b} }{ \prod_{ f \in \hat{F}} (-z_{f}) }
\right)^{a}
 \chi_{a,-s}^{\hat{B}_{1} ,\hat{B}_{2} , \hat{F}_{1} ,\hat{F}_{2} },
\end{align}
where $a \in {\mathbb Z}_{\ge 0}$ and $s \in {\mathbb Z}$. 
%%%
%
Let us define $\check{B}_{1}=\{ M+N+1-f_{i}^{(2)} \}_{i=1}^{n_{2}}$, 
$\check{B}_{2}=\{ M+N+1-f_{i}^{(1)} \}_{i=1}^{n_{1}}$, 
$\check{F}_{1}=\{ M+N+1-b_{i}^{(2)} \}_{i=1}^{m_{2}}$, 
$\check{F}_{2}=\{ M+N+1-b_{i}^{(1)} \}_{i=1}^{m_{1}}$, 
$\check{B}=\check{B}_{1} \sqcup \check{B}_{2} $, 
$\check{F}=\check{F}_{1} \sqcup \check{F}_{2} $. 
Then from \eqref{tauch}, \eqref{gf-right}, \eqref{gf-left}, 
\eqref{gf-right-du} and \eqref{gf-left-du}, 
we obtain
\begin{align}
\tau (\chi_{(s)}^{B_{1} ,\emptyset , F_{1} ,\emptyset}) &=(-1)^{s}
\left(
\frac{ \prod_{ f \in \check{F}_{2}} (-z_{f}) }{ \prod_{b \in \check{B}_{2}} z_{b} }
\right)
 \chi_{(1^{-s}) }^{\emptyset ,\check{B}_{2} , \emptyset ,\check{F}_{2} }, \\[6pt]
\tau (\chi_{(-s)}^{\emptyset, B_{2} ,\emptyset,  F_{2} }) &=(-1)^{s+\eta_{2}}
\left(
\frac{ \prod_{ f \in \check{F}_{1}} (-z_{f}) }{ \prod_{b \in \check{B}_{1}} z_{b} }
\right)
 \chi_{(1^{s})}^{\check{B}_{1},\emptyset ,\check{F}_{1} , \emptyset  },
\end{align}
where $s \in {\mathbb Z}_{\ge 0}$ and $\eta_{2}=m_{2}-n_{2}$. 
Therefore by \eqref{ch-infsym}, \eqref{Jacobi-Trudi}, \eqref{ch-infsym-du} and  \eqref{Jacobi-Trudi-du}, 
we have 
\begin{align}
\tau (\chi_{a,s}^{B_{1} ,B_{2} , F_{1} ,F_{2} }) =
(-1)^{(s+\eta_{2})a}
\left(
\frac{ \prod_{ f \in \check{F}} (-z_{f}) }{ \prod_{b \in \check{B}} z_{b} }
\right)^{a}
 \check{\chi}_{-s,a}^{\check{B}_{1} ,\check{B}_{2} , \check{F}_{1} ,\check{F}_{2} },
\end{align}
where $a \in {\mathbb Z}_{\ge 0}$ and $s \in {\mathbb Z}$. 
In the same way as above, we obtain 
\begin{align}
\sigma (\check{\chi}_{a,s}^{B_{1} ,B_{2} , F_{1} ,F_{2} }) =
 \left(
\frac{ \prod_{ f \in \hat{F}} (-z_{f}) }{ \prod_{b \in \hat{B}} z_{b} }
\right)^{s}
 \check{\chi}_{-a,s}^{\hat{B}_{1} ,\hat{B}_{2} , \hat{F}_{1} ,\hat{F}_{2} } 
\end{align}
and  
\begin{align}
\tau (\check{\chi}_{a,s}^{B_{1} ,B_{2} , F_{1} ,F_{2} }) =
(-1)^{(a+\eta_{2})s}
\left(
\frac{ \prod_{b \in \check{B}} z_{b} }{ \prod_{ f \in \check{F}} (-z_{f}) }
\right)^{s}
 \chi_{s,-a}^{\check{B}_{1} ,\check{B}_{2} , \check{F}_{1} ,\check{F}_{2} },
\end{align}
where $s \in {\mathbb Z}_{\ge 0}$ and $a \in {\mathbb Z}$. 
One can check that $\sigma $ and $\tau $ are involutions ($\sigma^{2}=\tau^{2}=1$). 
One can also check the commutativity $\sigma \tau =\tau \sigma $. Explicitly, we have 
\begin{align}
\sigma \tau (\chi_{a,s}^{B_{1} ,B_{2} , F_{1} ,F_{2} }) =\tau  \sigma (\chi_{a,s}^{B_{1} ,B_{2} , F_{1} ,F_{2} })=
(-1)^{(s+\eta_{2})a} 
 \check{\chi}_{s,a}^{\hat{\check{B}}_{1} ,\hat{\check{B}}_{2} , \hat{\check{F}}_{1} ,\hat{\check{F}}_{2} },
\end{align}
where $a \in {\mathbb Z}_{\ge 0}$ and $s \in {\mathbb Z}$ and 
$\hat{\check{B}}_{1}=\{f_{i}^{(1)} -M \}_{i=1}^{n_{1}}$, 
$\hat{\check{B}}_{2}=\{ f_{i}^{(2)}-M \}_{i=1}^{n_{2}}$, 
$\hat{\check{F}}_{1}=\{ b_{i}^{(1)} +N \}_{i=1}^{m_{1}}$, 
$\hat{\check{F}}_{2}=\{ b_{i}^{(2)} +N \}_{i=1}^{m_{2}}$. 
We also have 
\begin{align}
\sigma \tau (\check{\chi}_{a,s}^{B_{1} ,B_{2} , F_{1} ,F_{2} }) =\tau  \sigma (\check{\chi}_{a,s}^{B_{1} ,B_{2} , F_{1} ,F_{2} })=
(-1)^{(a+\eta_{2})s} 
 \chi_{s,a}^{\hat{\check{B}}_{1} ,\hat{\check{B}}_{2} , \hat{\check{F}}_{1} ,\hat{\check{F}}_{2} },
\end{align}
where $s \in {\mathbb Z}_{\ge 0}$ and $a \in {\mathbb Z}$. 
In this way, we obtain 4 type of solutions of the Q-system from the `seed' solution
\footnote{
For the full sets ($m_{1}=M_{1}, m_{2}=M_{2}, n_{1}=N_{1}, n_{2}=N_{2}$), we have 
$\hat{\mathfrak B}_{1}=\{i\}_{i=1}^{M_{2}}$, 
$\hat{\mathfrak B}_{2}=\{M_{2}+i\}_{i=1}^{M_{1}}$, 
$\hat{\mathfrak F}_{1}=\{M+i\}_{i=1}^{N_{2}}$, 
$\hat{\mathfrak F}_{2}=\{M+N_{2}+i\}_{i=1}^{N_{1}}$, 
$\check{\mathfrak B}_{1}=\{i\}_{i=1}^{N_{2}}$, 
$\check{\mathfrak B}_{2}=\{N_{2}+i\}_{i=1}^{N_{1}}$, 
$\check{\mathfrak F}_{1}=\{N+i\}_{i=1}^{M_{2}}$, 
$\check{\mathfrak F}_{2}=\{N+M_{2}+i\}_{i=1}^{M_{1}}$, 
$\hat{\check{\mathfrak B}}_{1}=\{i \}_{i=1}^{N_{1}}$, 
$\hat{\check{\mathfrak B}}_{2}=\{N_{1}+i \}_{i=1}^{N_{2}}$, 
$\hat{\check{\mathfrak F}}_{1}=\{N+i \}_{i=1}^{M_{1}}$, 
$\hat{\check{\mathfrak F}}_{2}=\{N+M_{1}+i \}_{i=1}^{M_{2}}$.
Note that 
$(\hat{\mathfrak B}_{1}, \hat{\mathfrak B}_{2}, \hat{\mathfrak F}_{1}, \hat{\mathfrak F}_{2})= 
({\mathfrak B}_{1}, {\mathfrak B}_{2},{\mathfrak F}_{1}, {\mathfrak F}_{2})$ if and only if 
$M_{1}=M_{2}$ and $N_{1}=N_{2}$; 
$(\check{\mathfrak B}_{1},\check{\mathfrak B}_{2}, \check{\mathfrak F}_{1}, \check{\mathfrak F}_{2})=
 ({\mathfrak B}_{1}, {\mathfrak B}_{2}, {\mathfrak F}_{1}, {\mathfrak F}_{2})$ if and only if 
$M_{1}=N_{2}$ and $M_{2}=N_{1}$; 
$(\hat{\check{\mathfrak B}}_{1}, \hat{\check{\mathfrak B}}_{2}, \hat{\check{\mathfrak F}}_{1},  \hat{\check{\mathfrak F}}_{2})
= ({\mathfrak B}_{1},{\mathfrak B}_{2}, {\mathfrak F}_{1}, {\mathfrak F}_{2})$ if and only if 
$M_{1}=N_{1}$ and $M_{2}=N_{2}$. 
Thus, all these coincide  if and only if $M_{1}=M_{2}=N_{1}=N_{2}$.
} 
$\chi_{a,s}=\chi_{a,s}^{{\mathfrak B}_{1},{\mathfrak B}_{2}, {\mathfrak F}_{1},{\mathfrak F}_{2}  }$ 
(see Figure \ref{4-T-hooks}). 
There is one to one correspondence among these solutions. 
Two of them $\{ \chi_{a,s}, \sigma (\chi_{a,s}) \}$ are solutions of the Q-system for 
$gl(M|N)$, which are `conjugate' one another, while the other two  $\{ \tau (\chi_{a,s}), \sigma \tau (\chi_{a,s}) \}$ 
are rather the ones for $gl(N|M)$. 
Only $M=N$ case, these are solutions for the same algebra  $gl(M|M)$. 
%%%%%%%%%%%%%%%%%%%%%%%%%%%%%%%%%%%%%%%%%%%%%%%%%%%
%%%%%%%%%%%%%%%%%
%%%%%%%%%%%%%%%%%%%%%%%%%%%%%%%%%%%%%%%%%%%%%%%%%%%%%%%
\section{T-system}
In this section, we propose Wronskian-like determinant solutions of the T-system for 
the generalized T-hook. Our formulas generalize formulas in our 
previous papers on $[M,N]$-hook for $gl(M|N)$ \cite{T09,BT08} and the T-hook for $gl(4|4)$ \cite{GKLT10}, 
and also a Wronskian-like determinant for the non-super  case $gl(M)$  \cite{KLWZ97} 
(see also \cite{BLZ97,Pronko-Stroganov00,BHK02,BDKM06,Kojima08,DM10}). 
%These also will unify various known results on Wronskian-like formulas in literatures 
% in the case of rectangular Young diagrams. 
To generalize the Wronskian-like formulas for the non-super  case $gl(M)$ in  \cite{KLWZ97} to 
the super case $gl(M|N)$  is a non-trivial task since matrix elements (Q-functions) of the determinants are 
non-trivially related one another by functional relations (QQ-relations) \eqref{QQb} and \eqref{QQf}. 
In addition, our results here and the ones in our previous papers \cite{T09,BT08} are conceptually 
closer to the one in \cite{BLZ97} than the one in \cite{KLWZ97}. 

In the context of the representation theory, 
our formulas will be examples of q-(super)characters \cite{FR99} 
(resp.\ Gelfand-Tsetlin (super)characters \cite{Knight95}) for infinite 
dimensional representations of $U_{q}(\widehat{gl}(M|N))$ (resp.\ $Y(gl(M|N))$). 
The q-character for finite dimensional representations of the quantum affine algebra  
 is a relatively well understood object. However, much is not known about the q-character for  infinite dimensional 
representations of the quantum affine algebra or 
 for even finite dimensional representations of the quantum affine superalgebra. 

Let us consider an arbitrary function $f(u)$ of  $u \in {\mathbb C}$ (the spectral parameter). 
Throughout this paper we use a notation on a shift of the spectral parameter such as  
$f^{[a]}=f(u+a \hbar)$ for an additive shift, and 
$f^{[a]}=f(uq^{a \hbar})$ for a multiplicative shift ($q$-difference), where 
$a \in {\mathbb Z}$. 
Here the unit of the shift $\hbar $ is any non-zero fixed complex number
\footnote{$ \hbar =\frac{i}{2}$ is often used for an additive shift in literatures.}. 
If there is no shift ($a=0$), we often omit $[0]$: $f^{[0]}=f=f(u)$. 
%%%%%%
\subsection{QQ-relations} 
\begin{figure}
  \begin{center}
    \setlength{\unitlength}{1.7pt}
    \begin{picture}(225,225)     
      {\linethickness{0.2pt} \put(7.6,193){\line(4,3){31}}}
       {\thicklines \put(45,194){\line(0,1){21}}}
       {\thicklines \put(82.4,193){\line(-4,3){31}}}
       {\thicklines \put(2.6,164){\line(0,1){21}}}
       {\linethickness{0.2pt} \put(7.6,163){\line(4,3){32}}}
       {\thicklines \put(40,163){\line(-4,3){31}}}
       {\linethickness{0.2pt} \put(50,163){\line(4,3){32}}}
       {\thicklines \put(82.4,163){\line(-4,3){31}}}
       {\thicklines \put(87.4,164){\line(0,1){21}}}
       {\thicklines \put(40,133){\line(-4,3){32}}}
       {\thicklines \put(45,134){\line(0,1){21}}}
       {\linethickness{0.2pt} \put(50,133){\line(4,3){32}}}
       \put(45,220){\makebox(1,0){$\Qb_{123}$}}
       \put(2.6,190){\makebox(1,0){$\Qb_{12}$}}
       \put(45,190){\makebox(1,0){$\Qb_{13}$}}
       \put(87,190){\makebox(1,0){$\Qb_{23}$}}
        \put(2.6,160){\makebox(1,0){$\Qb_{1}$}}
       \put(45,160){\makebox(1,0){$\Qb_{2}$}}
       \put(87,160){\makebox(1,0){$\Qb_{3}$}}
       \put(45,130){\makebox(1,0){$\Qb_{\emptyset }$}}
%%%%%%%%%%%%%%%%%%%%%%%%%%%%%%%%%%
       {\linethickness{0.2pt} \put(137.6,193){\line(4,3){31}}}
       {\thicklines \put(175,194){\line(0,1){21}}}
       {\thicklines \put(212.4,193){\line(-4,3){31}}}
       {\thicklines \put(132.6,164){\line(0,1){21}}}
       {\linethickness{0.2pt} \put(137.6,163){\line(4,3){32}}}
       {\thicklines \put(170,163){\line(-4,3){31}}}
       {\linethickness{0.2pt} \put(180,163){\line(4,3){32}}}
       {\thicklines \put(212.4,163){\line(-4,3){31}}}
       {\thicklines \put(217.4,164){\line(0,1){21}}}
       {\thicklines \put(169,133){\line(-4,3){31}}}
       {\thicklines \put(175,134){\line(0,1){21}}}
       {\linethickness{0.2pt} \put(180,133){\line(4,3){32}}}
       \put(175,220){\makebox(1,0){$\Qb_{\emptyset}$}}
       \put(132.6,190){\makebox(1,0){$\Qb_{3}$}}
       \put(175,190){\makebox(1,0){$\Qb_{1}$}}
       \put(217,190){\makebox(1,0){$\Qb_{2}$}}
        \put(132.6,160){\makebox(1,0){$\Qb_{13}$}}
       \put(175,160){\makebox(1,0){$\Qb_{23}$}}
       \put(217,160){\makebox(1,0){$\Qb_{12}$}}
       \put(175,130){\makebox(1,0){$\Qb_{123 }$}}
       %%%%%%%%%%%%%%%%%%%%  
       %%%%%%%%%%%%%%%%%%%%  
       {\thicklines \put(7.6,63){\line(4,3){31}}}
       {\linethickness{0.2pt}  \put(45,64){\line(0,1){21}}}
       {\linethickness{0.2pt}  \put(82.4,63){\line(-4,3){31}}}
       {\linethickness{0.2pt}  \put(2.6,34){\line(0,1){21}}}
       {\thicklines \put(7.6,33){\line(4,3){32}}}
       {\linethickness{0.2pt}  \put(40,33){\line(-4,3){31}}}
       {\thicklines \put(50,33){\line(4,3){32}}}
       {\linethickness{0.2pt}  \put(82.4,33){\line(-4,3){31}}}
       {\linethickness{0.2pt}  \put(87.4,34){\line(0,1){21}}}
       {\linethickness{0.2pt}  \put(40,3){\line(-4,3){32}}}
       {\linethickness{0.2pt}  \put(45,4){\line(0,1){21}}}
       {\thicklines \put(50,3){\line(4,3){32}}}
       \put(45,90){\makebox(1,0){$\Qb_{123}$}}
       \put(2.6,60){\makebox(1,0){$\Qb_{23}$}}
       \put(45,60){\makebox(1,0){$\Qb_{13}$}}
       \put(87,60){\makebox(1,0){$\Qb_{12}$}}
        \put(2.6,30){\makebox(1,0){$\Qb_{3}$}}
       \put(45,30){\makebox(1,0){$\Qb_{2}$}}
       \put(87,30){\makebox(1,0){$\Qb_{1}$}}
       \put(45,0){\makebox(1,0){$\Qb_{\emptyset }$}}
       %%%%%%%%%%%%%%%%%%%%    
       {\thicklines \put(137.6,63){\line(4,3){31}}}
       {\linethickness{0.2pt}  \put(175,64){\line(0,1){21}}}
       {\linethickness{0.2pt}  \put(212.4,63){\line(-4,3){31}}}
       {\linethickness{0.2pt}  \put(132.6,34){\line(0,1){21}}}
       {\thicklines \put(137.6,33){\line(4,3){32}}}
       {\linethickness{0.2pt}  \put(170,33){\line(-4,3){31}}}
       {\thicklines \put(180,33){\line(4,3){32}}}
       {\linethickness{0.2pt}  \put(212.4,33){\line(-4,3){31}}}
       {\linethickness{0.2pt}  \put(217.4,34){\line(0,1){21}}}
       {\linethickness{0.2pt}  \put(169,3){\line(-4,3){31}}}
       {\linethickness{0.2pt}  \put(175,4){\line(0,1){21}}}
       {\thicklines \put(180,3){\line(4,3){32}}}
       \put(175,90){\makebox(1,0){$\Qb_{\emptyset}$}}
       \put(132.6,60){\makebox(1,0){$\Qb_{1}$}}
       \put(175,60){\makebox(1,0){$\Qb_{3}$}}
       \put(217,60){\makebox(1,0){$\Qb_{2}$}}
        \put(132.6,30){\makebox(1,0){$\Qb_{13}$}}
       \put(175,30){\makebox(1,0){$\Qb_{12}$}}
       \put(217,30){\makebox(1,0){$\Qb_{23}$}}
       \put(175,0){\makebox(1,0){$\Qb_{123 }$}}
       %%%%%%%%%%%%%
       \put(108,47){$\sigma $}
       \put(105,43){$\Longrightarrow $}
       \put(108,177){$\sigma $}
       \put(105,173){$\Longrightarrow $}
       \put(42,113){\rotatebox{-90}{$\Longrightarrow $} } 
       \put(46,107){$\tau$}
       \put(172,113){\rotatebox{-90}{$\Longrightarrow $} } 
       \put(176,107){$\tau$}
       \put(0,85){(III)}
       \put(130,85){(IV)}
       \put(0,215){(I)}
       \put(130,215){(II)}
    \end{picture}
  \end{center}
  \caption{
Hasse diagrams for the Q-functions and discrete transformations on $gl(2|1) \sim gl(1|2)$:
$\tau$ changes the parity of the index sets. In (I) and (II),
$\{ 1,2 \} $ are bosonic and $\{ 3 \} $ is fermionic  ($gl(2|1)$), 
while in (III) and (IV), $\{ 1 \} $ is bosonic and
 $\{ 2,3 \} $ 
are fermionic  ($gl(1|2)$).
}
  \label{hasse}
\end{figure}
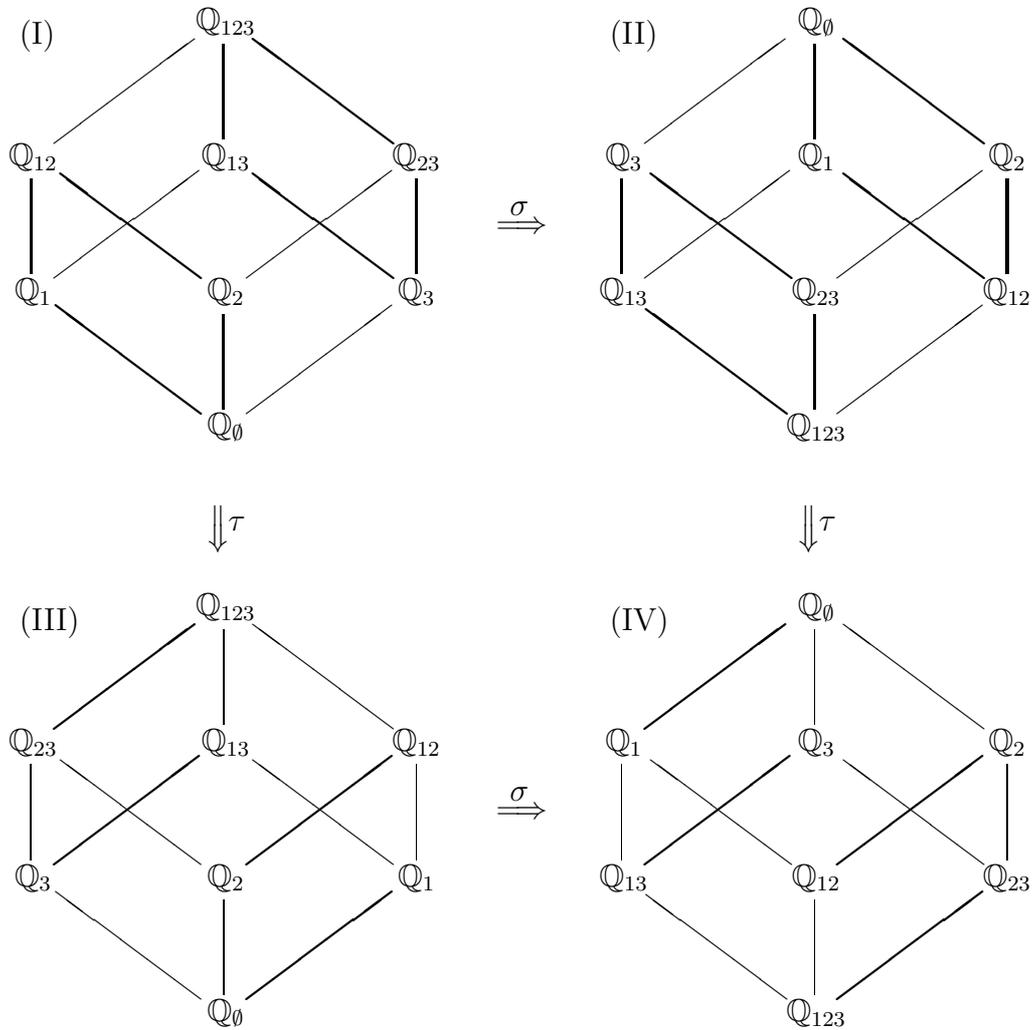

Let us consider complex functions 
$\Qb_{I}(u) $ of 
 $u \in {\mathbb C}$ (the spectral parameter), which are   
labeled by any subset $I$ of the full set ${\mathfrak I} $. 
We assume that these functions $\{ \Qb_{I}\}$ satisfy 
the following functional relations: 
\begin{align}
& (z_{i}-z_{j})\Qb_{I}\Qb_{I,ij}
=z_{i}\Qb_{I,i}^{[p_{i}]}
\Qb_{I,j}^{[-p_{i}]}-
z_{j}\Qb_{I,i}^{[-p_{i}]}
\Qb_{I,j}^{[p_{i}]}
\qquad 
\text{for} \qquad p_{i}=p_{j},   
\label{QQb}  \\[6pt]
& 
(z_{i}-z_{j})\Qb_{I,i}\Qb_{I,j}=
z_{i}\Qb_{I}^{[-p_{i}]}
\Qb_{I,ij}^{[p_{i}]}-
z_{j}\Qb_{I}^{[p_{i}]}
\Qb_{I,ij}^{[-p_{i}]}
\qquad \text{for} \qquad p_{i}=-p_{j}, 
\label{QQf} 
\end{align} 
where $i,j \in {\mathfrak I} \setminus I$ ($i\ne j$); 
$\{z_{i} \}_{i \in {\mathfrak I}}$ are complex parameters. 
Here we used an abbreviation $\Qb_{I,ij}=\Qb_{(i_{1},i_{2},\dots, i_{a},i,j)}$ 
for $I=(i_{1},i_{2},\dots, i_{a})$. In addition, 
we will not mind the order of the elements of the index set $I$ for $ \Qb_{I}$. 
For example, $\Qb_{(1,3,4)}=\Qb_{(4,3,1)}=\Qb_{(3,4,1)}=\Qb_{\{1,3,4\}}$. 
These functional relations (QQ-relations) are known as relations among 
the Baxter Q-operators or Q-functions for quantum integrable systems. 
Functional relations related to these can be seen, for example,
 in \cite{Pronko-Stroganov00,BHK02,GS03,BKSZ05,DDMST06,BDKM06,KSZ07,
Zabrodin07,BT08}. 
Here we used expressions based on index sets on Hasse diagram \cite{T09} (cf.\ Figure \ref{hasse}). 
The second equation \eqref{QQf}  is sometimes called `fermionic duality' in 
recent papers, which came from earlier papers on 
the `particle-hole transformation' in statistical physics
\footnote{It also appeared in the context of `Gauge/Bethe Correspondence' in \cite{OR10}.}
 \cite{Woynarovich83,T98}. 
And the parameters $\{z_{i} \}_{i \in {\mathfrak I}}$ correspond to the 
boundary twist of the transfer matrix in this context. 
These also correspond to the parameters for the supercharacters \eqref{formexp}. 
And in this paper, we will call the functions $\{\Qb_{I} \}$, the {\it Q-functions}
\footnote{The Baxter Q-functions have additional analytical structures on the 
spectral parameter such as polynomiality for some models. 
However in this paper, we will not assume such structures on 
  $\{\Qb_{I} \}$ and discuss functional relations among them, 
which are independent of detailed function form of  $\{\Qb_{I} \}$. 
Then the formulas in this paper are also valid for the Baxter Q-{\bf operators}. 
What is important for us is that 
 $\Qb_{I}(u) $, $\Qb_{J}(v) $ and the parameters $z_{i}$ commute each other for 
any $I,J \subset {\mathfrak I} $, $i \in {\mathfrak I}$ and 
$u,v \in {\mathbb C}$. }.  
As remarked in our previous paper \cite{T09}, there are $2^{M+N}$ Q-functions in total 
corresponding to the number of the choices of the subsets of the full set ${\mathfrak I}$. 
It is known that the shape of these equations is invariant under 
the following gauge transformation.
\begin{align}
%\Qb_{I}  \longrightarrow  
\widetilde{\Qb }_{I}=
g_{1}^{[\sum_{j \in I} p_{j}]}g_{2}^{[-\sum_{j \in I} p_{j}]} 
\Qb_{I}  ,
\label{gaugeQQ}
\end{align}
where $g_{1},g_{2}$ are arbitrary gauge functions of the spectral parameter. 
In this paper, we assume, without loss of generality, 
\begin{align}
\Qb_{\emptyset}=1. \label{Qnorm1}
\end{align}
Note that the $\{ \widetilde{\Qb_{I}} \}$ satisfies the QQ-relations with a generic 
$\widetilde{\Qb_{\emptyset}} $ if 
 the gauge functions are chosen as  
$g_{1}=\widetilde{\Qb}_{\emptyset} $, $g_{2}=1$ 
(or $g_{1}=1$, $g_{2}=\widetilde{\Qb} _{\emptyset} $) in 
\eqref{gaugeQQ}. 
Using the gauge freedom \eqref{gaugeQQ}, one may normalize 
\begin{align}
\Qb_{{\mathfrak I}}=1. \label{Qnorm2}
\end{align}
instead of \eqref{Qnorm1}. 
If both \eqref{Qnorm1} and \eqref{Qnorm2} are imposed, 
the Q-functions are considered to be in the normalization of the 
universal $R$-matrix. 
We have several determinant expressions of the 
solution of the QQ-relations. 
One of them is the following
\footnote{
The other determinant expressions based on `bosonization' 
or `fermionizatoin' trick (which correspond to changes of the basis) was proposed in 
Appendix A of \cite{GKLT10}.}: 
\begin{multline}  
\Qb_{I,B,F}  
=\frac{
\prod_{(b,f) \in B \times  F} (z_{b}-z_{f})
 } 
{
\prod_{b,b^{\prime} \in B, \atop 
b \prec b^{\prime} } (z_{b^{\prime}}-z_{b})
\prod_{f,f^{\prime} \in F, \atop 
f \prec f^{\prime} } (z_{f}-z_{f^{\prime}})
(\Qb_{I}^{[n-m]})^{n}
\prod_{k=1}^{m-n-1 }\Qb_{I}^{[n-m+2k ]} }   
\\ \times   
\begin{vmatrix}  
\left(
 \frac{ 
\Qb_{I,b,f}^{[n-m]}
}
{z_{b} -z_{f} }
\right)_{  
b \in B,  \atop 
f \in F} &   
\left(
z_{b}^{j-1}
\Qb_{I,b}^{[2j-1+n-m]}
\right)_{b \in B, \atop 1\le j \le m-n}   
  \\  
\end{vmatrix}   
\qquad \text{for} \quad  m \ge n, \label{detQQ1}  
\end{multline} 
%%%
\begin{multline}  
\Qb_{I,B,F}  
=\frac{
\prod_{(b,f) \in B \times  F} (z_{b}-z_{f})
 } 
{
\prod_{b,b^{\prime} \in B, \atop 
b \prec b^{\prime} } (z_{b^{\prime}}-z_{b})
\prod_{f,f^{\prime} \in F, \atop 
f \prec f^{\prime} } (z_{f}-z_{f^{\prime}})
(\Qb_{I}^{[n-m]})^{m}
\prod_{k=1}^{n-m-1 }\Qb_{I}^{[n-m-2k ]} }   
\\ \times   
\begin{vmatrix}  
\left(
 \frac{ 
\Qb_{I,b,f}^{[n-m]}
}
{z_{b} -z_{f} }
\right)_{  
b \in B,  \atop 
f \in F} 
\\ 
\left(
(-z_{f})^{j-1}
\Qb_{I,f}^{[-2j+1+n-m]}
\right)_{1\le j \le n-m, \atop f \in F}   
  \\  
\end{vmatrix}   
\qquad \text{for} \quad  m \le n, \label{detQQ2}  
\end{multline}   
where 
 $I \subset {\mathfrak I}$, $B \subset {\mathfrak B}$ and $F \subset {\mathfrak F}$ 
 ($m:=|B|$, $n:=|F|$) 
are mutually disjoint sets. 
Here we have to interpret the product 
as $\prod_{k=1}^{-1}f(k)=1/f(0)$,  $\prod_{k=1}^{0}f(k)=1$ 
for any complex function $f$ 
(the same remark should be applied for 
\eqref{Bglmmnn-4}-\eqref{Bglmmnn-5} and 
\eqref{Bglmmnnc-4}-\eqref{Bglmmnnc-5}). 
These determinant formulas \eqref{detQQ1}-\eqref{detQQ2} 
for $m,n \ge 1$ 
were introduced in  \cite{T09} for $I=\emptyset$ case and 
were generalized to $I \ne \emptyset$ case in \cite{GKLT10}. 
For any complement set $\overline{I}:={\mathfrak I} \setminus I$, we define 
$ \overline{\Qb}_{I}=\Qb_{\overline{I}}$. Note that $ \overline{\Qb}_{I}$ satisfy 
QQ-relations whose shift of the spectral parameter looks opposite  to the one for the original relations  
\eqref{QQb} and \eqref{QQf}: 
\begin{align}
& (z_{i}-z_{j})\overline{\Qb}_{I}\overline{\Qb}_{I,ij}
=z_{i}\overline{\Qb}_{I,i}^{[-p_{i}]}
\overline{\Qb}_{I,j}^{[p_{i}]}-
z_{j}\overline{\Qb}_{I,i}^{[p_{i}]}
\overline{\Qb}_{I,j}^{[-p_{i}]}
\qquad 
\text{for} \qquad p_{i}=p_{j},   
\label{barQQb}  \\[6pt]
& 
(z_{i}-z_{j})\overline{\Qb}_{I,i}\overline{\Qb}_{I,j}=
z_{i}\overline{\Qb}_{I}^{[p_{i}]}
\overline{\Qb}_{I,ij}^{[-p_{i}]}-
z_{j}\overline{\Qb}_{I}^{[-p_{i}]}
\overline{\Qb}_{I,ij}^{[p_{i}]}
\qquad \text{for} \qquad p_{i}=-p_{j}. 
\label{barQQf} 
\end{align} 
%%%
Now we want to lift the maps $\sigma$ \eqref{sigmadual} and $\tau$ \eqref{taudual} 
to the ones for the Q-functions so that the 
QQ-relations \eqref{Qnorm1} and \eqref{Qnorm2} are invariant under $\sigma$ and $\tau$.  
For any $ \{b_{i}\}_{i=1}^{m} 
\subset {\mathfrak B}, 
\{f_{i}\}_{i=1}^{n} \subset {\mathfrak F}$, 
we define 
\begin{align}
& \sigma ( \{b_{i}\}_{i=1}^{m}  \sqcup \{f_{i}\}_{i=1}^{n} )
={\mathfrak I} \setminus 
(\{ M+1-b_{i}\}_{i=1}^{m} \sqcup \{ 2M+N+1-f_{i} \}_{i=1}^{n} ), 
\label{sigma-set}
\\[6pt] 
& \tau (\{b_{i}\}_{i=1}^{m}  \sqcup \{f_{i}\}_{i=1}^{n} )
=
\{ M+N+1-f_{i} \}_{i=1}^{n} \sqcup  \{ M+N+1-b_{i}\}_{i=1}^{m}. 
\label{tau-set} 
\end{align} 
Then for $I=\{b_{i}\}_{i=1}^{m}  \sqcup \{f_{i}\}_{i=1}^{n} $, if we define
\begin{align}
\sigma (\Qb_{I}) 
= \Qb_{\sigma (I)} 
\label{sigmaQ},
\end{align} 
we find that the QQ-relations \eqref{QQb} and \eqref{QQf} 
are invariant 
 under \eqref{sigmach} and \eqref{sigmaQ}. 
 For simplicity, here we use the normalization \eqref{Qnorm1} and \eqref{Qnorm2} 
at the same time (or without using both of them). If not, 
one has to multiply with the Q-functions by the gauge functions (cf.\ \eqref{gaugeQQ}).
As for the map $\tau$, if we define 
\begin{align}
\tau (\Qb_{I}) 
= \Qb_{\tau (I)} 
\label{tauQ1},
\end{align} 
we find that the QQ-relations \eqref{QQb} and \eqref{QQf}  for $gl(M|N)$
are mapped to QQ-relations for $gl(N|M)$ 
  under \eqref{tauch} and \eqref{tauQ1}, and for $gl(M|M)$ ($M=N$) case,  
 they are invariant (see Figure \ref{hasse}). 
%%%%
%%%%%%%%%%%%%%%%%%%%%%%%%
%%%%%%%%%%%%%%%%%%%%%%%%%%%%%%%%%

The other option is to change \eqref{tau-set} as 
\begin{align}
\tau (\{b_{i}\}_{i=1}^{m}  \sqcup \{f_{i}\}_{i=1}^{n} )=
{\mathfrak I} \setminus 
( \{ M+N+1-f_{i} \}_{i=1}^{n} \sqcup  \{ M+N+1-b_{i}\}_{i=1}^{m})
\label{tau-set2} 
\end{align} 
and in stead of  \eqref{tauQ1}, to define for example 
\begin{align}
\tau (\Qb_{I}(u)) 
= \Qb_{\tau (I)}(-u)
\label{tauQ2}
\end{align} 
for an additive spectral parameter $u$, and 
\begin{align}
\tau (\Qb_{I}(x)) 
= \Qb_{\tau (I)}(x^{-1})
\label{tauQ3}
\end{align} 
for a multiplicative spectral parameter $x$. 
As we remarked for \eqref{barQQb} and \eqref{barQQf}, 
if we take a complement of the index set, the shift of the spectral parameter of 
the QQ-relations looks opposite to the original one. 
Then a manipulation to revert the shift of the spectral parameter is necessary 
to make the QQ-relation invariant
\footnote{for this, one may use complex conjugation if the unit of shift is a complex number} under $\tau $. 
A similar remark can be applied for $\sigma $. Namely, we may change 
\eqref{sigma-set} and \eqref{sigmaQ} as 
\begin{align}
& \sigma ( \{b_{i}\}_{i=1}^{m}  \sqcup \{f_{i}\}_{i=1}^{n} )
= \{ M+1-b_{i}\}_{i=1}^{m} \sqcup \{ 2M+N+1-f_{i} \}_{i=1}^{n} , 
\end{align}
\begin{align}
\sigma (\Qb_{I}(u)) 
= \Qb_{\sigma (I)}(-u)
\label{sigmaQ2}
\end{align} 
for an additive spectral parameter $u$, and 
\begin{align}
\sigma (\Qb_{I}(x)) 
= \Qb_{\sigma (I)}(x^{-1})
\label{sigmaQ3}
\end{align} 
for a multiplicative spectral parameter $x$. 
%%%%%%%%%%%%%%%%%%%%%%%%%%%%%%%%%%%%%%%%%%%%%%%%
\subsection{Baxterizing the supercharacter by Baxter Q-functions}
Let us introduce differential operators 
$d_{a}$ which evaluate the 
degrees of monomials on $\{z_{i}\}$. 
We will use these operators to produce shifts of the spectral 
parameter of the Baxter Q-functions:  
\begin{align}
 \Qb_{a}^{[p_{a} d_{a}]} 
 \cdot 
z_{1}^{k_{1}}z_{2}^{k_{2}}  \cdots z_{M+N}^{k_{M+N}}
&=
\Qb_{a}^{[p_{a}(2t_{a}\frac{\partial}{\partial t_{a}}+1)]}
t_{1}^{k_{1}} t_{2}^{k_{2}} \cdots t_{M+N}^{k_{M+N}}
|_{\{t_{i}=z_{i}\}_{i=1}^{M+N}}
\nonumber \\
&=
\Qb_{a}^{[(2k_{a}+1)p_{a}]}
z_{1}^{k_{1}} z_{2}^{k_{2}} \cdots z_{M+N}^{k_{M+N}}, 
 \label{Q-sh}
\end{align}
where $ k_{1},k_{2},\dots, k_{M+N} \in {\mathbb C}$. 
Consider an operator $ {\mathbf B^{I}}$ \cite{T09} which acts on any function  
$f(\{z_{i}\}) $ of $\{z_{i}\}_{i \in I}$: 
\begin{align}
{\mathbf B^{I}} 
\cdot  
f(\{z_{i}\})  =
\frac{1}{\Ds(I)}\prod_{a \in I}
\Qb_{a}^{[p_{a}d_{a}]}
\cdot 
\left[
\Ds(I)
f(\{z_{i}\}) 
\right] ,
 \label{baxterize}
\end{align} 
where $\Ds(I)=\Ds(I\cap {\mathfrak B} |I\cap {\mathfrak F})$. 
Our observation is that this operator produces 
T-functions (transfer matrices) or $q$-(super)characters when it acts on 
(super)characters. More specifically, it produces solutions of the T-system when it 
acts on the solutions of the Q-system. 
For example, 
the T-function for  the $(s+1)$ dimensional representation $V_{s}$ of $gl(2)$ can be obtained  
by acting ${\mathbf B^{I}} $ on the character 
$\chi_{(s)}=(z_{1}^{s+1}-z_{2}^{s+1})/(z_{1}-z_{2})$ as: 
\begin{align}
{\mathbf B}^{(1,2)} \cdot \chi_{(s)} &=
\frac{1}{z_{1}-z_{2}}
\Qb_{1}^{[d_{1}]}\Qb_{2}^{[d_{2}]} 
\cdot 
\left[
(z_{1}-z_{2})
 \chi_{(s)} 
\right] \nonumber \\
&=
\frac{1}{z_{1}-z_{2}}
\Qb_{1}^{[d_{1}]}\Qb_{2}^{[d_{2}]} 
\cdot 
(z_{1}^{s+1}-z_{2}^{s+1})
\nonumber \\
& =
\frac{z_{1}^{s+1} \Qb_{1}^{[2s+3]} \Qb_{2}^{[1]}-
z_{2}^{s+1} \Qb_{1}^{[1]} \Qb_{2}^{[2s+3]}}{z_{1}-z_{2}}
=\Tb_{s}.
\end{align}
$\Tb_{s}$ corresponds to the well known Wronskian-type formula on the 
T-operator in \cite{BLZ97} 
if $\Qb_{1}$ and $\Qb_{2}$ are the Baxter Q-operators for $U_{q}(\hat{sl_{2}})$ 
and $z_{1},z_{2}$ are boundary twist parameters with $z_{1}z_{2}=1$. 
One can apply the operator ${\mathbf B}^{I}$ 
to the supercharacter for the Verma modules, 
which leads
\begin{align}
\Tb_{\Lambda}^{+B,F}= 
\frac{1}{\Ds (B|F) }\prod_{i=1}^{m}z_{b_{i}}^{\Lambda_{b_{i}}+m-n-i}\Qb_{b_{i}}^{[2(\Lambda_{b_{i}}+m-n-i)+1]}
\prod_{i=1}^{n}(-z_{f_{i}})^{\Lambda_{f_{i}}+n-i}\Qb_{f_{i}}^{[-2(\Lambda_{f_{i}}+n-i)-1]}, 
 \label{T-Verma}
\end{align}
where $\Lambda$ is the highest weight of the form 
$\Lambda=\sum_{i=1}^{m}\Lambda_{b_{i}}\epsilon_{b_{i}}+\sum_{i=1}^{n}\Lambda_{f_{i}}\epsilon_{f_{i}}$ 
($\Lambda_{b_{i}} , \Lambda_{f_{i}} \in {\mathbb C}$).  
Most of the T-operators/functions can be given as summation
\footnote{As for the atypical representations of superalgebras, this summation can be 
infinite sum. In this case, one can use supercharacters of infinite dimensional representations which are 
smaller than Verma modules and obtain finite sum formulas. Examples can be seen in 
\eqref{soQQb-1}-\eqref{soQQb-3}.} over \eqref{T-Verma}. 
 In this way, (super)characters can be `Baxterized' by the Baxter Q-functions. 
 And obtained T-functions are `Q-deformation' of supercharacters. 
 The operator ${\mathbf B}^{I}$ works for both T-operators and T-functions 
in the sense that the definition of this operator does not depend on if 
$\Qb_{a}$ are operators or their eigenvalues. 
The way to Baxterize the supercharacter is not unique
\footnote{There are infinitely many ways to put spectral parameter into the supercharacters. 
However, we expect that the ways to produce solutions of the T-system from those of 
the Q-system will be classified up to the gauge transformations and trivial rescaling of the spectral parameter. 
 In this paper, we are only interested in the Baxterizations which produce solutions of the 
T-system. We speculate that the number of the right Baxterizations is related to the number of the 
automorphisms of the underlaying algebras. 
The notations bar and without bar are borrowed from notations for 
 two different kind of evaluation representations of a quantum affine algebra 
 (they came from automorphisms of the algebra) 
 in \cite{BHK02}. 
}. 
Instead of \eqref{Q-sh} and \eqref{baxterize}, we can use 
\begin{align}
\begin{split}
 \overline{\Qb}_{a}^{[-p_{a} d_{a}]} 
 \cdot 
z_{1}^{k_{1}}z_{2}^{k_{2}}  \cdots z_{M+N}^{k_{M+N}}
&=\overline{\Qb}_{a}^{[-(2k_{a}+1)p_{a}]}
z_{1}^{k_{1}} z_{2}^{k_{2}} \cdots z_{M+N}^{k_{M+N}},
\\[6pt]
\overline{{\mathbf B}}^{I} 
\cdot  
f(\{z_{i}\}) & =
\frac{1}{\Ds(I)}\prod_{a \in I}
\overline{\Qb}_{a}^{[-p_{a}d_{a}]}
\cdot 
\left[
\Ds(I) 
f(\{z_{i}\}) 
\right] .
\end{split}
\end{align} 
In this case, the formula looks nice if one uses the normalization \eqref{Qnorm2}. 
%${\mathbf B}^{I}$ can be applied for each term of the supercharacter seperatly. 
In the next section, we will Baxterize the solution of the Q-system in section 2 and 
obtain determinant solutions of the T-system. 
%%%%%%%%%%%%%%%%%%%%%%%%%%%%%%%%%
\subsection{Sparse expression of the determinants}  
Let us introduce a determinant  
labeled by the same tuples as \eqref{sparcedetch}: 
\begin{multline}  
\Delta^{B_{1},B_{2},R
,[\eta; \xi ]}_  
{F,S, T_{1}, T_{2}  
}  
=   
\begin{vmatrix}  
\left( \frac{\Qb_{b,f}^{[\xi ]}}{z_{b}- z_{f} } 
\right)_{  
\genfrac{}{}{0pt}{}{b\in B_{1}, }{f \in F} } 
&   
{\mathcal Q}_{B_{1}, S}^{[\xi]}   
&   
{\mathcal Q}_{B_{1},T_{1}}^{[\xi]}   
& (0)_{|B_{1}| \times |T_{2}| }   
  \\[6pt]  
\left( 
\frac{ \left(-\frac{z_{f}}{z_{b}} \right)^{\eta }
\Qb_{b,f}^{[\xi -2\eta]} }{z_{b}-z_{f} }
\right)_{ 
\genfrac{}{}{0pt}{}{b \in B_{2}, }{f \in F} }  
&   
{\mathcal Q}_{B_{2}, S}^{[\xi]}   
& (0)_{|B_{2}| \times |T_{1}| }   
&   
{\mathcal Q}_{B_{2}, T_{2} }^{[\xi]}  
   \\[6pt]  
\left(
(-z_{f})^{r-1}\Qb_{f}^{[\xi-2r+1 ]}
\right)_{  
\genfrac{}{}{0pt}{}{r \in R, }{f \in F} }  
& (0)_{|R|\times |S|}   
& (0)_{|R| \times |T_{1}| }  
& (0)_{|R| \times |T_{2}| }\\  
\end{vmatrix}  
,  \label{sparcedetbb}
\end{multline}  
where we used a notation for a matrix ${\mathcal Q}_{X, Y }^{[\xi]}= \left(
z_{b }^{s-1} \Qb_{b}^{[\xi+2s-1]}
\right)_{  
\genfrac{}{}{0pt}{}{b\in X, }{s \in Y} } $ and the number of the elements of the sets must satisfy 
$|B_{1}|+|B_{2}|+|R|=|F| + |S|+|T_{1}|+|T_{2}| $.
%where nonnegative integers $m_{1},m_{2},a,n,b,c_{1},c_{2}$ must satisfy  
%$m_{1}+m_{2}+a=n+b+c_{1}+c_{2}$.  
%%%%%%%%%%%%%%%%%%%%%%%%%%%%%%%%%%%%%  
Let us Baxterize the supercharacter formulas \eqref{sparsech1}-\eqref{sparsech8} 
by the operator ${\mathbf B}^{I} $ (for $I=B_{1} \sqcup B_{2} \sqcup  F_{1} \sqcup  F_{2}$, 
${\mathbf B}^{I} :={\mathbf B}^{B_{1},B_{2}, F_{1}, F_{2}}$). 
\begin{align}
 \Tb^{B_{1},B_{2}, F_{1}, F_{2}}_{a,s} := 
\left({\mathbf B}^{B_{1},B_{2}, F_{1}, F_{2}} 
\cdot 
\chi^{B_{1},B_{2}, F_{1}, F_{2}}_{a,s}
\right)^{[a-s-\eta_{1}-\eta_{2}]} , 
\qquad 
a,s \in {\mathbb Z},
 \label{baxt-t}
\end{align} 
where $\eta_{1}=m_{1}-n_{1}$, $\eta_{2}=m_{2}-n_{2}$ and an overall shift $(a-s-\eta_{1}-\eta_{2})$ on the spectral 
parameter is introduced just for a normalization
\footnote{For more general Young diagram $\lambda$, 
the T-function will be given by 
$  \Tb^{B_{1},B_{2}, F_{1}, F_{2}}_{\lambda } = 
{\mathbf B}^{B_{1},B_{2}, F_{1}, F_{2}} 
\cdot 
\chi^{B_{1},B_{2}, F_{1}, F_{2}}_{\lambda } $ up to the overall shift of the spectral parameter.}. 
From now on, we often abbreviate the index set dependence of the T-function:  
$\Tb_{a,s}=\Tb^{B_{1},B_{2}, F_{1}, F_{2}}_{a,s} $.
Explicitly, we obtain the following determinant formula (for $(a,s)$ in the generalized T-hook. 
cf.\ Figure \ref{T-hook-sparse},\ref{T-hook-sparse2})
\footnote{The T-function $\widetilde{\Tb}^{B_{1},B_{2}, F_{1}, F_{2}}_{a,s}$
in the normalization of the spectral parameter of our previous paper \cite{T09} 
corresponds to 
$\widetilde{\Tb}^{B_{1},B_{2}, F_{1}, F_{2}}_{a,s}=
\Tb^{B_{1},B_{2}, F_{1}, F_{2} [-\frac{\eta_{1}-\eta_{2}}{2}] }_{a,s}$ 
(for $B_{2}=F_{2}=\emptyset, \eta_{2}=0 $).}: 
\begin{align}  
\Tb_{a,s} &=   
(-1)^{(s+\eta_{1})(s+n_{2}) +\theta }  
\frac{
\Delta^{B_{1},B_{2},\emptyset,[a-s-\eta_{1} ;a-s-\eta_{1}-\eta_{2}]}_  
{F,\emptyset, \langle 1,s+\eta_{1} \rangle ,  
\langle s-a+\eta_{1}+1,-a+\eta_{1}+\eta_{2} \rangle }  
}{\Ds (B_{1},B_{2} | F_{1},F_{2})}
\nonumber  \\  
& \qquad \text{for} \quad a\ge \max\{s+\eta_{1}, -s+\eta_{2}, 0 \},  
\quad   
-\eta_{1} \le s \le \eta_{2},  
\label{sparse01}  
\\[6pt]
%%%% 
\Tb_{a,s} &=   
(-1)^{(s+\eta_{1})(s+n_{2})+\theta }    
\frac{
\Delta^{B_{1},B_{2}, \langle a-s-\eta_{1}+1, a-\eta_{1}-\eta_{2} \rangle,  
[a-s-\eta_{1};a-s-\eta_{1}-\eta_{2}]}_  
{F,\emptyset, \langle 1,s+\eta_{1} \rangle ,\emptyset}  
}{\Ds (B_{1},B_{2}|F_{1},F_{2})}
\nonumber  \\   
& \qquad \text{for} \quad a\ge \max\{s+\eta_{1}, -s+\eta_{2}, 0 \},  
\quad   
s \ge \max\{ -\eta_{1}, \eta_{2}\},  
\label{sparse02}  
\\[6pt]
%%%%%
\Tb_{a,s} &=   
(-1)^{(s+\eta_{1})m_{2}+\theta }   
\frac{
\Delta^{B_{1},B_{2},  
\langle 1,-s-\eta_{1} \rangle,  
[a-s-\eta_{1};a-s-\eta_{1}-\eta_{2}]}_  
{F,  
\emptyset, \emptyset, 
\langle s-a+\eta_{1}+1, -a+\eta_{1}+\eta_{2} \rangle}  
}{\Ds (B_{1},B_{2}|F_{1},F_{2})}
\nonumber \\ 
 & \qquad \text{for} \quad a\ge \max\{s+\eta_{1}, -s+\eta_{2}, 0 \},  
\quad   
s \le \min\{ -\eta_{1}, \eta_{2}\},  
\label{sparse03}  
\end{align}  
%%%%  
\begin{align}  
& \Tb_{a,s}  =  \nonumber \\
&= (-1)^{(s+\eta_{1})m_{2}+\theta }
\frac{
\Delta^{B_{1},B_{2},  
(1,2,\dots,-s-\eta_{1},a-s-\eta_{1}+1,a-s-\eta_{1}+2,\dots, a-\eta_{1}-\eta_{2}),  
[a-s-\eta_{1};a-s-\eta_{1}-\eta_{2}]}_  
{F,  
\emptyset, \emptyset,\emptyset}  
}{\Ds (B_{1},B_{2}|F_{1},F_{2})}
\nonumber  \\ 
& \qquad \text{for} \quad a\ge \max\{s+\eta_{1}, -s+\eta_{2}, 0 \},  
\quad   
\eta_{2}  \le s \le  -\eta_{1},   
\label{sparse04}  
\end{align}
%%%%%%%
\begin{align}
\Tb_{a,s} &=   (-1)^{m_{2}a+\theta }
\frac{
\Delta^{B_{1},B_{2},\emptyset ,  
[0;a-s-\eta_{1}-\eta_{2}]}_  
{F,  
\langle 1, -a+\eta_{1}+\eta_{2} \rangle, 
\langle s-a+\eta_{1}+1, s+\eta_{1}
\rangle,\emptyset}  
}{\Ds (B_{1},B_{2}|F_{1},F_{2})}
\nonumber \\
&
\qquad \text{for} \quad a\le \min \{ s+\eta_{1}, \eta_{1}+\eta_{2} \},   
\label{sparse05}  
\\[6pt]
%%%% 
\Tb_{a,s} &=   
(-1)^{a(\eta_{1}+n_{2}+1)+\theta }  
\frac{
\Delta^{B_{1},B_{2}, 
\langle 1, a-\eta_{1}-\eta_{2} \rangle ,  
[0;a-s-\eta_{1}-\eta_{2}]}_  
{F,  
\emptyset, 
\langle s-a+\eta_{1}+1, s+\eta_{1}
\rangle,\emptyset}  
}{\Ds (B_{1},B_{2}|F_{1},F_{2})}
\nonumber \\
&
\qquad \text{for} \quad \eta_{1}+\eta_{2} \le a \le s+\eta_{1},   
\label{sparse06}  
\\[6pt] 
%%%%%%%%%%%%%%%%%%%%%%%%%%%%%%%%%%%%    
\Tb_{a,s} &=   
(-1)^{a(m_{2}+1)+\theta }   
\frac{
\Delta^{B_{1},B_{2},\emptyset ,  
[0;a-s-\eta_{1}-\eta_{2}]}_  
{F,  
\langle 
1,-a+\eta_{1}+\eta_{2}
\rangle, 
\emptyset,
\langle s-a+\eta_{1}+1, s+\eta_{1} \rangle }  
}{\Ds (B_{1},B_{2}|F_{1},F_{2})}
\nonumber \\
&
\qquad \text{for} \quad a\le \min \{ -s+\eta_{2}, \eta_{1}+\eta_{2} \},   
\label{sparse07}  
\\[6pt]
%%%%%%%%%%%%%%%%%%%%%%%%%%%%%%%%%%%%  
\Tb_{a,s} &=   
(-1)^{a(\eta_{1}+n_{2})+\theta }  
\frac{
\Delta^{B_{1},B_{2},
\langle 1,a-\eta_{1}-\eta_{2} \rangle ,  
[0;a-s-\eta_{1}-\eta_{2}]}_  
{F,  
\emptyset,\emptyset, 
\langle s-a+\eta_{1}+1, s+\eta_{1} \rangle}  
}{\Ds (B_{1},B_{2}|F_{1},F_{2})}
\nonumber \\
& 
\qquad \text{for} \quad \eta_{1}+\eta_{2} \le a \le -s+\eta_{2},   
\label{sparse08}  
\end{align}  
where $\theta $ is defined in \eqref{sgnsp}. 
Due to \eqref{boundTch}, \eqref{baxt-t} naturally satisfies 
 the generalized T-hook  boundary condition: 
\begin{align}
\Tb_{a,s}=0
 \quad \text{if}  \quad 
\{a<0 \} \quad 
\text{or}  \quad 
\{a>m_{1},s>n_{1} \} \quad 
\text{or} \quad 
\{a>m_{2},s<-n_{2} \} .
\label{T-hookbound}
\end{align} 
At the boundaries of the T-hook, \eqref{sparse01}-\eqref{sparse08} 
take the following `boundary values':  
\begin{align} 
 & \Tb_{a,n_{1}} =   
\left( \frac{\prod_{b \in B_{2}}z_{b} }{\prod_{f \in F}(-z_{f}) } \right)^{m_{1}-a} 
\frac{\prod_{(b,f) \in B_{1} \times F } (z_{b}-z_{f}) }
{\prod_{(b^{\prime },b) \in B_{2} \times B_{1}  } (z_{b^{\prime}}-z_{b}) } 
\Qb^{[a-\eta_{2}]}_{B_{1}}  
\Qb^{[-a+\eta_{1}]}_{B_{2}, F_{1},F_{2}}  
  \nonumber \\
& 
\hspace{100pt}
\quad  \text{for} \quad a \ge  m_{1},  
\label{boundUR}  
\\[6pt]
& \Tb_{a,-n_{2}}  =   
\frac{ \prod_{b \in B_{2}}z_{b}^{\eta_{1}-a-n_{2}} 
\prod_{(b,f) \in B_{2} \times F } (z_{b}-z_{f}) }
{\prod_{(b^{\prime },b) \in B_{2} \times B_{1}  } (z_{b^{\prime}}-z_{b}) } 
\Qb^{[a-\eta_{2}]}_{B_{1}, F_{1}, F_{2}}  
\Qb^{[-a+\eta_{1}]}_{B_{2}}  
\nonumber \\
& 
\hspace{100pt}
 \text{for} \quad a\ge  m_{2},  
\label{boundUL}  
\\[6pt] 
& \Tb_{m_{1},s} =   
\frac{ \prod_{b \in B_{1}}z_{b}^{s-n_{1}} 
\prod_{(b,f) \in B_{1} \times F } (z_{b}-z_{f}) }
{\prod_{(b^{\prime },b) \in B_{2} \times B_{1}  } (z_{b^{\prime}}-z_{b}) } 
\Qb^{[s+\eta_{1}-\eta_{2}]}_{B_{1}}  
\Qb^{[-s]}_{  
B_{2}, F_{1}, F_{2}}   
\quad  \text{for} \quad s\ge  n_{1},  
\label{boundR}  
\\[6pt]  
&   
\Tb_{m_{2},s}   =
\frac{ \prod_{b \in B_{2}}z_{b}^{s-m_{2}+\eta_{1}} 
\prod_{(b,f) \in B_{2} \times F } (z_{b}-z_{f}) }
{\prod_{(b^{\prime },b) \in B_{2} \times B_{1}  } (z_{b^{\prime}}-z_{b}) } 
\Qb^{[s+\eta_{1}-\eta_{2}]}_{B_{2}}  
\Qb^{[-s]}_{  
B_{1}, F_{1},F_{2} }  
\nonumber 
\\ 
& \hspace{200pt}
  \text{for} \quad s\le  -n_{2}, \label{boundL}  
\\[6pt]  
& \Tb_{0,s} =    
\Qb^{[-s]}_{B_{1},  
B_{2}, F_{1}, F_{2}}  
\quad  \text{for} \quad s \in {\mathbb Z}, 
\label{boundD}  
\end{align}  
and satisfy 
\begin{align}
\begin{split}
\Tb_{m_{1},n_{1}+a}&=
\left( \frac{\prod_{b \in B}z_{b} }{\prod_{f \in F}(-z_{f}) } \right)^{a} 
\Tb_{m_{1}+a,n_{1}} ,
\\[6pt]
\Tb_{m_{2},-n_{2}-a}& =
\Tb_{m_{2}+a,-n_{2}} 
\quad \text{for} \quad 
a \in {\mathbb Z}_{\ge 0}. 
\end{split}
 \label{T=T}
\end{align} 
The  Baxterization of the Q-system \eqref{Q-system} is the following T-system 
(Hirota equation):
\begin{align}
  \Tb^{[-1]}_{a,s} \Tb^{[1]}_{a,s} 
  =
  \Tb_{a,s-1} \Tb_{a,s+1}+
  \Tb_{a-1,s} \Tb_{a+1,s} 
  \quad \text{for} \quad 
  a,s \in {\mathbb Z}.
  \label{hirota-BC}
\end{align}
In fact, we find
\begin{theorem}
Let $ \Tb_{a,s}$ be defined by \eqref{sparse01}-\eqref{T-hookbound}. 
Then  $ \Tb_{a,s}$ solves the T-system with the boundary conditions 
\eqref{boundUR}-\eqref{T=T}.
\end{theorem}
A proof of this theorem will be given in Appendix B, where the coefficient free form of 
the T- and Q-functions defined in section \ref{removetwist} and Appendix A will be used.  
The above `sparse' determinant solution is 
a natural generalization of the solution for the $[M,N]$-hook \cite{T09}. 
%%%%%%%%
At the boundaries of the T-hook, the Hirota equation  \eqref{hirota-BC} becomes 
discrete dfAlembert equations: 
\begin{align}
\begin{split} 
 \Tb^{[-1]}_{m_{1},s} \Tb^{[1]}_{m_{1},s}
& = \Tb_{m_{1},s-1} \Tb_{m_{1},s+1} 
 \quad  \text{for} \quad s > n_{1}, 
 %\label{dalem1}
 \\[6pt]
 \Tb^{[-1]}_{a,n_{1}} \Tb^{[1]}_{a,n_{1}}
& = \Tb_{a-1,n_{1}} \Tb_{a+1,n_{1}} 
 \quad  \text{for} \quad a > m_{1},
 \\[6pt]
 \Tb^{[-1]}_{m_{2},s} \Tb^{[1]}_{m_{2},s}
& = \Tb_{m_{2},s-1} \Tb_{m_{2},s+1} 
 \quad  \text{for} \quad s < -n_{2},
 \\[6pt]
 \Tb^{[-1]}_{a,-n_{2}} \Tb^{[1]}_{a,-n_{2}}
& = \Tb_{a-1,-n_{2}} \Tb_{a+1,-n_{2}} 
 \quad  \text{for} \quad a > m_{2},
 \\[6pt]
 \Tb^{[-1]}_{0,s} \Tb^{[1]}_{0,s}
 &= \Tb_{0,s-1} \Tb_{0,s+1} 
 \quad  \text{for} \quad s \in {\mathbb Z}. 
\end{split}
 \label{dalem}
\end{align}
One can check that \eqref{boundUR}-\eqref{boundD} satisfy 
the above equations \eqref{dalem}. 
It is known that the shape of the Hirota equation \eqref{hirota-BC} is invariant under 
the following gauge transformation (cf.\ \cite{KLWZ97}).
\begin{align}
%\Tb_{a,s}^{B_{1},B_{2},F_{1},F_{2}}  \longrightarrow  
\widetilde{\Tb }_{a,s}=
g_{3}^{[a+s]}g_{4}^{[a-s]} g_{5}^{[-a+s]} g_{6}^{[-a-s]}
\Tb_{a,s}  ,
\label{gaugeTsystem}
\end{align}
where $g_{3},g_{4},g_{5},g_{6}$ are arbitrary functions of the spectral parameter.  
The boundary conditions \eqref{boundUR}-\eqref{boundD} 
become more symmetric form (with a generic $\widetilde{\Qb}_{\emptyset} $) if 
the gauge functions are chosen as 
$g_{3}=\widetilde{\Qb}_{\emptyset}^{[\eta_{1}-\eta_{2}]}$,  
$g_{4}=g_{5}=1$, 
$g_{6}= \widetilde{\Qb}_{\emptyset}^{[\eta_{1}+\eta_{2} ]}$ 
 in 
\eqref{gaugeTsystem} and 
$g_{1}=\widetilde{\Qb}_{\emptyset} $, $g_{2}=1$ 
 in 
\eqref{gaugeQQ} 
(or 
$g_{4}=\widetilde{\Qb}_{\emptyset}^{[-\eta_{1}-\eta_{2} ]}$,  
$g_{5}= \widetilde{\Qb}_{\emptyset}^{[\eta_{1}-\eta_{2} ]}$, 
$g_{3}=g_{6}=1$ in \eqref{gaugeTsystem} and 
  $g_{1}=1$, $g_{2}=\widetilde{\Qb} _{\emptyset} $ in 
\eqref{gaugeQQ}).
In fact, \eqref{boundD} becomes
\begin{align}
& \widetilde{\Tb}_{0,s} =   
\widetilde{\Qb}^{[s+\eta_{1}-\eta_{2}]}_{\emptyset }  
\widetilde{\Qb}^{[-s]}_{B_{1},  
B_{2}, F_{1}, F_{2}}  
\quad  \text{for} \quad s \in {\mathbb Z}, 
\end{align}
while the shape of \eqref{boundUR}-\eqref{boundL} 
 is  unchanged. 
 $\Tb^{B_{1},B_{2},F_{1},F_{2}}_{a,s}$ for the full set 
$B_{1}={\mathfrak B}_{1}$,   
$B_{2}={\mathfrak B}_{2}$,   
$F_{1}={\mathfrak F}_{1}$ and   
$F_{2}={\mathfrak F}_{2}$ 
 gives the solution of the T-system for the generalized T-hook. 

We can also Baxterize the solution of the 
Q-system by the operator $\overline{\mathbf B}^{B_{1},B_{2}, F_{1}, F_{2}} $ in the same way as \eqref{baxt-t}: 
\begin{align}
 \overline{\Tb}^{B_{1},B_{2}, F_{1}, F_{2}}_{a,s} := 
\left(\overline{\mathbf B}^{B_{1},B_{2}, F_{1}, F_{2}} 
\cdot 
\chi^{B_{1},B_{2}, F_{1}, F_{2}}_{a,s}
\right)^{[-a+s+\eta_{1}+\eta_{2}]} , 
 \label{baxt-tbar}
\end{align} 
There is one to one correspondence between  $\Tb_{a,s}$ 
and $\overline{\Tb}_{a,s}$, and they can be transformed to one another 
by a discrete transformation. Then, 
we can repeat the same discussion for $\overline{\Tb}_{a,s}$ as 
 $\Tb_{a,s}$. 
%%%%%%%%%%%%%%%%%%%%%%%%%%%%%%%%%%%%%%%%%%%%%%%%%%%%%%%%%%%%%%%
\subsection{B\"{a}cklund transformations for the T-system}
The B\"{a}cklund transformations of the T-system for $gl(M)$ were 
introduced in \cite{KLWZ97}, and generalized  for 
$[M,N]$-hook of $gl(M|N)$ in \cite{KSZ07} and for the general T-hook in \cite{Hegedus09}.  
Hirota equation \eqref{hirota-BC} is a consistency condition for the above equations. 
For $a,s \in {\mathbb Z}$, $b \in B_{1}$, $f \in F_{1}$,  
$B_{1}^{\prime}:=B_{1}\setminus \{b \} $ and  $F_{1}^{\prime}:=F_{1}\setminus \{f \} $, 
B\"{a}cklund transformations for the `right wing' are   
\begin{align}   
& \Tb^{[-1]}_{a+1,s}  
\Tb^{B_{1}^{\prime},[1]}_{a,s}-  
\Tb_{a,s}  
\Tb^{B_{1}^{\prime}}_{a+1,s}  
 =    z_{b}
\Tb_{a+1,s-1}  
\Tb^{B_{1}^{\prime}}_{a,s+1},    
\label{bac1}   
\\[8pt]   
&  \Tb_{a,s+1}  
\Tb^{B_{1}^{\prime}}_{a,s}-  
\Tb^{[-1]}_{a,s}  
\Tb^{B_{1}^{\prime}[1]}_{a,s+1}  
 =  z_{b}
\Tb_{a+1,s}  
\Tb^{B_{1}^{\prime}}_{a-1,s+1},   
\label{bac2}  
\\[8pt]   
%%%   
&  \Tb^{F_{1}^{\prime}[-1]}_{a+1,s}  
\Tb^{[1]}_{a,s}-  
\Tb^{F_{1}^{\prime}}_{a,s}  
\Tb_{a+1,s}  
 =    z_{f}
\Tb^{F_{1}^{\prime}}_{a+1,s-1}  
\Tb_{a,s+1},   
\label{bac3}   
\\[8pt]   
& \Tb^{F_{1}^{\prime}}_{a,s+1}  
\Tb_{a,s}-  
\Tb^{F_{1}^{\prime}[-1]}_{a,s}  
\Tb^{[1]}_{a,s+1}  
=  z_{f}
\Tb^{F_{1}^{\prime}}_{a+1,s}  
\Tb_{a-1,s+1} , 
\label{bac4}  
\end{align}  
%%%%%%%%%%%%%%%%%%%%%%%%%%% 
Here we omitted the index sets whose size is unchanged: 
$\Tb^{B_{1}^{\prime}}_{a,s}=\Tb^{B_{1}^{\prime}, B_{2},F_{1},F_{2}}_{a,s}$, etc. 
We will use similar notation from now on. 
 For $b \in B_{2}$ and $f \in F_{2}$,   
$B_{2}^{\prime}:=B_{2}\setminus \{b \} $ and  $F_{2}^{\prime}:=F_{2}\setminus \{f \} $, 
B\"{a}cklund transformations for the `left wing' are 
\begin{align}   
& z_{b} \Tb^{[1]}_{a+1,s}  
\Tb^{B_{2}^{\prime}[-1]}_{a,s}  
-  
\Tb_{a,s}  
\Tb^{B_{2}^{\prime}}_{a+1,s}  
 = 
\Tb_{a+1,s+1}  
\Tb^{B_{2}^{\prime}}_{a,s-1},    
\label{bacL1}   
\\[8pt]   
&  \Tb_{a,s-1}  
\Tb^{B_{2}^{\prime}}_{a,s}  
-  
\Tb^{[1]}_{a,s}  
\Tb^{B_{2}^{\prime}[-1]}_{a,s-1}  
 = 
\Tb_{a+1,s}  
\Tb^{B_{2}^{\prime}}_{a-1,s-1},   
\label{bacL2}  
\\[8pt]   
%%%  
&  z_{f} \Tb^{F_{2}^{\prime}[1]}_{a+1,s}  
\Tb^{[-1]}_{a,s}  
+  
\Tb^{F_{2}^{\prime}}_{a,s}  
\Tb_{a+1,s}  
= 
\Tb^{F_{2}^{\prime}}_{a+1,s+1}  
\Tb_{a,s-1},   
\label{bacL3}  
\\[8pt]   
& \Tb^{F_{2}^{\prime}}_{a,s-1}  
\Tb_{a,s}  
-  
\Tb^{F_{2}^{\prime}[1]}_{a,s}  
\Tb^{[-1]}_{a,s-1}  
 =-
\Tb^{F_{2}^{\prime}}_{a+1,s}  
\Tb_{a-1,s-1}  . 
\label{bacL4}   
\end{align} 
These are Baxterization of the B\"{a}cklund transformations for the supercharacters 
\eqref{bacch1}-\eqref{bacchL4-2}. 
We find that our solution of the T-system \eqref{sparse01}-\eqref{sparse08} satisfies   
the above B\"{a}cklund transformations \eqref{bac1}-\eqref{bacL4}. 
We will not give a direct proof of this fact
\footnote{A direct proof for the case $m_{2}=n_{2}=0$ is available in Appendix C of our previous 
paper \cite{T09}.}. 
Rather, we have checked consistency 
of the boundary conditions \eqref{T-hookbound} and \eqref{boundUR}-\eqref{boundD} with 
these equations \eqref{bac1}-\eqref{bacL4}. 
Note that the solution of the 
B\"{a}cklund transformation of the T-system with a given boundary 
condition is unique as remarked in \cite{KSZ07} for the case $m_{2}=n_{2}=0$. 
 %%%%%%%%%%%%%%%%%%%%%%%%%%%%%%%%%%  
\subsection{Other expressions of the solutions}   
There are several Wronskian-like expressions for the 
solution of the T-system. 
This is because the Q-functions are, in general, not independent as there are 
QQ-relations \eqref{QQb}-\eqref{QQf}, 
and thus there are many choices of which Q-functions one uses to express 
the T-functions. 
One can obtain these by applying
\footnote{In practice, it will be easier to perform calculations 
by using 
 the coefficient free form given in section \ref{removetwist} and Appendix A. 
 In particular, one can obtain various equivalent determinant expressions from 
 \eqref{soQQbc-1}-\eqref{soQQbc-3} by using various determinant 
 solutions of the QQ-relations (summarized in Appendix A of \cite{GKLT10}) 
and the Laplace expansion formula. 
} the Laplace expansion on the determinant 
to  the sparse determinant expression \eqref{sparse01}-\eqref{sparse08} 
and determinant expressions for the solution of the QQ-relations such as 
\eqref{QQb}-\eqref{QQf}. 
As examples, here we only present two kind of expressions, which are equivalent to 
the sparse determinant expression \eqref{sparse01}-\eqref{sparse08}.  
These are the same quantity but different expressions (based on 
different basis on the Q-functions). 
%One should select an expression which is convenient for one's problem. 

Let us apply the Laplace expansion on the determinant 
to  the sparse determinant expression \eqref{sparse01}-\eqref{sparse08} 
and rewrite this based on 
the determinant expression \eqref{detQQ1}-\eqref{detQQ2} for $I=\emptyset$. 
Then we obtain the following simple expression
\footnote{In the representation theoretical context, these expressions should be interpreted in 
term of a kind of Bernstein-Gel'fand-Gel'fand (BGG) resolution of infinite dimensional modules. 
In this case, each term of these formulas 
of the form $\Tb_{J}^{+}= \chi^{+}_{J} \Qb_{J}^{[sh_{1}]} \Qb_{B \sqcup F \setminus J}^{[sh_{2}]}$
 ($J \subset B \sqcup F $; $ sh_{1},sh_{2} \in {\mathbb C}$; $ \chi^{+}_{J}$:  
 the character part) corresponds to 
 a supercharacter (or rather T-function or q-(super)character) of an 
infinite dimensional highest weight representation which is smaller than the Verma module. 
Examples for these expressions for T-{\bf operators} can be seen in \cite{BHK02} for $U_{q}(\hat{sl}(3))$, 
in \cite{BT08} for $U_{q}(\hat{sl}(2|1))$. 
In this way, 
the T-functions $ \Tb_{J}^{+} $ are building blocks of our solutions. 
%One of the questions is if  the T-function for the generalized Young diagram 
%is expressed  as a finite sum over $ \Tb_{J}^{+} $. 
One of the questions  is whether 
it is possible to construct solutions of the Hirota equation as summations over $ \Tb_{J}^{+} $ when 
 the boundary condition is more complicated than 
the T-hook, such as a `star hook' (a union of a T-hook and an upside-down T-hook).}
 of the solution. 
\begin{multline}    
\Tb_{a,s} =   
\sum_{I\sqcup J= F_{1}\sqcup  F_{2},  
|I|=n_{1}-s} 
\left( \frac{\prod_{b \in B_{2}}z_{b} }{\prod_{j \in J}(-z_{j}) } \right)^{s-a+\eta_{1}} 
\\
 \times 
\frac{\prod_{(b,j) \in B_{1} \times J } (z_{b}-z_{j}) 
\prod_{(b,i) \in B_{2} \times I } (z_{b}-z_{i}) }
{\prod_{(b^{\prime },b) \in B_{2} \times B_{1}  } (z_{b^{\prime}}-z_{b}) 
\prod_{(i,j) \in I \times J  } (z_{i}-z_{j}) }
\Qb^{[a-\eta_{2}]}_{B_{1},I}  
\Qb^{[-a+\eta_{1}]}_{B_{2},J}  
\\
 \qquad \qquad  \text{for} \quad a\ge \max\{s+\eta_{1}, -s+\eta_{2}, 0 \},  
\label{soQQb-1}  
\end{multline}
\begin{multline}
\Tb_{a,s} =   
\sum_{I\sqcup J=B_{1}, |I|=a} 
\frac{ \prod_{i \in I} z_{i}^{s-a+\eta_{1}}
\prod_{(i,f) \in I \times F } (z_{i}-z_{f})}
{\prod_{(b,i) \in B_{2} \times I } (z_{b}-z_{i})
\prod_{(i,j) \in I \times J  } (z_{i}-z_{j}) }  
\Qb^{[s+\eta_{1}-\eta_{2}]}_{I}  
\Qb^{[-s]}_{J,  
B_{2}, F_{1}, F_{2}}  
\\
\text{for} \quad  a \le s + \eta_{1}, \label{soQQb-2}  
\end{multline}
\begin{multline}
\Tb_{a,s} = 
\sum_{I\sqcup J= B_{2}, |I|=a} 
\frac{ \prod_{i \in I} z_{i}^{s-a+\eta_{1}}
\prod_{(i,f) \in I \times F } (z_{i}-z_{f})}
{\prod_{ (i,b) \in  I \times  B_{1}  } (z_{i}-z_{b})
\prod_{ (j,i) \in J  \times I } (z_{j}-z_{i}) }     
\Qb^{[s+\eta_{1}-\eta_{2}]}_{I}  
\Qb^{[-s]}_{  
B_{1}, J,F_{1},F_{2} }   
\\ \text{for} \quad a\le -s+ \eta_{2},
\label{soQQb-3}  
\end{multline} 
where the summation is taken over any possible decomposition of the original set 
into two disjoint sets $I $ and $J$ with fixed sizes. 
Note that 
the right hand side of \eqref{soQQb-1} is 
 well defined even for any $a \in {\mathbb C}$, and 
the right hand sides of \eqref{soQQb-2} and \eqref{soQQb-3} 
are well defined even for any $s \in {\mathbb C}$. 
Thus one can consider analytic continuation of these functions with respect to 
$a$ or $s$. 
%%%%%%%%%%%%%%%%%%%%%%%%%%%%%%%%%%%%%%  
%\subsection{Dense expression of the determinants}
We can further rewrite the above expression.  
Let us substitute 
the determinant expression \eqref{detQQ1}-\eqref{detQQ2} for non-trivial $I$ into 
\eqref{soQQb-1}-\eqref{soQQb-3}, and 
apply the Laplace expansion on the determinant. Then   
we obtain the following `dense determinant expression' of the solution. 
%%%%%%%%%%%%%%%%%%%%%%%%%%%  
\begin{multline}  
\Tb_{a,s}  
=\frac{(-1)^{m_{2}a} } 
{
\prod_{b,b^{\prime} \in B_{1}, \atop 
b \prec b^{\prime} } (z_{b^{\prime}}-z_{b})
\prod_{k=1}^{m_{1}-a-1}\Qb_{B_{2}, F_{1}, F_{2}}^{[  
a-s-m_{1}+2k]} }   
\\ \times   
\begin{vmatrix}  
\left(
z_{b}^{j-1} \Qb_{b,B_{2}, F_{1}, F_{2}}^{[-s+a-m_{1}+2j-1]}
\right)_{  
b \in B_{1},  \atop 
1\le j \le m_{1}-a} &   
\left(
\frac{
z_{b}^{j-1+s-a+\eta_{1}}
\prod_{f \in F}(z_{b}-z_{f})
}{
\prod_{b^{\prime} \in B_{2}}(z_{b}-z_{b^{\prime}})
}
\Qb_{b}^{[s-a+2j-1+\eta_{1}-\eta_{2}]}
\right)_{b \in B_{1} \atop 1\le j \le a}   
  \\  
\end{vmatrix}  
\\  
\quad \text{for} \quad  a \le s + \eta_{1}, \label{Bglmmnn-4}  
\end{multline}  
\begin{multline}  
%\hspace{-30pt}   
\Tb_{a,s} 
=\frac{(-1)^{(a+m_{1} +n)(s+n_{1}) }
\left( \frac{\prod_{b \in B_{2}}z_{b} }{\prod_{f \in F}(-z_{f}) } \right)^{s-a+\eta_{1}} 
}  
{
\prod_{k=1}^{n_{1}-s-1}  
\Qb_{B_{1}}^{[a-s+n_{1}-\eta_{2}-2k]}  
\prod_{k=1}^{n_{2}+s-1}  
\Qb_{B_{2}}^{[-a+s+n_{2}+\eta_{1}-2k]}
}   
\\ \times   
\frac{\prod_{b \in B, f \in F} (z_{b}-z_{f})}{
\prod_{ b^{\prime} \in B_{2}, \atop b \in B_{1} } (z_{b^{\prime}} - z_{b})
\prod_{f^{\prime},f \in F, \atop 
 f \prec f^{\prime}} (z_{f^{\prime}} -z_{f}) 
}
\begin{vmatrix}  
\left(
\frac{
z_{f}^{i-1+s-a+\eta_{1}}
}
{
\prod_{b \in B_{1}} (z_{b}-z_{f})
}
\Qb_{B_{1},f}^{[a-s+n_{1}-\eta_{2}-2i+1]}
\right)_{1\le i \le n_{1}-s,   \atop 
f \in F}  \\  
\left(
\frac{
z_{f}^{i-1}
}
{
\prod_{b \in B_{2}} (z_{b}-z_{f})
}
\Qb_{B_{2},f}^{[-a+s+n_{2}+\eta_{1}-2i+1]}
\right)_{1\le i \le n_{2}+s,   
\atop 
f \in F}   \\  
\end{vmatrix}  
\\  
\quad \text{for} \quad a\ge \max\{s+\eta_{1}, -s+\eta_{2}, 0 \},  \label{Bglmmnn-3}  
\end{multline}  
%%%%   
%  
\begin{multline}  
\Tb_{a,s}   
=\frac{(-1)^{m_{1}a}}  
{ 
\prod_{b^{\prime},b \in B_{2}, \atop 
b \prec b^{\prime} } (z_{b^{\prime}} -z_{b})
\prod_{k=1}^{m_{2}-a-1}  
\Qb_{B_{1},F_{1},F_{2} }^{[  
a-s-m_{2}+2k]} } \\  
%\hspace{-40pt}   
\times   
\begin{vmatrix}  
\left(
\frac{
z_{b}^{j-1+s-a+\eta_{1}}
\prod_{f \in F}(z_{b}-z_{f})
}{
\prod_{b^{\prime} \in B_{1}}(z_{b^{\prime}}-z_{b})
}
\Qb_{b}^{[s-a+2j-1+\eta_{1}-\eta_{2}]}
\right)_{  
b \in B_{2}, \atop 1\le j \le a}  &  
\left(
z_{b}^{j-1}
\Qb_{B_{1},b,F_{1},F_{2}}^{[-s+a-  
m_{2}+2j-1]}
\right)_{  
b \in B_{2}, \atop  
1\le j \le m_{2}-a}   \\  
\end{vmatrix}  
\\  
\quad \text{for} \quad a\le -s+ \eta_{2}. \label{Bglmmnn-5}  
\end{multline}  
These expressions \eqref{soQQb-1}-\eqref{Bglmmnn-5}  
of the solution of the T-system for 
$m_{1}=m_{2}=n_{1}=n_{2}=2$ case were  
previously reported in \cite{GKLT10} 
in the context of the T-system for AdS/CFT. 
Note that these expression satisfies 
the Q-system \eqref{Q-system} if we formally put Q-functions $\{\Qb \}$ to 1. 
Thus these are other new expressions of the supercharacters. 
In particular for $m_{1}=m_{2}=n_{1}=n_{2}=2$, 
\eqref{Bglmmnn-4} -\eqref{Bglmmnn-5} 
reduce to the determinant solution \cite{GKT10} of the 
Q-system for AdS/CFT. 
%%%%%%%%%%%%%%%%%%%%%%%%%%%%%%%%%%%%
\subsection{Discrete transformations on the solutions}
Let us consider how the solutions are transformed under the 
map
\footnote{discussion will be parallel to the one in this section even if we use the other definition of 
$\sigma $ and $\tau $ for the Q-functions 
\eqref{tau-set2}-\eqref{sigmaQ3}} 
$\sigma $ and $\tau$ defined in section 2.4 and \eqref{sigma-set}-\eqref{tauQ1}. 
For this purpose, we introduce other solutions of the T-systems  
defined on T-hooks and 90 degree rotated T-hooks (see Figure \ref{rot-T-hook}). 
For $a,s \in {\mathbb Z}$, we define: 
\begin{multline}    
\Tbb^{B_{1},B_{2}, F_{1}, F_{2}}_{a,s} =   
\sum_{I\sqcup J= F_{1}\sqcup  F_{2},  
|I|=n_{1}-s} 
\left( \frac{\prod_{b \in B_{2}}z_{b} }{\prod_{j \in J}(-z_{j}) } \right)^{s-a+\eta_{1}} 
\\
 \times 
\frac{\prod_{(b,j) \in B_{1} \times J } (z_{b}-z_{j}) 
\prod_{(b,i) \in B_{2} \times I } (z_{b}-z_{i}) }
{\prod_{(b^{\prime },b) \in B_{2} \times B_{1}  } (z_{b^{\prime}}-z_{b}) 
\prod_{(i,j) \in I \times J  } (z_{i}-z_{j}) }
\Qbb^{[-a+\eta_{2}]}_{B_{1},I}  
\Qbb^{[a-\eta_{1}]}_{B_{2},J}  
\\
 \qquad \qquad  \text{for} \quad a\ge \max\{s+\eta_{1}, -s+\eta_{2}, 0 \},  
\label{soQQb-1bar}  
\end{multline}
\begin{multline}
\Tbb^{B_{1},B_{2}, F_{1}, F_{2}}_{a,s} =   
\sum_{I\sqcup J=B_{1}, |I|=a} 
\frac{ \prod_{i \in I} z_{i}^{s-a+\eta_{1}}
\prod_{(i,f) \in I \times F } (z_{i}-z_{f})}
{\prod_{(b,i) \in B_{2} \times I } (z_{b}-z_{i})
\prod_{(i,j) \in I \times J  } (z_{i}-z_{j}) }  
\\
\times 
\Qbb^{[-s-\eta_{1}+\eta_{2}]}_{I}  
\Qbb^{[s]}_{J,  
B_{2}, F_{1}, F_{2}}  
\qquad \text{for} \quad  a \le s + \eta_{1}, \label{soQQb-2bar}  
\end{multline}
\begin{multline}
\Tbb^{B_{1},B_{2}, F_{1}, F_{2}}_{a,s} = 
\sum_{I\sqcup J= B_{2}, |I|=a} 
\frac{ \prod_{i \in I} z_{i}^{s-a+\eta_{1}}
\prod_{(i,f) \in I \times F } (z_{i}-z_{f})}
{\prod_{ (i,b) \in  I \times  B_{1}  } (z_{i}-z_{b})
\prod_{ (j,i) \in J  \times I } (z_{j}-z_{i}) }     
\\
\times 
\Qbb^{[-s-\eta_{1}+\eta_{2}]}_{I}  
\Qbb^{[s]}_{  
B_{1}, J,F_{1},F_{2} }   
\qquad \text{for} \quad a\le -s+ \eta_{2},
\label{soQQb-3bar}  
\end{multline} 
where 
$\overline{\Tb}_{a,s}^{B_{1} ,B_{2} , F_{1}, F_{2} }=0$ 
if $\{a<0 \}$, or $\{a>m_{1},s>n_{1} \}$ or $ \{a>m_{2},s<-n_{2} \}$.
%%%%%%%%%%%  
\begin{multline}    
\check{\Tb}^{B_{1},B_{2}, F_{1}, F_{2}}_{a,s} =   
\sum_{I\sqcup J= B_{1}\sqcup  B_{2},  
|I|=m_{2}+a} 
\left( \frac{\prod_{f \in F_{2}}(-z_{f}) }{\prod_{i \in I} z_{i} } \right)^{a-s-\eta_{1}} 
\\
 \times 
\frac{\prod_{(j,f) \in J \times F_{2} } (z_{j}-z_{f}) 
\prod_{(i,f) \in I \times F_{1} } (z_{i}-z_{f}) }
{\prod_{(f^{\prime },f) \in F_{1} \times F_{2}  } (z_{f^{\prime}}-z_{f}) 
\prod_{(i,j) \in I \times J  } (z_{i}-z_{j}) }
\Qb^{[s+\eta_{1}]}_{F_{2},I}  
\Qb^{[-s-\eta_{2}]}_{F_{1},J}  
\\
 \qquad \qquad  \text{for} \quad s \ge \max\{a-\eta_{1}, -a-\eta_{2}, 0 \},  
\label{soQQb-1ch}  
\end{multline}
\begin{multline}
{\check \Tb}^{B_{1},B_{2}, F_{1}, F_{2}}_{a,s} =   
\sum_{I\sqcup J=F_{1}, |I|=s} 
\frac{ \prod_{i \in I} (-z_{i})^{a-s-\eta_{1}}
\prod_{(b,i) \in B \times I } (z_{b}-z_{i})}
{\prod_{(i,f) \in I \times F_{2}  } (z_{i}-z_{f})
\prod_{(j,i) \in J \times I  } (z_{j}-z_{i}) }  
\\
\times 
\Qb^{[-a+\eta_{1}-\eta_{2}]}_{I}  
\Qb^{[a]}_{B_{1},  
B_{2}, J, F_{2}}  
\qquad \text{for} \quad  s \le a - \eta_{1}, \label{soQQb-2ch}  
\end{multline}
\begin{multline}
\check{\Tb}^{B_{1},B_{2}, F_{1}, F_{2}}_{a,s} = 
\sum_{I\sqcup J= F_{2}, |I|=s} 
\frac{ \prod_{i \in I} (-z_{i})^{a-s-\eta_{1}}
\prod_{(b,i) \in B \times I } (z_{b}-z_{i})}
{\prod_{ (f,i) \in  F_{1} \times I  } (z_{f}-z_{i})
\prod_{ (i,j) \in I  \times J } (z_{i}-z_{j}) }     
\\
\times 
\Qb^{[-a+\eta_{1}-\eta_{2}]}_{I}  
\Qb^{[a]}_{  
B_{1}, B_{1},F_{1}, J }   
\qquad \text{for} \quad s \le -a- \eta_{2}. 
\label{soQQb-3ch}  
\end{multline} 
%%%%%%%%%%%%%%%%%%%%%%%%%%%%%%  
\begin{multline}    
\check{\Tbb}^{B_{1},B_{2}, F_{1}, F_{2}}_{a,s} =   
\sum_{I\sqcup J= B_{1}\sqcup  B_{2},  
|I|=m_{2}+a} 
\left( \frac{\prod_{f \in F_{2}}(-z_{f}) }{\prod_{i \in I} z_{i} } \right)^{a-s-\eta_{1}} 
\\
 \times 
\frac{\prod_{(j,f) \in J \times F_{2} } (z_{j}-z_{f}) 
\prod_{(i,f) \in I \times F_{1} } (z_{i}-z_{f}) }
{\prod_{(f^{\prime },f) \in F_{1} \times F_{2}  } (z_{f^{\prime}}-z_{f}) 
\prod_{(i,j) \in I \times J  } (z_{i}-z_{j}) }
\Qbb^{[-s-\eta_{1}]}_{F_{2},I}  
\Qbb^{[s+\eta_{2}]}_{F_{1},J}  
\\
 \qquad \qquad  \text{for} \quad s \ge \max\{a-\eta_{1}, -a-\eta_{2}, 0 \},  
\label{soQQb-1chbar}  
\end{multline}
\begin{multline}
{\check \Tbb}^{B_{1},B_{2}, F_{1}, F_{2}}_{a,s} =   
\sum_{I\sqcup J=F_{1}, |I|=s} 
\frac{ \prod_{i \in I} (-z_{i})^{a-s-\eta_{1}}
\prod_{(b,i) \in B \times I } (z_{b}-z_{i})}
{\prod_{(i,f) \in I \times F_{2}  } (z_{i}-z_{f})
\prod_{(j,i) \in J \times I  } (z_{j}-z_{i}) }  
\\
\times 
\Qbb^{[a-\eta_{1}+\eta_{2}]}_{I}  
\Qbb^{[-a]}_{B_{1},  
B_{2}, J, F_{2}}  
\qquad \text{for} \quad  s \le a - \eta_{1}, \label{soQQb-2chbar}  
\end{multline}
\begin{multline}
\check{\Tbb}^{B_{1},B_{2}, F_{1}, F_{2}}_{a,s} = 
\sum_{I\sqcup J= F_{2}, |I|=s} 
\frac{ \prod_{i \in I} (-z_{i})^{a-s-\eta_{1}}
\prod_{(b,i) \in B \times I } (z_{b}-z_{i})}
{\prod_{ (f,i) \in  F_{1} \times I  } (z_{f}-z_{i})
\prod_{ (i,j) \in I  \times J } (z_{i}-z_{j}) }     
\\
\times 
\Qbb^{[a-\eta_{1}+\eta_{2}]}_{I}  
\Qbb^{[-a]}_{ B_{1}, B_{1},F_{1}, J }   
\qquad \text{for} \quad s \le -a- \eta_{2}, 
\label{soQQb-3chbar}  
\end{multline} 
where $\check{\Tb}^{B_{1},B_{2}, F_{1}, F_{2}}_{a,s}=
\check{\Tbb}^{B_{1},B_{2}, F_{1}, F_{2}}_{a,s}=0$ 
if $\{s<0 \}$, or $\{a>m_{1},s>n_{1} \}$ or $ \{a<-m_{2},s>n_{2} \}$. 
%%%%%%%%%%%%%%%%%%%%%%%%%%%%%%%
For the full sets of the index sets, these coincide with the T-functions without over-line 
up to the overall shift of the spectral parameter: 
 $\Tbb^{{\mathfrak B}_{1},{\mathfrak B}_{2}, {\mathfrak F}_{1}, {\mathfrak F}_{2}}_{a,s}=
\Tb^{{\mathfrak B}_{1},{\mathfrak B}_{2}, {\mathfrak F}_{1}, {\mathfrak F}_{2}[\eta_{2}-\eta_{1}]}_{a,s}$, 
 $\check{\Tbb}^{{\mathfrak B}_{1},{\mathfrak B}_{2}, {\mathfrak F}_{1}, {\mathfrak F}_{2}}_{a,s}=
\check{\Tb}^{{\mathfrak B}_{1},{\mathfrak B}_{2}, {\mathfrak F}_{1}, {\mathfrak F}_{2}[\eta_{2}-\eta_{1}]}_{a,s}$.
%%%%%%%
We find the following transformation property of the T-functions under $\sigma $ and $\tau $: 
\begin{align}
\sigma (\Tb_{a,s}^{B_{1} ,B_{2} , F_{1} ,F_{2} }) &=
 \left(
\frac{ \prod_{b \in \hat{B}} z_{b} }{ \prod_{ f \in \hat{F}} (-z_{f}) }
\right)^{a}
 \overline{\Tb}_{a,-s}^{\hat{B}_{1} ,\hat{B}_{2} , \hat{F}_{1} ,\hat{F}_{2} }, 
 \\[6pt]
\tau (\Tb_{a,s}^{B_{1} ,B_{2} , F_{1} ,F_{2} }) &=
(-1)^{(s+\eta_{2})a}
\left(
\frac{ \prod_{ f \in \check{F}} (-z_{f}) }{ \prod_{b \in \check{B}} z_{b} }
\right)^{a}
 \check{\Tb}_{-s,a}^{\check{B}_{1} ,\check{B}_{2} , \check{F}_{1} ,\check{F}_{2}},
\\[6pt]
\sigma (\Tbb_{a,s}^{B_{1} ,B_{2} , F_{1} ,F_{2} }) &=
 \left(
\frac{ \prod_{b \in \hat{B}} z_{b} }{ \prod_{ f \in \hat{F}} (-z_{f}) }
\right)^{a}
 \Tb_{a,-s}^{\hat{B}_{1} ,\hat{B}_{2} , \hat{F}_{1} ,\hat{F}_{2} }, 
 \\[6pt]
\tau (\Tbb_{a,s}^{B_{1} ,B_{2} , F_{1} ,F_{2} }) &=
(-1)^{(s+\eta_{2})a}
\left(
\frac{ \prod_{ f \in \check{F}} (-z_{f}) }{ \prod_{b \in \check{B}} z_{b} }
\right)^{a}
 \check{\Tbb}_{-s,a}^{\check{B}_{1} ,\check{B}_{2} , \check{F}_{1} ,\check{F}_{2}},
\end{align}
where $a \in {\mathbb Z}_{\ge 0}$ and $s \in {\mathbb Z}$, and 
\begin{align}
\sigma (\check{\Tb}_{a,s}^{B_{1} ,B_{2} , F_{1} ,F_{2} }) &=
 \left(
\frac{ \prod_{ f \in \hat{F}} (-z_{f}) }{ \prod_{b \in \hat{B}} z_{b} }
\right)^{s}
 \check{\overline{\Tb}}_{-a,s}^{\hat{B}_{1} ,\hat{B}_{2} , \hat{F}_{1} ,\hat{F}_{2} } 
\\[6pt]
\tau (\check{\Tb }_{a,s}^{B_{1} ,B_{2} , F_{1} ,F_{2} }) &=
(-1)^{(a+\eta_{2})s}
\left(
\frac{ \prod_{b \in \check{B}} z_{b} }{ \prod_{ f \in \check{F}} (-z_{f}) }
\right)^{s}
 \Tb_{s,-a}^{\check{B}_{1} ,\check{B}_{2} , \check{F}_{1} ,\check{F}_{2} }, 
\\[6pt]
\sigma (\check{\Tbb}_{a,s}^{B_{1} ,B_{2} , F_{1} ,F_{2} }) &=
 \left(
\frac{ \prod_{ f \in \hat{F}} (-z_{f}) }{ \prod_{b \in \hat{B}} z_{b} }
\right)^{s}
 \check{\Tb}_{-a,s}^{\hat{B}_{1} ,\hat{B}_{2} , \hat{F}_{1} ,\hat{F}_{2} } 
\\[6pt]
\tau (\check{\Tbb }_{a,s}^{B_{1} ,B_{2} , F_{1} ,F_{2} }) &=
(-1)^{(a+\eta_{2})s}
\left(
\frac{ \prod_{b \in \check{B}} z_{b} }{ \prod_{ f \in \check{F}} (-z_{f}) }
\right)^{s}
 \Tbb_{s,-a}^{\check{B}_{1} ,\check{B}_{2} , \check{F}_{1} ,\check{F}_{2} },
\end{align}
where $s \in {\mathbb Z}_{\ge 0}$ and $a \in {\mathbb Z}$.  
We also have 
\begin{align}
\sigma \tau (\Tb_{a,s}^{B_{1} ,B_{2} , F_{1} ,F_{2} }) &=\tau  \sigma (\Tb_{a,s}^{B_{1} ,B_{2} , F_{1} ,F_{2} })=
(-1)^{(s+\eta_{2})a} 
 \check{\Tbb}_{s,a}^{\hat{\check{B}}_{1} ,\hat{\check{B}}_{2} , \hat{\check{F}}_{1} ,\hat{\check{F}}_{2} }, 
 \\[6pt]
\sigma \tau (\Tbb_{a,s}^{B_{1} ,B_{2} , F_{1} ,F_{2} }) &=\tau  \sigma (\Tbb_{a,s}^{B_{1} ,B_{2} , F_{1} ,F_{2} })=
(-1)^{(s+\eta_{2})a} 
 \check{\Tb}_{s,a}^{\hat{\check{B}}_{1} ,\hat{\check{B}}_{2} , \hat{\check{F}}_{1} ,\hat{\check{F}}_{2} },
\end{align}
where $a \in {\mathbb Z}_{\ge 0}$ and $s \in {\mathbb Z}$, and 
\begin{align}
\sigma \tau (\check{\Tb}_{a,s}^{B_{1} ,B_{2} , F_{1} ,F_{2} }) &=\tau  \sigma (\check{\Tb}_{a,s}^{B_{1} ,B_{2} , F_{1} ,F_{2} })=
(-1)^{(a+\eta_{2})s} 
 \Tbb_{s,a}^{\hat{\check{B}}_{1} ,\hat{\check{B}}_{2} , \hat{\check{F}}_{1} ,\hat{\check{F}}_{2} }, 
 \\[6pt]
\sigma \tau (\check{\Tbb}_{a,s}^{B_{1} ,B_{2} , F_{1} ,F_{2} }) &=\tau  \sigma (\check{\Tbb}_{a,s}^{B_{1} ,B_{2} , F_{1} ,F_{2} })=
(-1)^{(a+\eta_{2})s} 
 \Tb_{s,a}^{\hat{\check{B}}_{1} ,\hat{\check{B}}_{2} , \hat{\check{F}}_{1} ,\hat{\check{F}}_{2} },
\end{align}
where $s \in {\mathbb Z}_{\ge 0}$ and $a \in {\mathbb Z}$. 
In this way, we can obtain 4 type of the solutions for the T-hook (see Figure \ref{4-T-hooks}). 
There is one to one correspondence among them. 
%%%%%%%%%%%%%%%%%%%%%%%%%%%%%%%%%%%%%%%%%%%%%%%%%
\subsection{Reductions of solutions by automorphisms}
In this section, we briefly announce our idea on how to obtain solutions of T-systems for other algebras. 
We will discuss details elsewhere. 

We find that reductions on the  QQ-relations by $\sigma$ or $\tau$ 
 (and some dualities among different superalgebras (cf.\ \cite{Z97})) 
 produce solutions of 
the T-systems for different algebras. 
The reductions here are basically accomplished by identifying the image of 
the Q-functions and the parameters $\{z_{a}\}$ by the maps $\sigma$ or $\tau $ 
with the original ones (up to the gauge and manipulations on the spectral parameter in some cases). 
Let us consider `$sl(M|N)^{(2)}$ type reduction' by $\sigma$: 
\begin{align}
\sigma (Q_{I})=Q_{I} \quad \text{for} \quad I \subset {\mathfrak I}, \quad 
\sigma (z_{a})=z_{a} \quad \text{for} \quad a \in {\mathfrak I} . 
\label{reduction-sigma}
\end{align} 
If $M$ or $N$ are odd, fixed points by $\sigma$ appear. 
For example for $N=2r+1$ case, we have 
$\sigma(z_{M+r+1})=z_{M+r+1}^{-1}=z_{M+r+1}$. Then $z_{M+r+1}=\pm 1$. 
The minus sign $z_{M+r+1}=-1$ effectively changes the sign of $p_{M+r+1}$ from the grading of the superalgebra, 
which induces a duality among a superalgebra ($z_{M+1}=1$) and an ordinary algebra ($z_{M+1}=-1$) for the case $N=1$. 
In particular, $sl(2r|1)^{(2)}$ type reduction for $z_{2r+1}=-1$ produces QQ-relations (and then 
Bethe equations) and Wronskian solutions 
\footnote{
For $s \in {\mathbb Z}_{\ge 1} $, $B=\{1,2,\dots, 2r \}$, $F=\{2r+1 \}$, we define 
$\Tb^{(a)}_{s}:=\Tb^{B, \emptyset, F, \emptyset }_{a,s} $ for $a \in \{1,2,\dots,r-1 \}$, 
$\Tb^{(r)}_{2s}:=\Tb^{B, \emptyset, F, \emptyset }_{r,s} $ and 
$\Tb^{(r)}_{2s-1}:=\prod_{k=1}^{r}(\sqrt{z_{k}} +\frac{1}{\sqrt{z_{k}}})\Tb^{B, \emptyset, \emptyset, \emptyset }_{r,s-1} $. 
Then we find that $\{ \Tb^{(a)}_{s} \}$ 
solve the T-system for ${\mathfrak g}=so(2r+1) $ up to some rescaling of the functions. 
It is interesting to see that the T-functions $\Tb^{(r)}_{2s-1}$ related to the spin representations 
are given by T-functions one level lower than the original one in the sense of the B\"{a}cklund transformations. 
Thus the T-function $\Tb^{(r)}_{1}$ is proportional to the Q-functions $\Qb_{B}=\Qb_{F}$. 
The T-system for  $so(2r+1) $ was proposed in \cite{KNS93}. However, the Wronskian like solution for it 
was not known in the literatures.
} 
of the T-system for $U_{q}({\mathfrak g}^{(1)})$ or $Y({\mathfrak g})$, where ${\mathfrak g}=so(2r+1)$. 
$sl(0|2r+1)^{(2)}$ type reduction for $z_{r+1}=-1$ corresponds to ${\mathfrak g}=osp(1|2r)$ (cf.\ \cite{T99}). 
$sl(2r+1|0)^{(2)}$ type reduction
\footnote{In this case, we have to add a shift of the spectral parameter to \eqref{reduction-sigma} as 
$\sigma (Q_{I})=Q_{I}^{[\frac{\pi i}{2\hbar}]}  $ ($\hbar \in {\mathbb C} \setminus \{0 \}$ 
cf. \cite{KS94,T02}).  
}
 for $z_{r+1}=1$ corresponds to $U_{q}(A_{2r}^{(2)})$, and 
$sl(2r|0)^{(2)}$ type reduction  corresponds to $U_{q}(A_{2r-1}^{(2)})$.  
These reduction produce additional functional relations among T-functions, which do not exist before the reductions. 
There will also be interesting reductions by $\tau$
\footnote{A question is whether it is possible to treat the Q-functions and the T-system for $AdS_{5}/CFT_{4}$ 
efficiently 
in relation to reductions similar to the ones based on $\sigma$ and $\tau $ for $gl(4|4)$. 
For this, one will need careful analysis on the analyticity on the spectral parameter. 
The reduction may accompany manipulation on 
the Riemann sheets on the spectral parameter. 
For the usual twisted quantum affine algebras case, the 
corresponding manipulation was just a shift of the spectral parameter. 
But in this case, it could be more involved (cf.\ \cite{GKLT10}).}. 
%%%%%%%%%%%%%%%%%%%%%%%%%%%%%%%%%%  
\subsection{Removing the twist} 
\label{removetwist} 
Bazhanov, Lukyanov, Zamolodchikov  \cite{BLZ97} defined 
the Baxter Q-operators as trace of the universal $R$-matrix 
over $q$-oscillator representations of the quantum affine algebra $U_{q}(\widehat{sl}(2))$. 
And  importance of boundary twists or horizontal fields to 
regularize the trace over the infinite dimensional space was recognized in 
\cite{BLZ97} for the first time. In this context, the parameters $\{z_{i} \}$ correspond to 
these boundary twists or horizontal fields. 
In this paper, these parameters were used to define supercharacters. 
We can eliminate these parameters $\{z_{i} \}$ by the following transformation: 
\begin{align}
\Qf_{I}^{[0]}
&=a_{I} f^{[0]}_{I}\Qb_{I}^{[0]}, \label{removtwistQ}
\\[6pt]
\Tf_{a,s}^{I [0]}
&=a_{I} f^{[ a-s]}_{I}\Tb_{a,s}^{I [0]}, \label{removtwistT}
\end{align}
where $I$ is a tuple (as a set, it is a subset of the full set ${\mathfrak I}$);  $a_{I}=\prod_{j,k \in I; j \prec k } 
\left(
\frac{z_{j} -z_{k}}{(z_{j} z_{k})^{\frac{1}{2}} } 
\right)^{p_{j} p_{k} }$ for $|I|\ge 2$, 
$a_{I}=1$ for $|I|=0,1$; 
$f^{[0]}_{I}=\prod_{i \in I} f^{[0]}_{i}$,  
$f^{[s]}_{i}=z_{i}^{\frac{p_{i} s}{2}} f^{[0]}_{i}$;  
$\Tb_{a,s}^{I}:=\Tb_{a,s}^{I\cap {\mathfrak B}_{1}, I\cap {\mathfrak B}_{2}, I\cap {\mathfrak F}_{2}, I\cap {\mathfrak F}_{2}}$. 
Now the order of the elements of the index set (a tuple $I$) of $ \Tf_{a,s}^{I}$ and $\Qf_{I}$ 
affects overall sign of the functions (as oppose to $ \Tb_{a,s}^{I}$ and $\Qb_{I}$). 
Then the QQ-relations \eqref{QQb} and \eqref{QQf} become a coefficient free form:
\begin{align}
& \Qf_{I}\Qf_{I,ij}
=\Qf_{I,i}^{[p_{i}]}
\Qf_{I,j}^{[-p_{i}]}-
\Qf_{I,i}^{[-p_{i}]}
\Qf_{I,j}^{[p_{i}]}
\qquad 
\text{for} \qquad p_{i}=p_{j},  \label{QQb-cf}  \\[6pt]
& 
\Qf_{I,i}\Qf_{I,j}=
\Qf_{I}^{[-p_{i}]}
\Qf_{I,ij}^{[p_{i}]}-
\Qf_{I}^{[p_{i}]}
\Qf_{I,ij}^{[-p_{i}]}
\qquad \text{for} \qquad p_{i}=-p_{j}.
\label{QQf-cf}
\end{align} 
We remark that the coefficient free form of \eqref{QQf-cf} for $(M+1)(N+1)$ Q-functions was discussed in 
detail in \cite{KSZ07}. 
In this form, one can see a
 $GL(M)\times GL(N)$ symmetry of the QQ-relations. In fact, 
the following transformation \eqref{Qpri1}-\eqref{Qpri2} preserve the shape of the $QQ$-relations. 
\begin{align}
\Qf_{b^{\prime}}^{\prime}
=
\sum_{b  \in {\mathfrak B}}
A_{b^{\prime}b}
\Qf_{b}, 
\qquad 
\Qf_{f^{\prime}}^{\prime}
=
\sum_{f  \in {\mathfrak F}}
B_{f^{\prime}f}
\Qf_{f}
\quad 
\text{for} \quad 
b^{\prime} \in {\mathfrak B}, 
\quad 
f^{\prime} \in {\mathfrak F}, 
\label{Qpri1}
\end{align}
where $(A_{i,j})_{1 \le i,j \le M} \in GL(M)$, 
$(B_{i+M,j+M})_{1 \le i,j \le N} \in GL(N)$. 
Here the matrix elements $A_{i,j}$ and  $B_{i,j}$ should be periodic functions: 
 $A_{i,j}^{[1]}=A_{i,j}^{[0]}, B_{i,j}^{[1]}=B_{i,j}^{[0]}$ 
if they depend on the spectral parameter.
The composite Q-functions transform as tensors. 
\begin{multline}
\Qf_{b_{1}^{\prime}\dots b_{m}^{\prime}f_{1}^{\prime}\dots f_{n}^{\prime}}^{\prime}
=
\sum_{b_{1}, \dots ,b_{m} \in {\mathfrak B}}
\sum_{f_{1}, \dots ,f_{n} \in {\mathfrak F}}
A_{b_{1}^{\prime}b_{1}}\dots A_{b_{m}^{\prime}b_{m}}
B_{f_{1}^{\prime}f_{1}}\dots B_{f_{n}^{\prime}f_{n}}
\Qf_{b_{1} \dots b_{m} f_{1} \dots f_{n} }
\\ 
\text{for} \quad 
b_{1}^{\prime},\dots ,b_{m}^{\prime} \in {\mathfrak B}, 
\quad 
f_{1}^{\prime}, \dots , f_{n}^{\prime} \in {\mathfrak F}. 
\end{multline}
In particular, 
\begin{align}
&\Qf_{\emptyset}^{\prime}
=
\Qf_{\emptyset}, 
\\[6pt]
&\Qf_{1\dots M,M+1,\dots ,M+N}^{\prime}
=
\det_{1 \le i,j \le M}(A_{i,j})
\det_{1 \le i,j \le N}(B_{i+M,j+M})
\Qf_{1\dots M,M+1,\dots ,M+N}. 
\label{Qpri2}
\end{align}
This $GL(M)\times GL(N)$ symmetry for $M=2,N=0$ case was used to take 
without twist limit ($z_{i} \to 1$) in \cite{BLMS10} (see also, \cite{DM10}). 
As in Figure \ref{hasse}, 
we put $2^{M+N}$ Q-functions on the Hasse diagram. 
Now the diagram is on a hyper-sphere or a hyper-oval sphere due to the above symmetry. 

In appendix A, we list the coefficient free form of the solutions of the T-system by 
the transformations \eqref{removtwistQ}-\eqref{removtwistT}. 
Once they are obtained, one can forget about the relations \eqref{removtwistQ} and \eqref{removtwistT},  
and regard $\{\Qf_{I} \}$ as any complex functions of the spectral parameter 
(with the normalization $\Qf_{\emptyset}=1$) which satisfy 
\eqref{QQb-cf} and \eqref{QQf-cf}. 
%%%%%%%%%%%%%%%%%%%%%%%%%%%%%%%%%%%%%

%%%%%%%%%%%%%%%%%%%%%%%%%%%%%%%%%%%%%%%%%%%%%%%%%
\section{$Y$-system} 
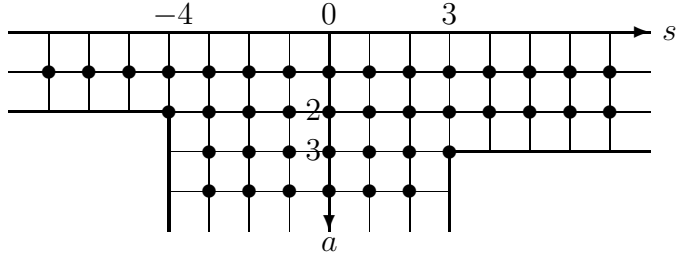
\begin{figure}
  \begin{center}
    \setlength{\unitlength}{1.5pt}
    \begin{picture}(160,55) 
      {\thicklines \put(80,50){\vector(0,-1){50}}}
      {\thicklines \put(40,0){\line(0,1){30}}}
       {\thicklines \put(110,0){\line(0,3){20}}}
       {\thicklines \put(110,20){\line(1,0){50}}}
       {\thicklines \put(0,30){\line(1,0){40}}}
       {\thicklines \put(0,50){\vector(1,0){160}}}
      {\linethickness{0.2pt}  \put(0,40){\line(1,0){160}}}
      {\linethickness{0.2pt}  \put(40,30){\line(1,0){120}}}
      {\linethickness{0.2pt}  \put(40,20){\line(1,0){70}}}
      {\linethickness{0.2pt}  \put(40,10){\line(1,0){70}}}
      {\linethickness{0.2pt}  \put(10,50){\line(0,-1){20}}}
      {\linethickness{0.2pt}  \put(20,50){\line(0,-1){20}}}
      {\linethickness{0.2pt}  \put(30,50){\line(0,-1){20}}}
      {\linethickness{0.2pt}  \put(40,50){\line(0,-1){20}}}
      {\linethickness{0.2pt}  \put(50,50){\line(0,-1){50}}}
      {\linethickness{0.2pt}  \put(60,50){\line(0,-1){50}}}
      {\linethickness{0.2pt}  \put(70,50){\line(0,-1){50}}}
      {\linethickness{0.2pt}  \put(90,50){\line(0,-1){50}}}
      {\linethickness{0.2pt}  \put(100,50){\line(0,-1){50}}}
      {\linethickness{0.2pt}  \put(110,50){\line(0,-1){30}}}
      {\linethickness{0.2pt}  \put(120,50){\line(0,-1){30}}}
      {\linethickness{0.2pt}  \put(130,50){\line(0,-1){30}}}
      {\linethickness{0.2pt}  \put(140,50){\line(0,-1){30}}}
      {\linethickness{0.2pt}  \put(150,50){\line(0,-1){30}}}
       \put(10,40){\circle*{3}}
       \put(20,40){\circle*{3}}
       \put(30,40){\circle*{3}}
       \put(40,40){\circle*{3}}
       \put(50,40){\circle*{3}}
       \put(60,40){\circle*{3}}
       \put(70,40){\circle*{3}}
       \put(80,40){\circle*{3}}
       \put(90,40){\circle*{3}}
       \put(100,40){\circle*{3}}
       \put(110,40){\circle*{3}}
       \put(120,40){\circle*{3}}
       \put(130,40){\circle*{3}}
       \put(140,40){\circle*{3}}
       \put(150,40){\circle*{3}}
       \put(40,30){\circle*{3}}
       \put(50,30){\circle*{3}}
       \put(60,30){\circle*{3}}
       \put(70,30){\circle*{3}}
       \put(80,30){\circle*{3}}
       \put(90,30){\circle*{3}}
       \put(100,30){\circle*{3}}
       \put(110,30){\circle*{3}}
       \put(120,30){\circle*{3}}
       \put(130,30){\circle*{3}}
       \put(140,30){\circle*{3}}
       \put(150,30){\circle*{3}}
       \put(50,20){\circle*{3}}
       \put(60,20){\circle*{3}}
       \put(70,20){\circle*{3}}
       \put(80,20){\circle*{3}}
       \put(90,20){\circle*{3}}
       \put(100,20){\circle*{3}}
       \put(110,20){\circle*{3}}
       \put(50,10){\circle*{3}}
       \put(60,10){\circle*{3}}
       \put(70,10){\circle*{3}}
       \put(80,10){\circle*{3}}
       \put(90,10){\circle*{3}}
       \put(100,10){\circle*{3}}
       \put(78,-5){$a$}
       \put(163,48){$s$}
       \put(74,18){$3$}
       \put(74,28){$2$}
       \put(108,52){$3$}
       \put(36,52){$-4$}
       \put(78,52){$0$}
    \end{picture}
  \end{center}
  \caption{ 
  A domain for the Y-system in a generalized T-hook (cf.\ \cite{GKV09,Hegedus09}): this T-hook is 
a union of $[3,3]$-hook and $[2,4]$-hook for
 $gl(3+2|3+4)$. The Y-system is defined on the dots.
}
  \label{T-Y}
\end{figure}

The Y-system is a system of functional relations related the thermodynamic Bethe ansatz. 
%\footnote{This section supplements  discussions in section 6 of \cite{T09}.}. 
The T-system \eqref{hirota-BC} is related to the Y-system by the following standard dependent 
variable transformation:
\begin{align}
\Yb_{a,s}:= 
\frac{ \Tb_{a,s-1}   \Tb_{a,s+1}  }
{ \Tb_{a-1,s}   \Tb_{a+1,s} } .
 \label{Y-fun}
\end{align}
Then the `Y-functions' $\Yb_{a,s} $ satisfy the following Y-system (cf.\ Figure \ref{T-Y}): 
\begin{align}
\Yb^{[-1]}_{a,s} \Yb^{[1]}_{a,s}
=
\frac{
(1+\Yb_{a,s-1})(1+\Yb_{a,s+1})
}
{
(1+(\Yb_{a-1,s})^{-1}) (1+(\Yb_{a+1,s})^{-1})
} .
\label{y-system}
\end{align}
%%%%%%%%%%%%%%%
%%%%%%%%%%%%%%%%%
Thus we can obtain the solution of the Y-system through \eqref{Y-fun} based on 
the solution of the T-system presented in section 3. 
It is known that the Y-system \eqref{y-system} is invariant under the gauge transformation \eqref{gaugeTsystem}. 
This Y-system for the T-hook for $M_{1}=M_{2}=N_{1}=N_{2}=2$ was proposed in \cite{GKV09} in the study of $AdS_{5}/CFT_{4}$ duality, 
and was generalized for the general T-hook in \cite{Hegedus09}. It was also studied \cite{Volin10-2}
 in relation to the string hypothesis 
on Bethe roots. 
These generalize Y-systems for  super spin chains \cite{JKS98} (cf. \cite{Saleur99}). 
We have also found Y-system like equations for the B\"{a}cklund transformations 
\eqref{bac1}-\eqref{bacL4}. 
Explicitly, they are written as
\footnote{There are several other definitions of the Y-functions. For example, we can define 
$ {\mathfrak A}^{(2)}_{a,s} := 
\frac{
z_{b}
\Tb_{a+1,s}  
\Tb^{B_{1}^{\prime}}_{a-1,s+1}
}
{\Tb^{[-1]}_{a,s}  
\Tb^{B_{1}^{\prime}[1]}_{a,s+1} } $ 
instead of \eqref{y-sys-bac2}. 
This satisfies 
$ {\mathfrak A}^{(2)}_{a,s+1}  
 {\mathfrak A}^{(2)B_{1}^{\prime}}_{a,s}
=
\frac{(1+ {\mathfrak A}^{(2)}_{a+1,s}  )
(1+ {\mathfrak A}^{(2)B_{1}^{\prime}}_{a-1,s+1}  )}
{(1+ ({\mathfrak A}^{(2)[-1]}_{a,s} )^{-1} )
(1+ ({\mathfrak A}^{(2)B_{1}^{\prime}[1]}_{a,s+1} )^{-1} )} 
$.  In the main text, we defined the Y-functions so that the shift of the spectral parameter 
appears in the left hand side of the equations in the same way as the Y-system \eqref{y-system}.}: 
\begin{align}   
 &{\mathfrak A}^{(1)[-1]}_{a+1,s}  
{\mathfrak A}^{(1) B_{1}^{\prime}[1]}_{a,s}
=
\frac{
(1+{\mathfrak A}^{(1)}_{a+1,s-1})
(1+{\mathfrak A}^{(1)B_{1}^{\prime}}_{a,s+1})
}
{
(1+({\mathfrak A}^{(1)}_{a,s})^{-1} )
(1+({\mathfrak A}^{(1)B_{1}^{\prime}}_{a+1,s})^{-1})
},
\nonumber \\[4pt]
& \hspace{120pt} 
{\mathfrak A}^{(1)}_{a,s} := 
\frac{
z_{b}
\Tb_{a+1,s-1}  
\Tb^{B_{1}^{\prime}}_{a,s+1}
}
{
\Tb_{a,s}  
\Tb^{B_{1}^{\prime}}_{a+1,s} 
}  ,
\\[8pt]   
%%%%%%%%%%%%%%%%%%%%%%%  
& {\mathfrak A}^{(2)[-1]}_{a,s}  
 {\mathfrak A}^{(2)B_{1}^{\prime}[1]}_{a,s+1}
=
\frac{(1+ {\mathfrak A}^{(2)}_{a+1,s}  )
(1+ {\mathfrak A}^{(2)B_{1}^{\prime}}_{a-1,s+1}  )}
{(1+ ({\mathfrak A}^{(2)}_{a,s+1} )^{-1} )
(1+ ({\mathfrak A}^{(2)B_{1}^{\prime}}_{a,s} )^{-1} )} , 
\nonumber  \\[4pt]   
&\hspace{120pt}  {\mathfrak A}^{(2)}_{a,s} := 
-\frac{
z_{b}
\Tb_{a+1,s}  
\Tb^{B_{1}^{\prime}}_{a-1,s+1}
}
{\Tb_{a,s+1}  
\Tb^{B_{1}^{\prime}}_{a,s} } ,  
 \label{y-sys-bac2}
\\[8pt]   
%%%%%%%%%%%%%%%%%%%%%%%%%  
& {\mathfrak A}^{(3)F_{1}^{\prime}[-1]}_{a+1,s}  
{\mathfrak A}^{(3)[1]}_{a,s} 
=
\frac{
(1+ {\mathfrak A}^{(3)F_{1}^{\prime}}_{a+1,s-1}  )
(1+ {\mathfrak A}^{(3)}_{a,s+1}  )
}
{
(1+ ({\mathfrak A}^{(3)F_{1}^{\prime}}_{a,s})^{-1}  )
(1+ ({\mathfrak A}^{(3)}_{a+1,s})^{-1}  )
}
\nonumber  \\[4pt]  
& \hspace{120pt} {\mathfrak A}^{(3)}_{a,s}  : = 
\frac{
z_{f}
\Tb^{F_{1}^{\prime}}_{a+1,s-1}  
\Tb_{a,s+1}
}
{
\Tb^{F_{1}^{\prime}}_{a,s}  
\Tb_{a+1,s} 
} , 
\\[8pt]   
%%%%%%%%%%%%%%%%%%%%%%%%%%%   
& {\mathfrak A}^{(4) F_{1}^{\prime}[-1]}_{a,s}  
{\mathfrak A}^{(4)[1]}_{a,s+1}
=
\frac{(1+{\mathfrak A}^{(4) F_{1}^{\prime}}_{a+1,s})
(1+{\mathfrak A}^{(4)}_{a-1,s+1})}
{(1+({\mathfrak A}^{(4) F_{1}^{\prime}}_{a,s+1})^{-1})
(1+({\mathfrak A}^{(4) }_{a,s})^{-1})}
,
\nonumber  \\[4pt]   
& \hspace{120pt} {\mathfrak A}^{(4)}_{a,s} : =  
-\frac{
z_{f}
\Tb^{F_{1}^{\prime}}_{a+1,s}  
\Tb_{a-1,s+1}  
}
{\Tb^{F_{1}^{\prime}}_{a,s+1}  
\Tb_{a,s} } . 
\end{align}  
%%%%%%%%%%%%%%%%%%%%%%%%%%%  
\begin{align}  
 &{\mathfrak A}^{(5)[1]}_{a+1,s}  
{\mathfrak A}^{(5)B_{2}^{\prime}[-1]}_{a,s}  
=
\frac{(1+ {\mathfrak A}^{(5)}_{a+1,s+1})
(1+ {\mathfrak A}^{(5)B_{2}^{\prime}}_{a,s-1})}
{(1+ ({\mathfrak A}^{(5)}_{a,s})^{-1} )
(1+ ({\mathfrak A}^{(5)B_{2}^{\prime }}_{a+1,s})^{-1} )} , 
\nonumber  \\[4pt]
& \hspace{120pt}
{\mathfrak A}^{(5)}_{a,s} : = 
\frac{
\Tb_{a+1,s+1}  
\Tb^{B_{2}^{\prime}}_{a,s-1}
}
{
\Tb_{a,s}  
\Tb^{B_{2}^{\prime}}_{a+1,s}  
}  ,
\\[8pt]   
%%%%%%%%%%%%% 
&  {\mathfrak A}^{(6)[1]}_{a,s}  
{\mathfrak A}^{(6)B_{2}^{\prime}[-1]}_{a,s-1}  
=\frac{(1+{\mathfrak A}^{(6)}_{a+1,s} )
(1+{\mathfrak A}^{(6)B_{2}^{\prime}}_{a-1,s-1} )}{
(1+({\mathfrak A}^{(6)}_{a,s-1})^{-1} )
(1+({\mathfrak A}^{(6)B_{2}^{\prime}}_{a,s})^{-1} )},
\nonumber  \\[4pt] 
& \hspace{120pt}
{\mathfrak A}^{(6)}_{a,s} := -
\frac{
\Tb_{a+1,s}  
\Tb^{B_{2}^{\prime}}_{a-1,s-1}
}
{
\Tb_{a,s-1}  
\Tb^{B_{2}^{\prime}}_{a,s} 
} ,
\\[8pt]   
%%%  
 & {\mathfrak A}^{(7)F_{2}^{\prime}[1]}_{a+1,s}  
{\mathfrak A}^{(7)[-1]}_{a,s}  
=\frac{(1+{\mathfrak A}^{(7)F_{2}^{\prime}}_{a+1,s+1} )
(1+{\mathfrak A}^{(7)}_{a,s-1} )}
{(1+({\mathfrak A}^{(7)F_{2}^{\prime}}_{a,s})^{-1} )
(1+({\mathfrak A}^{(7)}_{a+1,s})^{-1} )}, 
\nonumber  \\[4pt]   
& \hspace{120pt} {\mathfrak A}^{(7)}_{a,s} := - 
\frac{
\Tb^{F_{2}^{\prime}}_{a+1,s+1}  
\Tb_{a,s-1}
}
{
\Tb^{F_{2}^{\prime}}_{a,s}  
\Tb_{a+1,s}  
} ,
\\[8pt]   
%%%%%%
& {\mathfrak A}^{(8)F_{2}^{\prime}[1]}_{a,s}  
{\mathfrak A}^{(8)[-1]}_{a,s-1}  
=
\frac{
(1+ {\mathfrak A}^{(8)F_{2}^{\prime}}_{a+1,s} )
(1+ {\mathfrak A}^{(8)}_{a-1,s-1} )
}
{
(1+ ({\mathfrak A}^{(8)F_{2}^{\prime}}_{a,s-1})^{-1} )
(1+ ({\mathfrak A}^{(8)}_{a,s})^{-1} )
} ,
\nonumber  \\[4pt]
& \hspace{120pt}
{\mathfrak A}^{(8)}_{a,s} :=
\frac{
\Tb^{F_{2}^{\prime}}_{a+1,s}  
\Tb_{a-1,s-1}  
}
{\Tb^{F_{2}^{\prime}}_{a,s-1}  
\Tb_{a,s} } .
\end{align} 
Here notation on the index set is basically same as the one in \eqref{bac1}-\eqref{bacL4}. 
The double prime on the index set should be interpreted as follows. For example, 
${\mathfrak A}^{(1)B_{1}^{\prime}}_{a,s} = 
\frac{
z_{b}
\Tb^{B_{1}^{\prime}}_{a+1,s-1}  
\Tb^{B_{1}^{\prime \prime}}_{a,s+1}
}
{
\Tb^{B_{1}^{\prime}}_{a,s}  
\Tb^{B_{1}^{\prime \prime }}_{a+1,s} 
}  $ contains $B_{1}^{\prime \prime }$. 
Here we use notations 
$B_{1}^{\prime}:=B_{1}\setminus \{b \} $, $B_{1}^{\prime \prime}:=B_{1}\setminus \{b, b^{\prime} \} $ 
for any fixed elements $b,b^{\prime} \in B_{1}$ ($b \ne b^{\prime} $).  The double prime for the other sets 
 should be interpreted in the same manner. 

An integral form of the Y-system \eqref{y-system} is the thermodynamic Bethe ansatz equation. 
It is an infinite system of nonlinear integral equations with 
an infinite number of unknown functions for the free energy of quantum integrable systems at finite temperatures. 
Thus to find a finite system of nonlinear integral equations (NLIE) with a finite number of unknown functions which is equivalent to 
the thermodynamic Bethe ansatz equation will be an important problem. 
There are at least two type of  NLIE in  literatures. 
One of them is NLIE of Takahashi-type \cite{Takahashi00}. 
We derived NLIE of Takahashi-type for several integrable systems in statistical mechanics 
whose underlaying algebras have arbitrarily rank \cite{Tsuboi06}. 
 NLIE of Takahashi-type is suited for calculations of high temperature thermodynamics. 
 However, for the analysis of the low temperature thermodynamics, another type of NLIE 
 proposed by Destri-de Vega or Kl\"umper  \cite{DD92} seems to be better than Takahashi-type NLIE. 
 Attempts to generalize this type of NLIE have been made by several authors, and 
 in the case of integrable field theoretical models, NLIE of this type are known even for 
 models whose underlaying algebras have arbitrarily rank
\footnote{There are also NLIE related to algebras with arbitrarily rank \cite{nlie} 
in rather different context.} \cite{KL10}. 
 On the other hand for quantum integrable spin chains in statistical mechanics, this type of NLIE 
 for arbitrary rank has not been established yet. 
 This is because one needs a considerable trial and errors to find 
 auxiliary functions with good analytical properties which play a key role
(with a method to modify the Y-system proposed in \cite{Suzuki99})  
in the derivation of the NLIE. 
 In this context, 
we remark that the above `Y-functions' $\{ {\mathfrak A}^{(b)}_{a,s} \}_{b=1}^{8}$ for 
the B\"{a}cklund transformations are 
similar to such auxiliary functions for NLIE
\footnote{
Tableau sum expressions of the T-functions $\Tb_{a,s} $ for 
$B_{2}=F_{2}=\emptyset $ are available in \cite{T09}.  
In term of these, 
$\{ {\mathfrak A}^{(b)}_{a,s} \}_{b=1}^{8}$ 
have similar structure as auxiliary functions for the NLIE (cf. \cite{DK06}). 
}. 
It will be important to clarify exact relation among them. 
%%%%%%%%%%%%%%%%%%%%%%%%%%%%%%%%%%%%%%%%%%%%%%%%%%5
\section{Concluding remarks}
In this paper, 
we have continued our trials \cite{T09,BT08} (and also \cite{T97,T98,T98-2,T02,GKT10,GKLT10,KLT10}) to construct 
T-and Q-functions/operators for integrable models related to quantum affine (super)algebras 
and establish Wronskian-type formulas for any representations and functional relations among them. 
This paper is a small step toward our goal, and we focused our discussion on solutions of the 
T-system in terms of Q-functions 
in relation to infinite dimensional unitarizable modules of $gl(M|N)$. 
Most of our discussions here do not depend on whether the Q-functions are Baxter Q-operators or their eigenvalues. 
One of the next steps is to realize our formulas as operators. This will give us information on more precise algebraic and 
analytical structures 
of our formulas. 
There are several methods to perform this. 
One of them is to use the co-derivate \cite{KV07} on the generating function $w(t)$ 
of the supercharacters of the symmetric tensor representations 
%\footnote{One may also use the generating function of the anti-symmetric tensor representations $w(-t)^{-1}$.} 
of $gl(M|N)$, 
which produces a generating operator for T-operators. 
We used \cite{KLT10} it to define T-and Q-operators. 
Generating operators of the T-operators satisfy a kind of Hirota equation, 
which we call a master identity. Many functional relations among 
T-and Q-operators follow from the master identity by easy manipulations. 
Our T-operators in \cite{KLT10} were the ones for finite dimensional 
representations on both quantum and auxiliary spaces of the models 
(they are formulas on $[M,N]$-hook of $gl(M|N)$). 
It seems plausible that we can realize our formulas for infinite dimensional representations in the auxiliary 
space as operators just by changing the expansion point of the generating functions (as in \eqref{expan-inf}) 
and apply the co-derivative in the same way as \cite{KLT10}. 
Another approach relevant to our formulas is 
to use oscillator representations of the quantum affine superalgebras (or superYangians), 
which was proposed by \cite{BLZ97} for $U_{q}(\widehat{sl}(2))$ case
\footnote{As for T-operators related to the T-hook, 
there is an approach \cite{Benichou10} which is different but conceptually closer to \cite{BLZ97}.} 
and developed in various directions \cite{BHK02,Kojima08,BT08,osc-app,BFLMS10}. 
In this construction, Q-operators are defined as supertrace of monodromy matrices over some oscillator 
representations. 
To construct  a generating operator
\footnote{Note that the generating function of the characters is a kind of partition function of harmonic oscillators. 
The `co-derivative approach' and the `oscillator approach' will be unified in this context. 
In addition, a duality among two different groups seems to play a role to establish 
 the master identity. Here we mean a duality that parameters $\{t_{k} \}$ in the product $\prod_{k}w(t_{k})$ also generate group characters which are different from the original group characters generated by $w(t)$ itself.} 
for T-operators in this approach seems to be one of the key steps toward our goal. 
Although the Master identity in \cite{KLT10} was proposed for T-operators whose 
quantum space is fundamental representation on each site, notion of it 
is independent of the quantum space (and also independent of the coderivative), and will be generalizable to any quantum space
\footnote{In this context, the generating operators of the T-operators will be a kind of (super)determinant over a 
function of a L-operator which 
generates the super-Yangian or the quantum affine superalgebra. 
Some related quantities are known in literatures (for example, \cite{Talalaev04}).}.  
%However this needs further clarification. 
%One of the other interesting problems is to generalize our Wronskian like formulas to other algebras. 
%For this purpose, we find it quite useful to use reductions on the Q-functions by the automorphisms of the 
%algebra $gl(M|N)$ and dualities among different superalgebras (such as $sl(2n|1)^{(2)} $ and $so(2n+1)^{(1)} $ etc). 
%We have already obtained some Wronskian like solutions of the T-systems in terms of Q-functions for some other 
%algebras including $U_{q}(so(2n+1)^{(1)})$.  
%based on automorphism of $gl(M|N)$ and 
%dualities among different algebras (such as $sl(2n|1)^{(2)} $ and $so(2n+1)^{(1)} $ etc). 
%They will be published elsewhere.  
We hope to address these issues step by step near future. 
%%%%%%%%%%%%%%%%%%%%%%%%%%%%%%
\section*{Acknowledgments}
The author thanks Nikolay Gromov, Vladimir Kazakov, Sebastien Leurent for 
collaboration in \cite{GKLT10}, Ivan Cherednik for communication, and Dmytro Volin for 
 a critical reading of the manuscript and useful comments. 
The work of ZT is supported by 
SFB647 ``Space-Time-Matter'' and 
was partially supported by Nishina Memorial Foundation and by
Grant-in-Aid for Young Scientists, B \#19740244 from
The Ministry of Education, Culture, Sports, Science and Technology in Japan.
 He also thanks 
Ecole Normale Superieure, LPT, where a part of this work was done,
for the hospitality. 
The main part of this manuscript was written when he was at 
Max-Planck-Institut f\"{u}r Gravitationsphysik, 
Albert-Einstein-Institut. 
He thanks Niklas Beisert for giving an opportunity to stay there. 
The main results of this paper 
have been previously announced on a 
number of conferences or seminars 
\footnote{first, as a poster at 
`Integrability in Gauge and String Theory 2010', 
Nordita, Sweden, 28 June 2010 - 2 July, 
and seminars at AEI, Potsdam, 9 August 2010;  
at IPMU, Japan, 7 September 2010},  
and also presented at meetings of a group of Matthias Staudacher at Humboldt university
 when he visited him in August 2010. He thanks  
 Matthias Staudacher and people in his group for their interest to this work.
%%%%%%%%%%%%%%%%%%%%%%%%%%%%%%%%%%%%%%%%%

\appendix
\section{Coefficient free form of the solutions}
%\app{Coefficient free form of the solutions}
In this section, we list the coefficient free form of the solutions of the T-system given in 
section 3.3 and 3.5 by the transformations \eqref{removtwistQ}-\eqref{removtwistT}. 
For any sets $I$ and $J$, we will use a notation   
  $\epsilon_{I;J}:=(-1)^{\mathrm{Card} \{(i,j) \in I \times J| i>j\}}$. 
Then \eqref{soQQb-1}-\eqref{soQQb-3} reduce to
\footnote{
%We were pointed out that 
V.\ Kazakov, S.\ Leurent, D.\ Volin rewrote 
 some of our formulas, which we gave to them,  
%can be easily rewritten
 in a compact form in terms of the differential form \cite{KLV10}. 
 This is based on the fact that any determinant can be expressed in terms of the differential form. 
 And the differential form is particularly useful to express Wronskian type determinant formulas. 
We used Laplace expansion on determinants and Pl\"{u}cker identities in our proof of the formulas (in appendix B). 
One can prove Laplace expansion formula and Pl\"{u}cker identities elegantly based on the differential form. 
Thus, the differential form expression will be useful to make our proof  of our formulas transparent. 
%It will also be useful to discuss symmetry of the solutions under the automorphisms 
%(such as $\sigma$ and $\tau$ in section 2.) of the underlying algebra.  
On the other hand, when one evaluates the formulas (for example, numerically), 
one has to extract coefficients of Grassmann numbers, and  reproduces Wronskian like determinant formulas.}: 
\begin{align}    
\Tf_{a,s} &=   
(-1)^{(a+m_{1})(s+n_{2}) +m_{1}m_{2} }   
\sum_{I\sqcup J= F_{1}\sqcup  F_{2},  
|I|=n_{1}-s} \epsilon_{J;I}   
\Qf^{[a-\eta_{2}]}_{B_{1},I}  
\Qf^{[-a+\eta_{1}]}_{B_{2},J}  
\nonumber \\
&
 \qquad \qquad  \text{for} \quad a\ge \max\{s+\eta_{1}, -s+\eta_{2}, 0 \},  
\label{soQQbc-1}  
\\[6pt]
\Tf_{a,s} &=   (-1)^{m_{2}a}
\sum_{I\sqcup J=B_{1}, |I|=a} \epsilon_{I;J}   
\Qf^{[s+\eta_{1}-\eta_{2}]}_{I}  
\Qf^{[-s]}_{J,  
B_{2}, F_{1}, F_{2}}  
\qquad \text{for} \quad  a \le s + \eta_{1}, \label{soQQbc-2}  
\\[6pt]
\Tf_{a,s} &= (-1)^{m_{1}a}  
\sum_{I\sqcup J= B_{2}, |I|=a} \epsilon_{J;I}   
\Qf^{[s+\eta_{1}-\eta_{2}]}_{I}  
\Qf^{[-s]}_{  
B_{1}, J,F_{1},F_{2} }   
\qquad \text{for} \quad a\le -s+ \eta_{2}, 
\label{soQQbc-3}  
\end{align}  
where the summation is taken over any possible decomposition of the original set 
into two disjoint sets $I $ and $J$ with fixed sizes. And we have to keep 
the order of the original set in $I $ and $J$ (because of the sign). 
For example in \eqref{soQQbc-1}, suppose 
$i_{\alpha }=f_{\alpha^{\prime}}$ and $i_{\beta }=f_{\beta^{\prime}}$ 
(or $j_{\alpha }=f_{\alpha^{\prime}}$ and $j_{\beta }=f_{\beta^{\prime}}$) hold for 
$\alpha < \beta$, then $\alpha^{\prime} < \beta^{\prime}$ must be valid, where 
$F_{1}\sqcup  F_{2}=(f_{1},f_{2},\dots, f_{n})$,  
$I=(i_{1},i_{2},\dots, i_{n_{1}-s})$, $J=(j_{1},j_{2},\dots, j_{n_{2}+s})$. 
In particular at the boundaries of the T-hook, the above formulas become
\begin{align}  
& \Tf_{a,n_{1}} =   
(-1)^{(a+m_{1})(n_{1} - n_{2})+m_{1}m_{2}}
\Qf^{[a-\eta_{2}]}_{B_{1}}  
\Qf^{[-a+\eta_{1}]}_{B_{2}, F_{1},F_{2}}  
\quad  \text{for} \quad a \ge  m_{1},  
\label{boundURc}  
\\[6pt]   
& \Tf_{a,-n_{2}} =   
(-1)^{m_{1}m_{2}}   
\Qf^{[a-\eta_{2}]}_{B_{1}, F_{1}, F_{2}}  
\Qf^{[-a+\eta_{1}]}_{B_{2}}  
\quad  \text{for} \quad a\ge  m_{2},  
\label{boundULc}  
\\[6pt]  
& \Tf_{m_{1},s} =   
(-1)^{m_{1} m_{2}}  
\Qf^{[s+\eta_{1}-\eta_{2}]}_{B_{1}}  
\Qf^{[-s]}_{  
B_{2}, F_{1}, F_{2}}   
\quad  \text{for} \quad s\ge  n_{1},  
\label{boundRc}  
\\[6pt]  
&   
\Tf_{m_{2},s}   
= (-1)^{m_{1}m_{2}}  
\Qf^{[s+\eta_{1}-\eta_{2}]}_{B_{2}}  
\Qf^{[-s]}_{  
B_{1}, F_{1},F_{2} }  
\quad  \text{for} \quad s\le  -n_{2}, \label{boundLc}  
\\[6pt]  
& \Tf_{0,s} =   
\Qf^{[-s]}_{B_{1},  
B_{2}, F_{1}, F_{2}}  
\quad  \text{for} \quad s \in {\mathbb Z}.
\label{boundDc}  
\end{align}  
%%%%%%%%%%%%%%%%%%%%%%%%%%%%%%%%%%%%%%  
The sparse determinant \eqref{sparcedetbb} becomes 
in the following form. 
\begin{multline}  
\Def^{B_{1},B_{2},R
,[\eta; \xi ]}_  
{F,S, T_{1}, T_{2}  
}  
=
\\  
\begin{vmatrix}  
\left( \Qf_{b,f}^{[\xi ]}
\right)_{  
\genfrac{}{}{0pt}{}{b\in B_{1}, }{f \in F} } 
&   
\left(
\Qf_{b}^{[\xi+2s-1]}
\right)_{  
\genfrac{}{}{0pt}{}{b\in B_{1}, }{s \in S} }   
&   
\left(
 \Qf_{b}^{[\xi+2t-1]}
\right)_{  
\genfrac{}{}{0pt}{}{b\in B_{1}, }{t \in T_{1}} }   
& (0)_{|B_{1}| \times |T_{2}| }   
  \\[6pt]  
\left( 
(-1)^{\eta} \Qf_{b,f}^{[\xi -2\eta]}
\right)_{ 
\genfrac{}{}{0pt}{}{b \in B_{2}, }{f \in F} }  
&   
\left(
 \Qf_{b}^{[\xi+2s-1]}
\right)_{  
\genfrac{}{}{0pt}{}{b\in B_{2}, }{s \in S} }   
& (0)_{|B_{2}| \times |T_{1}| }   
&   
\left(
 \Qf_{b}^{[\xi+2t-1]}
\right)_{  
\genfrac{}{}{0pt}{}{b\in B_{2}, }{t \in T_{2}} }  
   \\[6pt]  
\left(
(-1)^{r-1} \Qf_{f}^{[\xi-2r+1 ]}
\right)_{  
\genfrac{}{}{0pt}{}{r \in R, }{f \in F} }  
& (0)_{|R|\times |S|}   
& (0)_{|R| \times |T_{1}| }  
& (0)_{|R| \times |T_{2}| }\\  
\end{vmatrix}  
,  \label{sparcedetfc}
\end{multline}  
where the number of the elements of the sets must satisfy 
$|B_{1}|+|B_{2}|+|R|=|F| + |S|+|T_{1}|+|T_{2}| $.
%%%%%%%%%%%%%%%%%%%%%%%%%%%%%%%%%%%%%  
Then \eqref{sparse01}-\eqref{sparse08} reduce to :   
\begin{align}  
\Tf_{a,s} &=   
(-1)^{(s+\eta_{1})(s+n_{2})+\Theta }  
\Def^{B_{1},B_{2},\emptyset,[a-s-\eta_{1} ;a-s-\eta_{1}-\eta_{2}]}_  
{F_{1}, F_{2},\emptyset,\langle 1,s+\eta_{1} \rangle,  
\langle s-a+\eta_{1}+1,-a+\eta_{1}+\eta_{2} \rangle}  
\nonumber  \\  
& \qquad \text{for} \quad a\ge \max\{s+\eta_{1}, -s+\eta_{2}, 0 \},  
\quad   
-\eta_{1} \le s \le \eta_{2},  
\label{sparsec1}  
\\[6pt]
%%%% 
\Tf_{a,s} &=   
(-1)^{(s+\eta_{1})(s+n_{2})+\Theta }    
\Def^{B_{1},B_{2},\langle a-s-\eta_{1}+1,a-\eta_{1}-\eta_{2} \rangle,  
[a-s-\eta_{1};a-s-\eta_{1}-\eta_{2}]}_  
{F_{1}, F_{2},\emptyset, \langle 1,s+\eta_{1} \rangle ,\emptyset}  
\nonumber  \\   
& \qquad \text{for} \quad a\ge \max\{s+\eta_{1}, -s+\eta_{2}, 0 \},  
\quad   
s \ge \max\{ -\eta_{1}, \eta_{2}\},  
\label{sparsec2}  
\\[6pt]
%%%%%
\Tf_{a,s} &=   
(-1)^{(s+\eta_{1})m_{2}+\Theta }   
\Def^{B_{1},B_{2},  
\langle 1,-s-\eta_{1} \rangle ,  
[a-s-\eta_{1};a-s-\eta_{1}-\eta_{2}]}_  
{F_{1}, F_{2},  
\emptyset, \emptyset, \langle s-a+\eta_{1}+1,-a+\eta_{1}+\eta_{2} \rangle }  
\nonumber \\ 
 & \qquad \text{for} \quad a\ge \max\{s+\eta_{1}, -s+\eta_{2}, 0 \},  
\quad   
s \le \min\{ -\eta_{1}, \eta_{2}\},  
\label{sparsec3}  
\end{align}  
%%%%  
\begin{align}  
& \Tf_{a,s}  =  \nonumber \\
&= (-1)^{(s+\eta_{1})m_{2}+\Theta }
\Def^{B_{1},B_{2},  
(1,2,\dots,-s-\eta_{1},a-s-\eta_{1}+1,a-s-\eta_{1}+2,\dots, a-\eta_{1}-\eta_{2}),  
[a-s-\eta_{1};a-s-\eta_{1}-\eta_{2}]}_  
{F_{1}, F_{2},  
\emptyset, \emptyset,\emptyset}  
\nonumber  \\ 
& \qquad \text{for} \quad a\ge \max\{s+\eta_{1}, -s+\eta_{2}, 0 \},  
\quad   
\eta_{2}  \le s \le  -\eta_{1},   
\label{sparsec4}  
\end{align}  
%%%%  
\begin{align}  
\Tf_{a,s} &=   (-1)^{m_{2}a+\Theta }
\Def^{B_{1},B_{2},\emptyset ,  
[0;a-s-\eta_{1}-\eta_{2}]}_  
{F_{1}, F_{2},  
\langle 1,-a+\eta_{1}+\eta_{2} \rangle , \langle s-a+\eta_{1}+1,s+\eta_{1} \rangle ,\emptyset}  
\nonumber \\
&
\qquad \text{for} \quad a\le \min \{ s+\eta_{1}, \eta_{1}+\eta_{2} \},   
\label{sparsec5}  
\\[6pt]
%%%% 
\Tf_{a,s} &=   
(-1)^{a(\eta_{1}+n_{2}+1)+\Theta }  
\Def^{B_{1},B_{2},\langle 1,a-\eta_{1}-\eta_{2} \rangle ,  
[0;a-s-\eta_{1}-\eta_{2}]}_  
{F_{1}, F_{2},  
\emptyset, \langle s-a+\eta_{1}+1, s+\eta_{1}\rangle ,\emptyset}  
\nonumber \\
&
\qquad \text{for} \quad \eta_{1}+\eta_{2} \le a \le s+\eta_{1},   
\label{sparsec6}  
\\[6pt] 
%%%%%%%%%%%%%%%%%%%%%%%%%%%%%%%%%%%%    
\Tf_{a,s} &=   
(-1)^{a(m_{2}+1)+\Theta }   
\Def^{B_{1},B_{2},\emptyset ,  
[0;a-s-\eta_{1}-\eta_{2}]}_  
{F_{1}, F_{2},  
\langle 1,-a+\eta_{1}+\eta_{2}\rangle ,\emptyset,
\langle s-a+\eta_{1}+1, s+\eta_{1}\rangle }  
\nonumber \\
&
\qquad \text{for} \quad a\le \min \{ -s+\eta_{2}, \eta_{1}+\eta_{2} \},   
\label{sparsec7}  
\\[6pt]
%%%%%%%%%%%%%%%%%%%%%%%%%%%%%%%%%%%%  
\Tf_{a,s} &=   
(-1)^{a(\eta_{1}+n_{2}) +\Theta }  
\Def^{B_{1},B_{2}, \langle 1,a-\eta_{1}-\eta_{2} \rangle  ,  
[0;a-s-\eta_{1}-\eta_{2}]}_  
{F_{1}, F_{2},  
\emptyset,\emptyset, \langle s-a+\eta_{1}+1, s+\eta_{1} \rangle }  
\nonumber \\
& 
\qquad \text{for} \quad \eta_{1}+\eta_{2} \le a \le -s+\eta_{2},   
\label{sparsec8}  
\end{align}  
where we introduced a symbol 
$\Theta:= \frac{(|B_{1}|+|B_{2}|)(|B_{1}|+|B_{2}|-1)}{2} $. 
%%%%%%%%%%%%%%%%%%%%%%%%%%%%%%%%%%%%%%%%%%%%%%%%%%%
We remark that we can unify \eqref{sparsec1}-\eqref{sparsec4}, \eqref{sparsec5}-\eqref{sparsec6} and 
\eqref{sparsec7}-\eqref{sparsec8} as
\begin{align}
\Tf_{a,s} &=   
(-1)^{\min\{s+\eta_{1}, 0\} (s+\eta_{2})+(s+\eta_{1})(s+n_{2})+\Theta }   
\nonumber \\
& \quad \times
\Def^{B_{1},B_{2},  
\langle 1,-s-\eta_{1} \rangle  \sqcup \langle a-s-\eta_{1}+1,a-\eta_{1}-\eta_{2} \rangle ,  
[a-s-\eta_{1};a-s-\eta_{1}-\eta_{2}]}_  
{F_{1}, F_{2},  
\emptyset, \langle 1,s+\eta_{1} \rangle, \langle s-a+\eta_{1}+1,-a+\eta_{1}+\eta_{2} \rangle }  
\nonumber \\ 
 & \qquad \text{for} \quad a\ge \max\{s+\eta_{1}, -s+\eta_{2}, 0 \},  
 \label{sparsec14}  
 \\[6pt]
%%%%
\Tf_{a,s} &=   (-1)^{a(\max\{a-\eta_{1}-\eta_{2},0 \} +m_{2})+\Theta }
\Def^{B_{1},B_{2}, \langle 1,a-\eta_{1}-\eta_{2} \rangle ,  
[0;a-s-\eta_{1}-\eta_{2}]}_  
{F_{1}, F_{2},  
\langle 1,-a+\eta_{1}+\eta_{2} \rangle , \langle s-a+\eta_{1}+1,s+\eta_{1} \rangle ,\emptyset}  
\nonumber \\
&
\qquad \text{for} \quad a\le s+\eta_{1},   
\label{sparsec56}  
\\[6pt]
%%%%
\Tf_{a,s} &=   
(-1)^{a(\min\{a-\eta_{1}-\eta_{2}, 0 \} +\eta_{1}+ n_{2})+\Theta }   
\Def^{B_{1},B_{2},\langle 1,a-\eta_{1}-\eta_{2}\rangle ,  
[0;a-s-\eta_{1}-\eta_{2}]}_  
{F_{1}, F_{2},  
\langle 1,-a+\eta_{1}+\eta_{2}\rangle ,\emptyset,
\langle s-a+\eta_{1}+1, s+\eta_{1}\rangle }  
\nonumber \\
&
\qquad \text{for} \quad a\le  -s+\eta_{2}. 
\label{sparsec78}  
\end{align}
Nevertheless, we will keep \eqref{sparsec1}-\eqref{sparsec8} since 
we need some case study in Appendix B.

The dense determinant expression of the solution
 \eqref{Bglmmnn-4}-\eqref{Bglmmnn-5} becomes in the following form. 
\begin{multline}  
\Tf_{a,s}  
=\frac{(-1)^{\frac{m_{1}(m_{1}-1)}{2} +m_{2}a} } 
{\prod_{k=1}^{m_{1}-a-1}\Qf_{B_{2}, F_{1}, F_{2}}^{[  
a-s-m_{1}+2k]} }   
\\ \times   
\begin{vmatrix}  
\left(
\Qf_{b,B_{2}, F_{1}, F_{2}}^{[-s+a-m_{1}+2j-1]}
\right)_{  
b \in B_{1}, 1\le j \le m_{1}-a} &   
\left(
\Qf_{b}^{[s-a+2j-1+\eta_{1}-\eta_{2}]}
\right)_{b \in B_{1}, 1\le j \le a}   
  \\  
\end{vmatrix}  
\\  
\quad \text{for} \quad  a \le s + \eta_{1}, \label{Bglmmnnc-4}  
\end{multline}  
\begin{multline}  
%\hspace{-30pt}   
\Tf_{a,s} 
=\frac{(-1)^{\frac{(n_{1}+n_{2})(n_{1}+n_{2}-1)}{2}+(a+m_{1})(s+n_{2}) +m_{1}m_{2}}}  
{\prod_{k=1}^{n_{1}-s-1}  
\Qf_{B_{1}}^{[a-s+n_{1}-\eta_{2}-2k]}  
\prod_{k=1}^{n_{2}+s-1}  
\Qf_{B_{2}}^{[-a+s+n_{2}+\eta_{1}-2k]}}   
\\ \times   
\begin{vmatrix}  
\left(
\Qf_{B_{1},f}^{[a-s+n_{1}-\eta_{2}-2i+1]}
\right)_{1\le i \le n_{1}-s,   
f \in F}  \\  
\left(
\Qf_{B_{2},f}^{[-a+s+n_{2}+\eta_{1}-2i+1]}
\right)_{1\le i \le n_{2}+s,   
f \in F }   \\  
\end{vmatrix}  
\\  
\quad \text{for} \quad a\ge \max\{s+\eta_{1}, -s+\eta_{2}, 0 \},  \label{Bglmmnnc-3}  
\end{multline}  
%%%%   
%  
\begin{multline}  
\Tf_{a,s}   
=\frac{(-1)^{\frac{m_{2}(m_{2}-1)}{2}+m_{1}a}}  
{\prod_{k=1}^{m_{2}-a-1}  
\Qf_{B_{1},F_{1},F_{2} }^{[  
a-s-m_{2}+2k]} } \\  
\hspace{-40pt}   
\times   
\begin{vmatrix}  
\left(
\Qf_{b }^{[s-a+2j-1+\eta_{1}-\eta_{2}]}
\right)_{  
b \in B_{2}, 1\le j \le a}  &  
\left(
\Qf_{B_{1},b, F_{1},F_{2}}^{[-s+a-  
m_{2}+2j-1]}
\right)_{  
b \in B_{2}, 1\le j \le m_{2}-a}   \\  
\end{vmatrix}  
\\  
\quad \text{for} \quad a\le -s+ \eta_{2}. \label{Bglmmnnc-5}  
\end{multline}  

%%%%%%%%%%%%%%%%%%%%%%%%%%%%%%%%%%%%%%%%%%%%%%%%%%%%
\section{Proof of the functional relations}
%\app{Proof of the functional relations}
In this appendix, we will prove that the T-function 
$\Tf_{a,s}^{B_{1},B_{2},F_{1},F_{2}}$ in Appendix A 
satisfies the 
Hirota equation: 
\begin{align}
  \Tf^{[-1]}_{a,s} \Tf^{[1]}_{a,s} 
  =
  \Tf_{a,s-1} \Tf_{a,s+1}+
  \Tf_{a-1,s} \Tf_{a+1,s} 
  \label{hirota-BCb}
\end{align}
on the generalized T-hook. 
We will use QQ-relation for $I=\emptyset$
 with a normalization $\Qf_{\emptyset}=1$. 
\begin{align}
\Qf_{b}\Qf_{f}=\Qf_{bf}^{[1]}-\Qf_{bf}^{[-1]} ,
\qquad b \in {\mathfrak B}, 
\quad f \in {\mathfrak F}.
\label{QQ-bf1} 
\end{align} 
We will use the following lemma in the proof. 
\begin{lemma}\label{shift-lem}
The following relations are valid for the determinant \eqref{sparcedetfc} 
under the relation \eqref{QQ-bf1}, 
where $|B_{1}|+|B_{2}|+a=|F|+b+c_{1}+c_{2}$. 

\noindent
(i) If there are components such that $t^{(1)}_{i}=0$ for some $i \in \{ 1,2,\dots, c_{1} \}$ 
or $r_{j}=1$ for some $j \in \{ 1,2,\dots, a \}$ , 
the following relation holds 
\begin{multline}
 \Def^{B_{1},B_{2},(r_{1},r_{2},\dots, r_{a}),[\eta; \xi ]}_
{F,\emptyset ,
(t^{(1)}_{1},t^{(1)}_{2},\dots, t^{(1)}_{c_{1}}),
(t^{(2)}_{1},t^{(2)}_{2},\dots, t^{(2)}_{c_{2}})
}
\\
=(-1)^{a+m_{2}+c_{2}}
\Def^{B_{1},B_{2},(r_{1}-1,r_{2}-1,\dots, r_{a}-1),[\eta-1; \xi-2 ]}_
{F,\emptyset ,
(t^{(1)}_{1}+1,t^{(1)}_{2}+1,\dots, t^{(1)}_{c_{1}}+1),
(t^{(2)}_{1}+1,t^{(2)}_{2}+1,\dots, t^{(2)}_{c_{2}}+1)
}. 
\label{shl1}
\end{multline}
(ii) If there is a component such that $s_{i}=0$ for some $i \in \{ 1,2,\dots, b \}$ , 
the following relation holds 
\begin{multline}
 \Def^{B_{1},B_{2},(r_{1},r_{2},\dots, r_{a}),[0; \xi ]}_
{F,(s_{1},s_{2},\dots, s_{b}),
(t^{(1)}_{1},t^{(1)}_{2},\dots, t^{(1)}_{c_{1}}),
(t^{(2)}_{1},t^{(2)}_{2},\dots, t^{(2)}_{c_{2}})
}
\\
=(-1)^{a}
\Def^{B_{1},B_{2},(r_{1}-1,r_{2}-1,\dots, r_{a}-1),[0; \xi-2 ]}_
{F,(s_{1}+1,s_{2}+1,\dots, s_{b}+1),
(t^{(1)}_{1}+1,t^{(1)}_{2}+1,\dots, t^{(1)}_{c_{1}}+1),
(t^{(2)}_{1}+1,t^{(2)}_{2}+1,\dots, t^{(2)}_{c_{2}}+1)
}.
\label{shl2}
\end{multline}
(iii) If there are components such that $t^{(2)}_{i}=-\eta +1$ for some $i \in \{ 1,2,\dots, c_{2} \}$ 
or $r_{j}=\eta $ for some $j \in \{ 1,2,\dots, a \}$ , 
the following relation holds 
\begin{multline}
 \Def^{B_{1},B_{2},(r_{1},r_{2},\dots, r_{a}),[\eta; \xi ]}_
{F, \emptyset ,
(t^{(1)}_{1},t^{(1)}_{2},\dots, t^{(1)}_{c_{1}}),
(t^{(2)}_{1},t^{(2)}_{2},\dots, t^{(2)}_{c_{2}})
}
=(-1)^{m_{2}+c_{2}}
\Def^{B_{1},B_{2},(r_{1},r_{2},\dots, r_{a}),[\eta-1; \xi ]}_
{F,\emptyset ,
(t^{(1)}_{1},t^{(1)}_{2},\dots, t^{(1)}_{c_{1}}),
(t^{(2)}_{1},t^{(2)}_{2},\dots, t^{(2)}_{c_{2}})
}. 
\label{shl3}
\end{multline}
\end{lemma}
%%%%
For any ${\mathcal N} \times ({\mathcal N}+2) $ matrix, 
we will write a minor determinant whose 
 $a$-th and $b$-th columns are removed from 
it as 
$D\begin{bmatrix}
a ,& b 
\end{bmatrix}
$. 
We will use the following Pl\"ucker identity among determinants. 
\begin{align}
&
D\begin{bmatrix}
k_{1} ,&  k_{2}
\end{bmatrix}
D\begin{bmatrix}
k_{3}, &  k_{4}
\end{bmatrix}
-
D\begin{bmatrix}
k_{1}, &  k_{3} 
\end{bmatrix}
D\begin{bmatrix}
k_{2}, &  k_{4}
\end{bmatrix}
+
D\begin{bmatrix}
k_{1}, &  k_{4}
\end{bmatrix}
D\begin{bmatrix}
k_{2} ,&  k_{3}
\end{bmatrix}
=0,  \label{plucker2}
\end{align}
where $k_{1}<k_{2} <k_{3}<k_{4}$. 
%%%%%%%%%%%%%%%%%%%%%%%%%%%%%%%%%%%%%%%%%%%%%

At first, let us check the determinant formulas in Appendix A satisfy the 
boundary conditions \eqref{boundURc}-\eqref{boundDc}.  
\eqref{boundURc}-\eqref{boundLc} trivially follow from \eqref{soQQbc-1}-\eqref{soQQbc-3}. 
\eqref{boundDc} for $ s \ge -\eta_{1}$ or  $ s \le \eta_{2}$ 
also trivially follows from \eqref{soQQbc-2}-\eqref{soQQbc-3}. 
For $ \eta_{1} +\eta_{2} \ge 0$, this covers the boundary condition at $a=0$. 
Let us consider  the case $ \eta_{1} +\eta_{2} < 0$. All we have to do is to check  
\eqref{boundDc} for $ \eta_{2}< s < -\eta_{1}$. In this case, 
\begin{align}
 \Tf_{0,s}  =   (-1)^{(s+\eta_{1})m_{2}+\Theta }
\Def^{B_{1},B_{2},  
\langle 1, -\eta_{1}-\eta_{2} \rangle ,  
[-s-\eta_{1};-s-\eta_{1}-\eta_{2}]}_  
{F_{1}, F_{2},  
\emptyset, \emptyset,\emptyset}   \label{azero}
\end{align} 
follows from \eqref{sparsec4}. 
One can rewrite \eqref{detQQ1} and \eqref{detQQ2} for $I=\emptyset$, 
$B=B_{1} \sqcup B_{2}$ and $F=F_{1} \sqcup F_{2}$ in the following form
\footnote{One may unify these as 
$  \Qf_{B_{1},B_{2}, F_{1}, F_{2}} =   (-1)^{\frac{m(m-1)}{2}}
\Def^{B_{1},B_{2},  
 \langle 1, -\eta_{1}-\eta_{2} \rangle,  [0;-\eta_{1}-\eta_{2}]}_{F_{1}, F_{2},  
\langle 1, \eta_{1}+\eta_{2} \rangle , \emptyset,\emptyset}$.}. 
\begin{align}
 \Qf_{B_{1},B_{2}, F_{1}, F_{2}} &=   (-1)^{\frac{m(m-1)}{2}}
\Def^{B_{1},B_{2},  
 \emptyset ,  [0;-\eta_{1}-\eta_{2}]}_{F_{1}, F_{2},  
\langle 1, \eta_{1}+\eta_{2} \rangle , \emptyset,\emptyset}  
\quad \text{for} \quad \eta_{1}+\eta_{2} \ge 0, 
\label{Qdetc1}
\\[6pt]
 \Qf_{B_{1},B_{2}, F_{1}, F_{2}} &=   (-1)^{\frac{m(m-1)}{2} }
\Def^{B_{1},B_{2},  
\langle 1, -\eta_{1}-\eta_{2} \rangle ,  [0;-\eta_{1}-\eta_{2}]}_{F_{1}, F_{2},  
\emptyset, \emptyset,\emptyset}  
\quad \text{for} \quad 
\quad \eta_{1}+\eta_{2} \le 0. 
\label{Qdetc2}
\end{align} 
Applying  \eqref{shl3} to \eqref{azero} and comparing this with \eqref{Qdetc2}, 
we find that \eqref{boundDc} is valid. 

Except on the lines defied by $a=s+\eta_{1}$ or $a=-s+\eta_{2}$ and 
 the boundaries in the T-hook,  
that \eqref{Bglmmnnc-4}-\eqref{Bglmmnnc-5} satisfy the 
Hirota equation 
\eqref{hirota-BCb} reduces to the Pl\"{u}cker identity \eqref{plucker2}. 

Then the rest of our task is to prove \eqref{hirota-BCb} on the lines 
defied by $a=s+\eta_{1}$ or $a=-s+\eta_{2}$ in the T-hook. 
For this, we consider the following five cases depending on the values of 
$m_{1}, m_{2},n_{1}, n_{2}$ (cf.\ Figure \ref{T-hook-sparse}, \ref{T-hook-sparse2}). 
We remark that some of them become void for some specific 
values of $m_{1}, m_{2},n_{1}, n_{2}$. 
From now on, we will abbreviate the index sets $B_{1},B_{2},F_{1},F_{2}$ except when the size of them changes. 
For example, we use abbreviation: 
$\Def^{B_{1}\setminus b_{\alpha}^{(1)}, B_{2},\langle 1,s-\eta_{2}\rangle ,[0;0]}_{F_{1},F_{2},\emptyset, \langle 2,s+\eta_{1}\rangle, \emptyset}
=
\Def^{B_{1}\setminus b_{\alpha}^{(1)}, \langle 1,s-\eta_{2}\rangle ,[0;0]}_{\emptyset, \langle 2,s+\eta_{1}\rangle, \emptyset}$, 
$ B_{1}\setminus b_{\alpha}^{(1)}= (b_{1}^{(1)},b_{2}^{(1)},\dots,b_{\alpha -1}^{(1)},b_{\alpha +1}^{(1)},\dots, b_{m_{1}}^{(1)})$. 
%%%%%%%%%%%%%%%%%%%%%%%%%%%
\\\\
{\bf (i) the case $a=s+\eta_{1}$, $\max \{-\eta_{1}, \eta_{2} \}+1 \le s \le n_{1}$ ($m_{1}, n \ge 1 $)} \\
That \eqref{sparsec2} and \eqref{sparsec6} satisfy \eqref{hirota-BCb} is equivalent to 
\begin{multline}
\Def^{\langle 1,s-\eta_{2}\rangle ,[0;0]}_{\emptyset, \langle 1,s+\eta_{1}\rangle, \emptyset}
\Def^{\langle 1,s-\eta_{2}\rangle ,[0;2]}_{\emptyset, \langle 1,s+\eta_{1}\rangle, \emptyset}
\\
=
(-1)^{\eta_{1}+n_{2}+1}
\left(
\Def^{\langle 2,s-\eta_{2}\rangle ,[1;2]}_{\emptyset, \langle 1,s+\eta_{1}-1 \rangle, \emptyset}
\Def^{\langle 1,s-\eta_{2}\rangle ,[0;0]}_{\emptyset, \langle 2,s+\eta_{1}+1 \rangle, \emptyset}
+
\Def^{\langle 2,s-\eta_{2}+1\rangle ,[1;2]}_{\emptyset, \langle 1,s+\eta_{1}\rangle, \emptyset}
\Def^{\langle 1,s-\eta_{2}-1 \rangle ,[0;0]}_{\emptyset, \langle 2,s+\eta_{1} \rangle, \emptyset}
\right) .
\label{topro1}
\end{multline}
Let us apply \eqref{shl3} for the left hand side of \eqref{topro1} and 
expand the determinants, and apply \eqref{QQ-bf1}:
\begin{multline}
\Def^{\langle 1,s-\eta_{2}\rangle ,[0;0]}_{\emptyset, \langle 1,s+\eta_{1}\rangle, \emptyset}
\Def^{\langle 1,s-\eta_{2}\rangle ,[0;2]}_{\emptyset, \langle 1,s+\eta_{1}\rangle, \emptyset}
=
(-1)^{m_{2}}
\Def^{\langle 1,s-\eta_{2}\rangle ,[0;0]}_{\emptyset, \langle 1,s+\eta_{1}\rangle, \emptyset}
\Def^{\langle 1,s-\eta_{2}\rangle ,[1;2]}_{\emptyset, \langle 1,s+\eta_{1}\rangle, \emptyset}
\\
=(-1)^{m_{2}} 
\sum_{\alpha=1}^{m_{1}} (-1)^{\alpha +n+1}\Qf_{b^{(1)}_{\alpha}}^{[1]}
\Def^{B_{1}\setminus b_{\alpha}^{(1)}, \langle 1,s-\eta_{2}\rangle ,[0;0]}_{\emptyset, \langle 2,s+\eta_{1}\rangle, \emptyset}
\\
\times 
\sum_{\beta=1}^{n} (-1)^{\beta +m_{1}+m_{2}+1} \Qf_{f_{\beta}}^{[1]}
\Def^{\langle 2,s-\eta_{2}\rangle ,[1;2]}_
{F \setminus f_{\beta}, \emptyset, \langle 1,s+\eta_{1}\rangle, \emptyset} 
\\
=(-1)^{m_{2}}
\sum_{\alpha=1}^{m_{1}} \sum_{\beta=1}^{n}
(-1)^{\alpha +n+1} (-1)^{\beta +m_{1}+m_{2}+1}
(\Qf_{b^{(1)}_{\alpha}f_{\beta}}^{[2]}- \Qf_{b^{(1)}_{\alpha}f_{\beta}}^{[0]})
\\ \times 
\Def^{B_{1}\setminus b_{\alpha}^{(1)}, 
\langle 1,s-\eta_{2}\rangle ,[0;0]}_{\emptyset, \langle 2,s+\eta_{1}\rangle, \emptyset}
\Def^{\langle 2,s-\eta_{2} \rangle ,[1;2]}_
{F \setminus f_{\beta}, \emptyset, \langle 1,s+\eta_{1}\rangle, \emptyset} 
\\
=-(-1)^{m_{2}}
\sum_{\beta=1}^{n} (-1)^{\beta +m_{1}+m_{2}+1}
\\ 
\times 
\begin{vmatrix}
(\Qf_{b_{i}^{(1)} f_{j}}^{[0]})_{
\genfrac{}{}{0pt}{}{1\le i \le m_{1}, }{1\le j \le n} }
& (\Qf_{b_{i}^{(1)}f_{\beta}}^{[0]})_{1\le i \le m_{1} }
& (\Qf_{b_{i}^{(1)}}^{[2j-1]})_{
\genfrac{}{}{0pt}{}{1\le i \le m_{1}, }{2\le j \le s+\eta_{1}} } 
  \\[6pt]
(\Qf_{b_{i}^{(2)} f_{j}}^{[0]})_{
\genfrac{}{}{0pt}{}{1\le i \le m_{2}, }{1\le j \le n} }
& (0)_{m_{2}\times 1}
& (0)_{m_{2}\times (s+\eta_{1}-1)}
   \\[6pt]
((-1)^{i-1}\Qf_{f_{j}}^{[-2i+1]})_{
\genfrac{}{}{0pt}{}{1\le i \le s-\eta_{2}, }{1\le j \le n} }
& (0)_{(s-\eta_{2})\times 1}
& (0)_{(s-\eta_{2})\times (s+\eta_{1}-1)} \\
\end{vmatrix}
\Def^{\langle 2,s-\eta_{2} \rangle ,[1;2]}_
{F \setminus f_{\beta}, \emptyset, \langle 1,s+\eta_{1}\rangle, \emptyset} 
\\
+ (-1)^{m_{2}}
\sum_{\alpha=1}^{m_{1}} (-1)^{\alpha +n+1}
\Def^{B_{1}\setminus b_{\alpha}^{(1)}, 
\langle 1,s-\eta_{2}\rangle ,[0;0]}_{\emptyset, \langle 2,s+\eta_{1}\rangle, \emptyset} 
\\
\times 
\begin{vmatrix}
(\Qf_{b_{i}^{(1)} f_{j}}^{[2]})_{
\genfrac{}{}{0pt}{}{1\le i \le m_{1}, }{1\le j \le n} }
& (\Qf_{b_{i}^{(1)}}^{[2j+1]})_{
\genfrac{}{}{0pt}{}{1\le i \le m_{1}, }{1\le j \le s+\eta_{1}} }
  \\[6pt]
(-\Qf_{b_{i}^{(2)} f_{j}}^{[0]})_{
\genfrac{}{}{0pt}{}{1\le i \le m_{2}, }{1\le j \le n} }
& (0)_{m_{2}\times (s+\eta_{1})}
  \\[6pt]
(\Qf_{b_{\alpha}^{(1)} f_{j}}^{[2]})_{1\le j \le n} 
& (0)_{1\times (s+\eta_{1})}
   \\[6pt]
((-1)^{i-1}\Qf_{f_{j}}^{[-2i+3]})_{
\genfrac{}{}{0pt}{}{2\le i \le s-\eta_{2}, }{1\le j \le n} }
& (0)_{(s-\eta_{2}-1)\times (s+\eta_{1})}
 \\
\end{vmatrix}.
 \label{topro2}
\end{multline}
Subtracting the $\beta $-th column from the $(n+1)$-th column in the determinant 
in the first summand in the right hand side of \eqref{topro2}, we obtain
\begin{multline}
\begin{vmatrix}
(\Qf_{b_{i}^{(1)} f_{j}}^{[0]})_{
\genfrac{}{}{0pt}{}{1\le i \le m_{1}, }{1\le j \le n} }
& (\Qf_{b_{i}^{(1)}f_{\beta}}^{[0]})_{1\le i \le m_{1} }
& (\Qf_{b_{i}^{(1)}}^{[2j-1]})_{
\genfrac{}{}{0pt}{}{1\le i \le m_{1}, }{2\le j \le s+\eta_{1}} } 
  \\[6pt]
(\Qf_{b_{i}^{(2)} f_{j}}^{[0]})_{
\genfrac{}{}{0pt}{}{1\le i \le m_{2}, }{1\le j \le n} }
& (0)_{m_{2}\times 1}
& (0)_{m_{2}\times (s+\eta_{1}-1)}
   \\[6pt]
((-1)^{i-1}\Qf_{f_{j}}^{[-2i+1]})_{
\genfrac{}{}{0pt}{}{1\le i \le s-\eta_{2}, }{1\le j \le n} }
& (0)_{(s-\eta_{2})\times 1}
& (0)_{(s-\eta_{2})\times (s+\eta_{1}-1)} \\
\end{vmatrix}
\\
= - 
\begin{vmatrix}
(\Qf_{b_{i}^{(1)} f_{j}}^{[0]})_{
\genfrac{}{}{0pt}{}{1\le i \le m_{1}, }{1\le j \le n} }
& (0)_{m_{1}\times 1}
& (\Qf_{b_{i}^{(1)}}^{[2j-1]})_{
\genfrac{}{}{0pt}{}{1\le i \le m_{1}, }{2\le j \le s+\eta_{1}} } 
  \\[6pt]
(\Qf_{b_{i}^{(2)} f_{j}}^{[0]})_{
\genfrac{}{}{0pt}{}{1\le i \le m_{2}, }{1\le j \le n} }
& (\Qf_{b_{i}^{(2)} f_{\beta}}^{[0]})_{1\le i \le m_{2}}
& (0)_{m_{2}\times (s+\eta_{1}-1)}
   \\[6pt]
((-1)^{i-1}\Qf_{f_{j}}^{[-2i+1]})_{
\genfrac{}{}{0pt}{}{1\le i \le s-\eta_{2}, }{1\le j \le n} }
& ((-1)^{i-1}\Qf_{f_{\beta}}^{[-2i+1]})_{1\le i \le s-\eta_{2}}
& (0)_{(s-\eta_{2})\times (s+\eta_{1}-1)} \\
\end{vmatrix}
\\
=-\sum_{\gamma =1}^{m_{2}} (-1)^{\gamma +m_{1}+n+1}
\Qf_{b_{\gamma}^{(2)} f_{\beta}}^{[0]}
\Def^{B_{2} \setminus b^{(2)}_{\gamma}, \langle 1,s-\eta_{2} \rangle ,[0;0]}_
{\emptyset, \langle 2,s+\eta_{1}\rangle, \emptyset} 
\\
-\sum_{\gamma =1}^{s-\eta_{2}} (-1)^{\gamma +m_{1}+m_{2}+n+1}
(-1)^{\gamma-1}\Qf_{f_{\beta}}^{[-2\gamma+1]}
\Def^{\langle 1,s-\eta_{2} \rangle \setminus \gamma ,[0;0]}_
{\emptyset, \langle 2,s+\eta_{1}\rangle, \emptyset} .
 \label{topro3}
\end{multline}
Subtracting the $\alpha $-th row from the $(m_{1}+m_{2}+1)$-th row in the determinant 
in the second summand in the right hand side of \eqref{topro2}, we find: 
\begin{multline}
\begin{vmatrix}
(\Qf_{b_{i}^{(1)} f_{j}}^{[2]})_{
\genfrac{}{}{0pt}{}{1\le i \le m_{1}, }{1\le j \le n} }
& (\Qf_{b_{i}^{(1)}}^{[2j+1]})_{
\genfrac{}{}{0pt}{}{1\le i \le m_{1}, }{1\le j \le s+\eta_{1}} }
  \\[6pt]
(-\Qf_{b_{i}^{(2)} f_{j}}^{[0]})_{
\genfrac{}{}{0pt}{}{1\le i \le m_{2}, }{1\le j \le n} }
& (0)_{m_{2}\times (s+\eta_{1})}
  \\[6pt]
(\Qf_{b_{\alpha}^{(1)} f_{j}}^{[2]})_{1\le j \le n} 
& (0)_{1\times (s+\eta_{1})}
   \\[6pt]
((-1)^{i-1}\Qf_{f_{j}}^{[-2i+3]})_{
\genfrac{}{}{0pt}{}{2\le i \le s-\eta_{2}, }{1\le j \le n} }
& (0)_{(s-\eta_{2}-1)\times (s+\eta_{1})}
 \\
\end{vmatrix}
\\
=- 
\begin{vmatrix}
(\Qf_{b_{i}^{(1)} f_{j}}^{[2]})_{
\genfrac{}{}{0pt}{}{1\le i \le m_{1}, }{1\le j \le n} }
& (\Qf_{b_{i}^{(1)}}^{[2j+1]})_{
\genfrac{}{}{0pt}{}{1\le i \le m_{1}, }{1\le j \le s+\eta_{1}} }
  \\[6pt]
(-\Qf_{b_{i}^{(2)} f_{j}}^{[0]})_{
\genfrac{}{}{0pt}{}{1\le i \le m_{2}, }{1\le j \le n} }
& (0)_{m_{2}\times (s+\eta_{1})}
  \\[6pt]
(0)_{1 \times n}
&(\Qf_{b_{\alpha}^{(1)}}^{[2j+1]})_{1\le j \le s+\eta_{1}} 
   \\[6pt]
((-1)^{i-1}\Qf_{f_{j}}^{[-2i+3]})_{
\genfrac{}{}{0pt}{}{2\le i \le s-\eta_{2}, }{1\le j \le n} }
& (0)_{(s-\eta_{2}-1)\times (s+\eta_{1})}
 \\
\end{vmatrix}
\\
=-\sum_{\gamma =1}^{s+\eta_{1}} (-1)^{\gamma +n+m_{1}+m_{2}+1}
\Qf_{b_{\alpha}^{(1)}}^{[2\gamma +1]}
\Def^{\langle 2,s-\eta_{2} \rangle ,[1;2]}_
{\emptyset, \langle 1,s+\eta_{1}\rangle \setminus \gamma , \emptyset} .
 \label{topro4}
\end{multline}
Substituting \eqref{topro3} and \eqref{topro4} into the right 
hand side of \eqref{topro2}, and taking summation over $\alpha$ and $\beta$, 
we find that only a few terms in the summation over $\gamma $ are non-zero 
(since there are the same rows or columns in determinants except 
for a few values of $\gamma $), and finally 
we obtain the right hand side of \eqref{topro1}. 
%%%%%%%%%%%%%%%%%%%%%%%%%%%%%%%%

{\bf (ii) the case 
 $a=s+\eta_{1}$, $\frac{\eta_{2}-\eta_{1}}{2} < s \le \eta_{2}$ ($\eta_{1}+\eta_{2} \ge 1 $)} \\
  At first, we assume $\eta_{1}+\eta_{2} \ge 2 $ and $s<\eta_{2} $. 
Due to \eqref{shl2} and \eqref{shl2}, 
that \eqref{sparsec1} and \eqref{sparsec5} satisfy \eqref{hirota-BCb} is equivalent to 
\begin{multline}
\Def^{\emptyset ,[0;0]}_{\langle 0,\eta_{2}-s-1 \rangle, \langle 0,s+\eta_{1}-1 \rangle, \emptyset }
\Def^{\emptyset, [0;0]}_{\langle 1,\eta_{2}-s \rangle, \langle 1,s+\eta_{1} \rangle, \emptyset}
\\
=
(-1)^{(s+\eta_{1})(s+\eta_{2}+1)}
\Def^{\emptyset ,[0;0]}_{\emptyset, \langle 1,\eta_{1}+s-1 \rangle, \langle 0,\eta_{2}-s \rangle }
\Def^{\emptyset, [0;0]}_{\langle 0,\eta_{2}-s-1 \rangle, \langle 1,s+\eta_{1} \rangle, \emptyset }
\\ 
+(-1)^{(s+\eta_{1}+1)(s+\eta_{2})}
\Def^{\emptyset ,[0;0]}_{\langle 0,\eta_{2}-s \rangle, \langle 1,s+\eta_{1}-1 \rangle, \emptyset }
\Def^{\emptyset, [0;0]}_{\emptyset, \langle 1,\eta_{1}+s \rangle, \langle 0,\eta_{2}-s-1 \rangle } .
\label{topro5}
\end{multline}
Due to the following identities follow from manipulations on columns of the determinants:
\begin{align}
\Def^{\emptyset ,[0;0]}_{\langle 0,\eta_{2}-s-1 \rangle, \langle 0,s+\eta_{1}-1 \rangle, \emptyset }
&=(-1)^{\eta_{1}+s}\Def^{\emptyset ,[0;0]}_{\langle 0,\eta_{2}-s-1 \rangle, \langle 1,s+\eta_{1}-1 \rangle, (0)}, 
\nonumber \\[6pt]
\Def^{\emptyset ,[0;0]}_{\emptyset, \langle 1,\eta_{1}+s-1 \rangle, \langle 0,\eta_{2}-s \rangle }
&=
(-1)^{(\eta_{1}+s)(\eta_{2}+s)}
\Def^{\emptyset, [0;0]}_{\langle 1,\eta_{2}-s \rangle, \langle 1,s+\eta_{1}-1 \rangle, (0)} ,
\\[6pt]
\Def^{\emptyset, [0;0]}_{\emptyset, \langle 1,\eta_{1}+s \rangle, \langle 0,\eta_{2}-s-1 \rangle }
&=
(-1)^{(\eta_{1}+s)(\eta_{2}+s)+\eta_{1}+\eta_{2}+1}
\Def^{\emptyset, [0;0]}_{\langle 1,\eta_{2}-s-1 \rangle, \langle 1,s+\eta_{1} \rangle, (0)}, 
\nonumber 
\end{align}
\eqref{topro5} is equivalent to 
\begin{multline}
\Def^{\emptyset ,[0;0]}_{\langle 0,\eta_{2}-s-1 \rangle, \langle 1,s+\eta_{1}-1 \rangle, (0)}
\Def^{\emptyset, [0;0]}_{\langle 1,\eta_{2}-s \rangle, \langle 1,s+\eta_{1} \rangle, \emptyset}
=
\Def^{\emptyset ,[0;0]}_{\langle 0,\eta_{2}-s-1 \rangle, \langle 1,s+\eta_{1} \rangle, \emptyset}
\Def^{\emptyset, [0;0]}_{\langle 1,\eta_{2}-s \rangle, \langle 1,s+\eta_{1}-1 \rangle, (0)}
\\
-
\Def^{\emptyset ,[0;0]}_{\langle 0,\eta_{2}-s \rangle, \langle 1,s+\eta_{1}-1 \rangle, \emptyset }
\Def^{\emptyset, [0;0]}_{\langle 1,\eta_{2}-s-1 \rangle, \langle 1,s+\eta_{1} \rangle, (0)} .
\label{topro5-2}
\end{multline}
This is nothing but the Pl\"{u}cker identity \eqref{plucker2}. 
%%%%%%%%%%%%%

Next we consider the case
\footnote{Note that the condition $\eta_{1}+\eta_{2}=1$ induces 
 $a=1$, $s=\eta_{2}$.} 
$s=\eta_{2}$ and $a=\eta_{1}+\eta_{2} \ge 1 $. 
  After manipulation on rows and columns  of the determinants based on \eqref{QQ-bf1}, 
that \eqref{sparsec1} and \eqref{sparsec5} satisfy \eqref{hirota-BCb}  
 reduces to 
 the Pl\"ucker identity \eqref{plucker2} for 
 the following $ (m_{1}+m_{2}+1) \times (m_{1}+m_{2}+3)$ matrix
\footnote{
This matrix also suggests us yet another determinant expression of the solution.} 
\begin{align}
\begin{pmatrix}
1  &  1 &  1 & (\Qf_{f_{j}}^{[-1]})_{1\le j \le n}  & (0)_{1 \times (\eta_{1}+\eta_{2})} \\
 (0)_{m_{1} \times 1} &    (0)_{m_{1} \times 1}   & 
(\Qf_{b_{i}^{(1)}}^{[-1]})_{1\le i \le m_{1}}  &   (\Qf_{b_{i}^{(1)},f_{j} })_{ 1\le i \le m_{1} \atop 1\le j \le n}  & 
(\Qf_{b_{i}^{(1)}}^{[2j-1]})_{1\le i \le m_{1} \atop 1 \le j \le \eta_{1}+\eta_{2} } 
 \\
  (0)_{m_{2} \times 1} &     (\Qf_{b_{i}^{(2)}}^{[-1]})_{1\le i \le m_{2}} & 
(\Qf_{b_{i}^{(2)}}^{[-1]})_{1\le i \le m_{2}}  &   (\Qf_{b_{i}^{(2)},f_{j} })_{ 1\le i \le m_{2} \atop 1\le j \le n}  & 
 (0)_{m_{2} \times (\eta_{1}+\eta_{2})} 
 \\
\end{pmatrix} 
\end{align}
at  $k_{1}=1,k_{2}=2,k_{3}=3,k_{4}=m_{1}+m_{2}+3$. 
 %%%%%%%%%%%%%%%%%%%%%%%%%%%%%%%%%%%%%%%%%%%

{\bf (iii) the case $a=\frac{\eta_{1}+\eta_{2}}{2}$, $s=\frac{\eta_{2}-\eta_{1}}{2}$ 
($\eta_{1}+\eta_{2} \in 2 {\mathbb Z}_{\ge 1}$)} \\
Due to \eqref{shl2} and \eqref{shl3}, 
that \eqref{sparsec1}, \eqref{sparsec5} and \eqref{sparsec7} satisfy \eqref{hirota-BCb} is equivalent to 
\begin{multline}
\Def^{\emptyset ,[0;0]}_{\langle 0,\frac{\eta_{1}+\eta_{2}}{2}-1 \rangle, 
\langle 0,\frac{\eta_{1}+\eta_{2}}{2} -1 \rangle, \emptyset}
\Def^{\emptyset, [0;0]}_{\langle 1,\frac{\eta_{1}+\eta_{2}}{2}  \rangle, 
\langle 1,\frac{\eta_{1}+\eta_{2}}{2}  \rangle, \emptyset}
\\
= (-1)^{\frac{\eta_{1}+\eta_{2}}{2}}
\Def^{\emptyset ,[0;0]}_{\langle 1,\frac{\eta_{1}+\eta_{2}}{2} \rangle, 
\emptyset, \langle 0, \frac{\eta_{1}+\eta_{2}}{2}-1 \rangle}
\Def^{\emptyset, [0;0]}_{\langle 0,\frac{\eta_{1}+\eta_{2}}{2}-1 \rangle, 
\langle 1,\frac{\eta_{1}+\eta_{2}}{2} \rangle, \emptyset}
\\
+
\Def^{\emptyset ,[0;0]}_{\langle 0,\frac{\eta_{1}+\eta_{2}}{2} \rangle, 
\langle 1,\frac{\eta_{1}+\eta_{2}}{2}-1 \rangle, \emptyset }
\Def^{\emptyset, [0;0]}_{\emptyset, \langle 1,\frac{\eta_{1}+\eta_{2}}{2} \rangle, 
\langle 0,\frac{\eta_{1}+\eta_{2}}{2}-1 \rangle }.
\label{topro6}
\end{multline}
Then we find \eqref{topro6} is equivalent
\footnote{Note that \eqref{topro5} is not equivalent to 
\eqref{topro5-2} at $s=\frac{\eta_{2}-\eta_{1}}{2}$.} 
to \eqref{topro5-2} at $s=\frac{\eta_{2}-\eta_{1}}{2}$ since 
 there are  identities follow from 
manipulations on columns of the determinants: 
\begin{align}
\Def^{\emptyset, [0;0]}_{\langle 0,\frac{\eta_{1}+\eta_{2}}{2}-1 
\rangle, \langle 0,\frac{\eta_{1}+\eta_{2}}{2}-1 \rangle, \emptyset}
&=
(-1)^{\frac{\eta_{1}+\eta_{2}}{2}}
\Def^{\emptyset, [0;0]}_{\langle 0,\frac{\eta_{1}+\eta_{2}}{2}-1 
\rangle, \langle 1,\frac{\eta_{1}+\eta_{2}}{2}-1 \rangle, (0)},
\nonumber 
\\[6pt]
\Def^{\emptyset, [0;0]}_{\langle 1,\frac{\eta_{1}+\eta_{2}}{2}
\rangle,\emptyset, \langle 0,\frac{\eta_{1}+\eta_{2}}{2}-1 \rangle }
&=
\Def^{\emptyset, [0;0]}_{\langle 1,\frac{\eta_{1}+\eta_{2}}{2} 
\rangle, \langle 1,\frac{\eta_{1}+\eta_{2}}{2}-1 \rangle, (0)}, 
\label{manuco}
\\[6pt]
\Def^{\emptyset, [0;0]}_{\emptyset, \langle 1,\frac{\eta_{1}+\eta_{2}}{2}
\rangle, \langle 0,\frac{\eta_{1}+\eta_{2}}{2}-1 \rangle }
&=(-1)^{\frac{\eta_{1}+\eta_{2}}{2}-1}
\Def^{\emptyset, [0;0]}_{\langle 1,\frac{\eta_{1}+\eta_{2}}{2} -1
\rangle, \langle 1,\frac{\eta_{1}+\eta_{2}}{2} \rangle, (0)}.
\nonumber 
\end{align}
%%%%%%%%%%%%%%%%%%

{\bf (iv) the case
 $a=-s+\eta_{2}$, $-\eta_{1} \le s < \frac{\eta_{2}-\eta_{1}}{2} $ ($\eta_{1}+\eta_{2} \ge 1 $)} \\
 At first, we assume $\eta_{1}+\eta_{2} \ge 2 $ and $s > -\eta_{1} $. 
Using \eqref{shl1} and  \eqref{shl3}  repeatedly, we obtain 
\begin{align}
\begin{split}
\Def^{\emptyset ,[-2s+\eta_{2}-\eta_{1}-1;-2s-2\eta_{1}]}_{\emptyset , 
\langle 1, s+\eta_{1}+1 \rangle, 
 \langle 2s+\eta_{1}-\eta_{2}+2,s+\eta_{1} \rangle} 
 &=(-1)^{(m_{2}-s+\eta_{2}-1)(s+\eta_{1})}
\Def^{\emptyset ,[-s+\eta_{2}-1;0]}_{\emptyset , 
\langle 1-s-\eta_{1}, 1 \rangle, 
 \langle s-\eta_{2}+2,0 \rangle} 
\\
 &=(-1)^{(n_{2}+s+1)(\eta_{1}+\eta_{2}+1)}
\Def^{\emptyset ,[0;0]}_{\emptyset , \langle -s-\eta_{1}+1,1 \rangle, 
 \langle s-\eta_{2}+2,0 \rangle} ,
\\[6pt]
\Def^{\emptyset ,[-2s+\eta_{2}-\eta_{1}+1;-2s-2\eta_{1}+2]}_{\emptyset , 
\langle 1, s+\eta_{1} \rangle, 
 \langle 2s+\eta_{1}-\eta_{2},s+\eta_{1}-1 \rangle} 
 &=(-1)^{(n_{2}+s)(\eta_{1}+\eta_{2}+1)}
\Def^{\emptyset ,[0;0]}_{\emptyset , \langle -s-\eta_{1}+2,1 \rangle, 
 \langle s-\eta_{2}+1,0 \rangle} .
\end{split} 
\label{au1}
\end{align}
We can also obtain the following relations based on \eqref{shl2}: 
\begin{align}
\begin{split}
\Def^{\emptyset ,[0;-2s-2\eta_{1}]}_{\langle 1, s+\eta_{1} \rangle, 
\emptyset , \langle 2s+\eta_{1}-\eta_{2}+1,s+\eta_{1} \rangle}
&=
\Def^{\emptyset ,[0;0]}_{\langle -s-\eta_{1}+1,0 \rangle, 
\emptyset , \langle s-\eta_{2}+1,0 \rangle},
\\[6pt]
\Def^{\emptyset ,[0;-2s-2\eta_{1}+2]}_{\langle 1, s+\eta_{1} \rangle, 
\emptyset , \langle 2s+\eta_{1}-\eta_{2}, s+\eta_{1}-1 \rangle}
&=
\Def^{\emptyset ,[0;0]}_{\langle -s-\eta_{1}+2,1 \rangle, 
\emptyset , \langle s-\eta_{2}+1,0 \rangle},
\\[6pt]
\Def^{\emptyset ,[0;-2s-2\eta_{1}]}_{\langle 1,s+\eta_{1}+1 \rangle, 
\emptyset , \langle 2s+\eta_{1}-\eta_{2}+2,s+\eta_{1} \rangle}
&=
\Def^{\emptyset ,[0;0]}_{\langle -s-\eta_{1}+1,1 \rangle, 
\emptyset , \langle s-\eta_{2}+2,0 \rangle}. 
\end{split}
 \label{au2}
\end{align}
 Then, based on \eqref{au1}-\eqref{au2}, we can show 
that \eqref{sparsec1} and \eqref{sparsec7} satisfy \eqref{hirota-BCb}  
is equivalent to 
\begin{multline}
\Def^{\emptyset ,[0;0]}_{\langle -s-\eta_{1}+1,0 \rangle, 
\emptyset , \langle s-\eta_{2}+1,0 \rangle}
\Def^{\emptyset ,[0;0]}_{\langle -s-\eta_{1}+2,1 \rangle, 
\emptyset , \langle s-\eta_{2}+2,1 \rangle}
\\
=(-1)^{\eta_{2}-s}
\Def^{\emptyset ,[0;0]}_{\langle -s-\eta_{1}+2,1 \rangle, 
\emptyset , \langle s-\eta_{2}+1,0 \rangle}
\Def^{\emptyset ,[0;0]}_{\emptyset , \langle -s-\eta_{1}+1,1 \rangle, 
 \langle s-\eta_{2}+2,0 \rangle}
 \\
-(-1)^{\eta_{2}-s}
\Def^{\emptyset ,[0;0]}_{\langle -s-\eta_{1}+1,1 \rangle, 
\emptyset , \langle s-\eta_{2}+2,0 \rangle}
\Def^{\emptyset ,[0;0]}_{\emptyset , \langle -s-\eta_{1}+2,1 \rangle, 
 \langle s-\eta_{2}+1,0 \rangle}.
\label{topro7}
\end{multline}
After manipulations on columns of determinants similar to \eqref{manuco}, 
we find \eqref{topro7} reduces to 
\begin{multline}
\Def^{\emptyset ,[0;0]}_{\langle -s-\eta_{1}+1,0 \rangle, 
\emptyset , \langle s-\eta_{2}+1,0 \rangle}
\Def^{\emptyset ,[0;0]}_{\langle -s-\eta_{1}+2,1 \rangle, 
(1) , \langle s-\eta_{2}+2,0 \rangle}
\\
=
\Def^{\emptyset ,[0;0]}_{\langle -s-\eta_{1}+2,1 \rangle, 
\emptyset , \langle s-\eta_{2}+1,0 \rangle}
\Def^{\emptyset ,[0;0]}_{\langle -s-\eta_{1}+1,0 \rangle, (1),
 \langle s-\eta_{2}+2,0 \rangle}
 \\
-
\Def^{\emptyset ,[0;0]}_{\langle -s-\eta_{1}+1,1 \rangle, 
\emptyset , \langle s-\eta_{2}+2,0 \rangle}
\Def^{\emptyset ,[0;0]}_{ \langle -s-\eta_{1}+2,0 \rangle, (1),
 \langle s-\eta_{2}+1,0 \rangle}.
\label{topro8}
\end{multline}
This is nothing but the Pl\"{u}cker identity \eqref{plucker2}. 
%%%%%%%%%%%%%%%%%%

Next we consider the case
\footnote{Note that the condition $\eta_{1}+\eta_{2}  =1 $ 
induces 
 $a=1$, $s=-\eta_{1}$.}
 $s=-\eta_{1}$ and $a=\eta_{1}+\eta_{2}  \ge 1 $.
 Using \eqref{shl3}  repeatedly, we obtain 
\begin{align}
\Def^{\emptyset ,[\eta_{1}+\eta_{2}+1;2]}_{\emptyset , 
\emptyset, 
 \langle -\eta_{1}-\eta_{2},-1 \rangle} 
 &=(-1)^{(\eta_{1}+n_{2})(\eta_{1}+\eta_{2}+1)}
\Def^{\emptyset ,[2;2]}_{\emptyset , \emptyset , 
 \langle -\eta_{1}-\eta_{2},-1 \rangle} .
\label{au1-1}
\end{align}
 Using \eqref{au1-1} and  the second equality in \eqref{au1} at $s=-\eta_{1}$, and 
  manipulating rows and columns  of the determinants based on \eqref{QQ-bf1}, 
  we find 
that \eqref{sparsec1} and \eqref{sparsec7} satisfy \eqref{hirota-BCb}  
 reduces to 
 the Pl\"ucker identity \eqref{plucker2} for 
 the following $ (m_{1}+m_{2}+1) \times (m_{1}+m_{2}+3) $ matrix 
\begin{align}
\begin{pmatrix}
-1  &  -1 &  -1 & (\Qf_{f_{j}}^{[1]})_{1\le j \le n}  & (0)_{1 \times (\eta_{1}+\eta_{2})} \\
 (0)_{m_{1} \times 1} &   (\Qf_{b_{i}^{(1)}}^{[1]})_{1\le i \le m_{1}}  & 
(\Qf_{b_{i}^{(1)}}^{[1]})_{1\le i \le m_{1}}  &   (\Qf_{b_{i}^{(1)},f_{j} })_{ 1\le i \le m_{1} \atop 1\le j \le n}  & 
 (0)_{m_{1} \times (\eta_{1}+\eta_{2})}
 \\
  (0)_{m_{2} \times 1} &     (0)_{m_{2} \times 1} & 
(\Qf_{b_{i}^{(2)}}^{[1]})_{1\le i \le m_{2}}  &   (\Qf_{b_{i}^{(2)},f_{j} })_{ 1\le i \le m_{2} \atop 1\le j \le n}  & 
(\Qf_{b_{i}^{(2)}}^{[2j-1]})_{1\le i \le m_{2} , \atop  -\eta_{1}-\eta_{2}+1 \le j \le 0  } 
 \\
\end{pmatrix} 
\end{align}
at  $k_{1}=1,k_{2}=2,k_{3}=3,k_{4}=n+4$. 
 %%%%%%%%%%%%%%%%%%%%%%%%%%%%%%%%%%%%%%%%%%%

{\bf (v) the case $a=-s+\eta_{2}$, $-n_{2} \le s \le \min \{ -\eta_{1}, \eta_{2} \}-1$ 
($n,m_{2} \ge 1 $) } \\
Using \eqref{shl1} and \eqref{shl3} repeatedly, we obtain: 
\begin{align}
\begin{split}
\Def^{\langle 1, -s-\eta_{1} \rangle ,[-2s-\eta_{1}+\eta_{2} ;-2(s+\eta_{1})]}_{\emptyset, 
\emptyset , \langle 2s+\eta_{1} -\eta_{2}+1,s+\eta_{1} \rangle}
&=
(-1)^{(n_{2}+s)(s+\eta_{2})}
\Def^{\langle 1, -s-\eta_{1} \rangle ,[-s-\eta_{1} ;-2(s+\eta_{1})]}_{\emptyset, 
\emptyset , \langle 2s+\eta_{1} -\eta_{2}+1,s+\eta_{1} \rangle}
%\nonumber 
\\
&=
(-1)^{(n_{2}+s)(s+\eta_{2})+(\eta_{1}+n_{2})(s+\eta_{1}+1)}
\Def^{\langle s+\eta_{1}+2,1\rangle ,[1;2]}_{\emptyset, 
\emptyset , \langle s-\eta_{2},-1 \rangle},
%\label{au3}
\\[6pt]
\Def^{\langle 1, -s-\eta_{1} \rangle ,[0;-2(s+\eta_{1}-1)]}_{\emptyset, 
\emptyset , \langle 2s+\eta_{1}-\eta_{2}+1,s+\eta_{1} \rangle}
&=
(-1)^{s+\eta_{1}}
\Def^{\langle s+\eta_{1}+1,0\rangle ,[0;2]}_{\emptyset, 
\emptyset , \langle s-\eta_{2}+1,0 \rangle},
\\[6pt]
\Def^{\langle 1, -s-\eta_{1} \rangle ,[0;-2(s+\eta_{1}-1)]}_{\emptyset, 
\emptyset , \langle 2s+\eta_{1} -\eta_{2}, s+\eta_{1}-1  \rangle}
&= 
(-1)^{s+\eta_{1}}
\Def^{\langle s+\eta_{1}+1,0\rangle ,[0;2]}_{\emptyset, 
\emptyset , \langle s-\eta_{2},-1 \rangle},
\\[6pt]
\Def^{\langle 1, -s-\eta_{1}-1 \rangle ,[-2s-\eta_{1}+\eta_{2}-1;-2(s+\eta_{1})]}_{\emptyset, 
\emptyset , \langle 2s+\eta_{1}-\eta_{2}+2,s+\eta_{1} \rangle}
&=
(-1)^{(s+n_{2}+1)(s+\eta_{2}+1)+(\eta_{1}+n_{2})(s+\eta_{1}+1)} 
\\ 
& \qquad  \times 
\Def^{\langle s+\eta_{1}+2,0\rangle ,[1;2]}_{\emptyset, 
\emptyset , \langle s-\eta_{2}+1,-1 \rangle},
\\[6pt]
\Def^{\langle 1,-s-\eta_{1}-1 \rangle ,[0;-2(s+\eta_{1})]}_{\emptyset, 
\emptyset , \langle 2s+\eta_{1}-\eta_{2}+2,s+\eta_{1} \rangle}
&=
(-1)^{s+\eta_{1}+1}
\Def^{\langle s+\eta_{1}+2,0\rangle ,[0;2]}_{\emptyset, 
\emptyset , \langle s-\eta_{2}+1,-1 \rangle},
\\[6pt]
\Def^{\langle 1, -s-\eta_{1} \rangle ,[-2s-\eta_{1}+\eta_{2}+1;-2(s+\eta_{1}-1)]}_{\emptyset, 
\emptyset , \langle 2s+\eta_{1}-\eta_{2},s+\eta_{1}-1 \rangle}
&=
(-1)^{(s+n_{2})(s+\eta_{2})+(\eta_{1}+n_{2})(s+\eta_{1})}
\Def^{\langle s+\eta_{1}+1,0\rangle ,[1;2]}_{\emptyset, 
\emptyset , \langle s-\eta_{2},-1 \rangle}. 
\end{split}
\label{au3}
\end{align}
Due to the above relations \eqref{au3}, 
that \eqref{sparsec3} and \eqref{sparsec8} satisfy \eqref{hirota-BCb} is equivalent to 
\begin{multline}
\Def^{\langle s+\eta_{1}+2,1\rangle ,[1;2]}_{\emptyset, 
\emptyset , \langle s-\eta_{2},-1 \rangle}
\Def^{\langle s+\eta_{1}+1,0\rangle ,[0;2]}_{\emptyset, 
\emptyset , \langle s-\eta_{2}+1,0 \rangle}
\\
= -
\Def^{\langle s+\eta_{1}+1,0\rangle ,[0;2]}_{\emptyset, 
\emptyset , \langle s-\eta_{2},-1 \rangle}
\Def^{\langle s+\eta_{1}+2,0\rangle ,[1;2]}_{\emptyset, 
\emptyset , \langle s-\eta_{2}+1,-1 \rangle}
-
\Def^{\langle s+\eta_{1}+2,0\rangle ,[0;2]}_{\emptyset, 
\emptyset , \langle s-\eta_{2}+1,-1 \rangle}
\Def^{\langle s+\eta_{1}+1,0\rangle ,[1;2]}_{\emptyset, 
\emptyset , \langle s-\eta_{2},-1 \rangle} .
\label{topro9}
\end{multline}
Let us expand the determinants in the left hand side of \eqref{topro9}, and apply \eqref{QQ-bf1}:
\begin{multline}
\Def^{\langle s+\eta_{1}+2,1\rangle ,[1;2]}_{\emptyset, 
\emptyset , \langle s-\eta_{2},-1 \rangle}
\Def^{\langle s+\eta_{1}+1,0\rangle ,[0;2]}_{\emptyset, 
\emptyset , \langle s-\eta_{2}+1,0 \rangle}
\\
=\sum_{\alpha=1}^{n} (-1)^{\alpha +m_{1}+m_{2}-s-\eta_{1}}\Qf_{f_{\alpha}}^{[1]}
\Def^{\langle s+\eta_{1}+2,0 \rangle ,[1;2]}_
{F \setminus f_{\alpha}, \emptyset,\emptyset, \langle s-\eta_{2},-1 \rangle } 
\\
\times 
\sum_{\beta=1}^{m_{2}} (-1)^{\beta +m_{1}+n-s+\eta_{2}} \Qf_{b_{\beta}^{(2)}}^{[1]}
\Def^{B_{2}\setminus b_{\beta}^{(2)},\langle s+\eta_{1}+1,0 \rangle ,[0;2]}_
{\emptyset, \emptyset, \langle s-\eta_{2}+1,-1 \rangle } 
\\
=
\sum_{\alpha=1}^{n} \sum_{\beta=1}^{m_{2}} 
(-1)^{\alpha +m_{1}+m_{2}-s-\eta_{1}} (-1)^{\beta +m_{1}+n-s+\eta_{2}}
(\Qf_{b^{(2)}_{\beta}f_{\alpha}}^{[2]}- \Qf_{b^{(2)}_{\beta}f_{\alpha}}^{[0]})
\\ \times 
\Def^{\langle s+\eta_{1}+2,0 \rangle ,[1;2]}_
{F \setminus f_{\alpha}, \emptyset,\emptyset, \langle s-\eta_{2},-1 \rangle }
\Def^{B_{2}\setminus b_{\beta}^{(2)},\langle s+\eta_{1}+1,0 \rangle ,[0;2]}_
{\emptyset, \emptyset, \langle s-\eta_{2}+1,-1 \rangle } 
\\
=-
\sum_{\beta=1}^{m_{2}} (-1)^{\beta +m_{1}+n-s+\eta_{2}}
\Def^{B_{2}\setminus b_{\beta}^{(2)},\langle s+\eta_{1}+1,0 \rangle ,[0;2]}_
{\emptyset, \emptyset, \langle s-\eta_{2}+1,-1 \rangle } 
\\ 
\times 
\begin{vmatrix}
(\Qf_{b_{i}^{(1)} f_{j}}^{[2]})_{
\genfrac{}{}{0pt}{}{1\le i \le m_{1}, }{1\le j \le n} }
& (0)_{m_{1} \times (\eta_{2}-s)}
  \\[6pt]
(-\Qf_{b_{i}^{(2)} f_{j}}^{[0]})_{
\genfrac{}{}{0pt}{}{1\le i \le m_{2}, }{1\le j \le n} }
& (\Qf_{b_{i}^{(2)}}^{[2j+1]})_{
\genfrac{}{}{0pt}{}{1\le i \le m_{2}, }{s-\eta_{2} \le j \le -1} }
   \\[6pt]
((-1)^{i-1}\Qf_{f_{j}}^{[-2i+3]})_{
\genfrac{}{}{0pt}{}{s+\eta_{1}+2 \le i \le 0, }{1\le j \le n} }
& (0)_{(-s-\eta_{1}-1) \times (\eta_{2}-s)}
\\[6pt]
(\Qf_{b_{\beta}^{(2)} f_{j}}^{[0]})_{1\le j \le n}
& (0)_{1 \times (\eta_{2}-s)}
 \\
\end{vmatrix}
\\
+ \sum_{\alpha=1}^{n}
(-1)^{\alpha +m_{1}+m_{2}-s-\eta_{1}} 
\Def^{\langle s+\eta_{1}+2,0 \rangle ,[1;2]}_
{F \setminus f_{\alpha}, \emptyset,\emptyset, \langle s-\eta_{2},-1 \rangle }
\\ 
\times 
\begin{vmatrix}
(\Qf_{b_{i}^{(1)} f_{j}}^{[2]})_{
\genfrac{}{}{0pt}{}{1\le i \le m_{1}, }{1\le j \le n} }
& (0)_{m_{1} \times (\eta_{2}-s-1)}
& (0)_{m_{1} \times 1}
  \\[6pt]
(\Qf_{b_{i}^{(2)} f_{j}}^{[2]})_{
\genfrac{}{}{0pt}{}{1\le i \le m_{2}, }{1\le j \le n} }
& (\Qf_{b_{i}^{(2)}}^{[2j+1]})_{
\genfrac{}{}{0pt}{}{1\le i \le m_{2}, }{s-\eta_{2}+1 \le j \le -1} }
& (\Qf_{b_{i}^{(2)}f_{\alpha}}^{[2]})_{1\le i \le m_{2}}
   \\[6pt]
((-1)^{i-1}\Qf_{f_{j}}^{[-2i+3]})_{
\genfrac{}{}{0pt}{}{s+\eta_{1}+1 \le i \le 0, }{1\le j \le n} }
& (0)_{(-s-\eta_{1}) \times (\eta_{2}-s-1)}
& (0)_{(-s-\eta_{1}) \times 1}
 \\
\end{vmatrix} .
\label{topro10}
\end{multline}
%%%
Adding the $(\beta +m_{1})$-th row to the $(m_{1}+m_{2}-s-\eta_{1})$-th row in the determinant 
in the first summand in the right hand side of \eqref{topro10}, we obtain
\begin{multline}
\begin{vmatrix}
(\Qf_{b_{i}^{(1)} f_{j}}^{[2]})_{
\genfrac{}{}{0pt}{}{1\le i \le m_{1}, }{1\le j \le n} }
& (0)_{m_{1} \times (\eta_{2}-s)}
  \\[6pt]
(-\Qf_{b_{i}^{(2)} f_{j}}^{[0]})_{
\genfrac{}{}{0pt}{}{1\le i \le m_{2}, }{1\le j \le n} }
& (\Qf_{b_{i}^{(2)}}^{[2j+1]})_{
\genfrac{}{}{0pt}{}{1\le i \le m_{2}, }{s-\eta_{2} \le j \le -1} }
   \\[6pt]
((-1)^{i-1}\Qf_{f_{j}}^{[-2i+3]})_{
\genfrac{}{}{0pt}{}{s+\eta_{1}+2 \le i \le 0, }{1\le j \le n} }
& (0)_{(-s-\eta_{1}-1) \times (\eta_{2}-s)}
\\[6pt]
(\Qf_{b_{\beta}^{(2)} f_{j}}^{[0]})_{1\le j \le n}
& (0)_{1 \times (\eta_{2}-s)}
 \\
\end{vmatrix}
\\
=
\begin{vmatrix}
(\Qf_{b_{i}^{(1)} f_{j}}^{[2]})_{
\genfrac{}{}{0pt}{}{1\le i \le m_{1}, }{1\le j \le n} }
& (0)_{m_{1} \times (\eta_{2}-s)}
  \\[6pt]
(-\Qf_{b_{i}^{(2)} f_{j}}^{[0]})_{
\genfrac{}{}{0pt}{}{1\le i \le m_{2}, }{1\le j \le n} }
& (\Qf_{b_{i}^{(2)}}^{[2j+1]})_{
\genfrac{}{}{0pt}{}{1\le i \le m_{2}, }{s-\eta_{2} \le j \le -1} }
   \\[6pt]
((-1)^{i-1}\Qf_{f_{j}}^{[-2i+3]})_{
\genfrac{}{}{0pt}{}{s+\eta_{1}+2 \le i \le 0, }{1\le j \le n} }
& (0)_{(-s-\eta_{1}-1) \times (\eta_{2}-s)}
\\[6pt]
(0)_{1 \times n}
& (\Qf_{b_{\beta}^{(2)}}^{[2j+1]})_{s-\eta_{2} \le j \le -1}
 \\
\end{vmatrix}
\\
=
\sum_{\gamma =s-\eta_{2}}^{-1}
(-1)^{m_{1}+m_{2}-s-\eta_{1}+n+\gamma- s+\eta_{2}+1}
\Qf_{b_{\beta}^{(2)}}^{[2\gamma +1]}
\Def^{\langle s+\eta_{1}+2,0\rangle ,[1;2]}_{\emptyset, 
\emptyset , \langle s-\eta_{2},-1 \rangle \setminus \gamma}.
\label{topro11}
\end{multline}
%%%%
Subtracting the $\alpha $-th column from the $(n-s+\eta_{2})$-th column in the determinant 
in the second summand in the right hand side of \eqref{topro10}, we obtain
\begin{multline}
\begin{vmatrix}
(\Qf_{b_{i}^{(1)} f_{j}}^{[2]})_{
\genfrac{}{}{0pt}{}{1\le i \le m_{1}, }{1\le j \le n} }
& (0)_{m_{1} \times (\eta_{2}-s-1)}
& (0)_{m_{1} \times 1}
  \\[6pt]
(\Qf_{b_{i}^{(2)} f_{j}}^{[2]})_{
\genfrac{}{}{0pt}{}{1\le i \le m_{2}, }{1\le j \le n} }
& (\Qf_{b_{i}^{(2)}}^{[2j+1]})_{
\genfrac{}{}{0pt}{}{1\le i \le m_{2}, }{s-\eta_{2}+1 \le j \le -1} }
& (\Qf_{b_{i}^{(2)}f_{\alpha}}^{[2]})_{1\le i \le m_{2}}
   \\[6pt]
((-1)^{i-1}\Qf_{f_{j}}^{[-2i+3]})_{
\genfrac{}{}{0pt}{}{s+\eta_{1}+1 \le i \le 0, }{1\le j \le n} }
& (0)_{(-s-\eta_{1}) \times (\eta_{2}-s-1)}
& (0)_{(-s-\eta_{1}) \times 1}
 \\
\end{vmatrix}
\\
= - 
\begin{vmatrix}
(\Qf_{b_{i}^{(1)} f_{j}}^{[2]})_{
\genfrac{}{}{0pt}{}{1\le i \le m_{1}, }{1\le j \le n} }
& (0)_{m_{1} \times (\eta_{2}-s-1)}
& (\Qf_{b_{i}^{(1)} f_{\alpha}}^{[2]})_{1\le i \le m_{1} }
  \\[6pt]
(\Qf_{b_{i}^{(2)} f_{j}}^{[2]})_{
\genfrac{}{}{0pt}{}{1\le i \le m_{2}, }{1\le j \le n} }
& (\Qf_{b_{i}^{(2)}}^{[2j+1]})_{
\genfrac{}{}{0pt}{}{1\le i \le m_{2}, }{s-\eta_{2}+1 \le j \le -1} }
& (0)_{m_{2}\times 1}
   \\[6pt]
((-1)^{i-1}\Qf_{f_{j}}^{[-2i+3]})_{
\genfrac{}{}{0pt}{}{s+\eta_{1}+1 \le i \le 0, }{1\le j \le n} }
& (0)_{(-s-\eta_{1}) \times (\eta_{2}-s-1)}
& ((-1)^{i-1}\Qf_{f_{\alpha}}^{[-2i+3]})_{s+\eta_{1}+1 \le i \le 0 }
 \\
\end{vmatrix}
\\
= -
\sum_{\gamma=1}^{m_{1}}(-1)^{\gamma + n+\eta_{2}-s}
\Qf_{b_{\gamma}^{(1)} f_{\alpha}}^{[2]}
\Def^{B_{1}\setminus b^{(1)}_{\gamma}, \langle s+\eta_{1}+1,0\rangle ,[0;2]}_{\emptyset, 
\emptyset , \langle s-\eta_{2}+1,-1 \rangle}
\\
-
\sum_{\gamma=s+\eta_{1}+1}^{0}(-1)^{m_{1}+m_{2}+\gamma-s-\eta_{1} + n+\eta_{2}-s}
(-1)^{\gamma -1}\Qf_{f_{\alpha}}^{[-2\gamma +3]}
\Def^{\langle s+\eta_{1}+1,0\rangle \setminus \gamma,[0;2]}_{\emptyset, 
\emptyset , \langle s-\eta_{2}+1,-1 \rangle}.
\label{topro12}
\end{multline}
Substituting \eqref{topro11} and \eqref{topro12} into the right 
hand side of \eqref{topro10}, and taking summation over $\alpha$ and $\beta$, 
we find that only a few terms in the summation over $\gamma $ are non-zero 
(since there are the same rows or columns in determinants except 
for a few values of $\gamma $), and finally 
we obtain the right hand side of \eqref{topro9}.
%%%%%%%%%%%%%%%%%%%%%%%%%%%%%%%
%\newpage 
%\setlength{\oddsidemargin}{2cm}
 

\begin{thebibliography}{99}

\bibitem{T09}
Z. Tsuboi, 
Solutions of the $T$-system and Baxter equations for supersymmetric spin chains, 
Nucl. Phys. B 826 [PM] (2010) 399-455; arXiv:0906.2039 [math-ph]. 

\bibitem{GKLT10}
N.\ Gromov, V.\ Kazakov, S.\ Leurent, Z.\ Tsuboi, 
Wronskian Solution for AdS/CFT Y-system, 
JHEP 1101 (2011) 155 [arXiv:1010.2720 [hep-th]]. 

\bibitem{KNS10}
A.\ Kuniba, T.\ Nakanishi, J.\ Suzuki, 
T-systems and Y-systems in integrable systems, 
J.Phys.A44 (2011) 103001 [arXiv:1010.1344 [hep-th]].

\bibitem{KR87} 
A.N.\ Kirillov, N. Yu. Reshetikhin, 
Exact solution of the integrable XXZ Heisenberg model with
arbitrary spin: I. The ground state and the excitation spectrum, 
J. Phys. A: Math. Gen. 20 (1987) 1565-1585;\\
%\bibitem{KP92} 
A.\ Kl\"umper, P.A.\ Pearce, 
Conformal weights of RSOS lattice models and their fusion hierarchies,
Physica A183 (1992) 304-350.
%

\bibitem{KNS93}
A.\ Kuniba, T.\ Nakanishi, J.\ Suzuki, 
Functional Relations in Solvable Lattice Models I: Functional
Relations and Representation Theory, 
Int. J. Mod. Phys. A9 (1994) 5215-5266 [arXiv:hep-th/9309137]. 

\bibitem{T97}
Z.\ Tsuboi, Analytic Bethe Ansatz and functional equations 
 for Lie superalgebra $sl(r+1|s+1)$, 
J.Phys.A: Math. Gen. 30 (1997) 7975-7991[arXiv:0911.5386 [math-ph]].

\bibitem{T98} 
Z.\ Tsuboi, 
Analytic Bethe Ansatz and functional equations associated with 
 any simple root systems of the Lie superalgebra 
 $sl(r+1|s+1)$, Physica A 252 (1998) 565-585 [arXiv:0911.5387 [math-ph]].
 
%%
\bibitem{Hirota81}
R.\ Hirota, 
Discrete Analogue of a Generalized Toda Equation, 
J. Phys. Soc. Jpn. 50 (1981) 3785-3791.

\bibitem{KSZ07}
V.\ Kazakov, A.\ Sorin, A.\ Zabrodin,
Supersymmetric Bethe Ansatz and Baxter Equations from Discrete Hirota
Dynamics,  Nucl. Phys. B 790 (2008) 345-413 
[arXiv:hep-th/0703147].

\bibitem{GKV09}
N.\ Gromov, V.\ Kazakov, P.\ Vieira,  
Integrability for the Full Spectrum of Planar AdS/CFT, arXiv:0901.3753 [hep-th]; 
Exact Spectrum of Anomalous Dimensions of Planar N=4 Supersymmetric Yang-Mills Theory, 
Phys.Rev.Lett.103:131601,2009.

\bibitem{Hegedus09}
A.\ Hegedus,
Discrete Hirota dynamics for AdS/CFT, 
Nucl.\ Phys.\ B825 (2010) 341-365 [arXiv:0906.2546 [hep-th]].

\bibitem{Volin10-1}
D.\ Volin, 
Quantum integrability and functional equations, 
J. Phys. A: Math. Theor. 44 (2011) 124003 [arXiv:1003.4725 [hep-th]]. 

\bibitem{Serban10}
D.\ Serban, 
Integrability and the AdS/CFT correspondence, 
J.Phys.A: Math. Theor. 44 (2011) 124001 [arXiv:1003.4214 [hep-th]].

\bibitem{GK10}
N.\ Gromov, V.\ Kazakov, 
Review of AdS/CFT Integrability, Chapter III.7: Hirota Dynamics for Quantum Integrability, 
arXiv:1012.3996v2 [hep-th]. 

\bibitem{Beisert10}
N.\ Beisert et al,
Review of AdS/CFT Integrability: An Overview, 
arXiv:1012.3982v3 [hep-th].

\bibitem{BFT09}
D.\ Bombardelli, D.\ Fioravanti, R.\ Tateo, 
Thermodynamic Bethe Ansatz for planar AdS/CFT: a proposal, 
J. Phys. A: Math. Theor. 42 (2009) 375401  
[arXiv:0902.3930 [hep-th]];
\\
%
N.\ Gromov, V.\ Kazakov, A.\ Kozak, P.\ Vieira, 
Integrability for the Full Spectrum of Planar AdS/CFT II, 
arXiv:0902.4458 [hep-th]; 
Exact Spectrum of Anomalous Dimensions of Planar $N = 4$ Supersymmetric Yang-Mills Theory: TBA and excited states, 
 Lett. Math. Phys. 91(2010) 265-287.
\\
%
G.\ Arutyunov, S.\ Frolov, 
Thermodynamic Bethe Ansatz for the ${\rm AdS_5 \times S^5}$ Mirror Model, 
JHEP05(2009)068 [arXiv:0903.0141 [hep-th]].


\bibitem{GKT10}
N.\ Gromov, V.\ Kazakov, Z.\ Tsuboi, 
$PSU(2,2|4)$ Character of Quasiclassical AdS/CFT, 
JHEP 1007 (2010) 097  
[arXiv:1002.3981 [hep-th]]. 


\bibitem{CLZ03}
S.-J. Cheng, N. Lam and R.B. Zhang,
\textit{ ``Character Formula for Infinite Dimensional Unitarizable Modules of the General Linear Superalgebra,''}
J. Algebra {\bf 273}  (2004) 780-805
[arXiv:math/0301183 [math.RT]].




\bibitem{KLWZ97}
I.\ Krichever, O.\ Lipan, P.\ Wiegmann, A.\ Zabrodin, 
 Quantum Integrable Models and Discrete Classical Hirota Equations, 
Commun. Math.Phys. 188 (1997) 267-304 [arXiv:hep-th/9604080].

%
\bibitem{BR90}
V.V.\ Bazhanov, N.\ Reshetikhin, 
Restricted solid-on-solid models connected with simply laced algebras
 and conformal field theory, J. Phys. A: Math. Gen. 23 (1990) 1477-1492; \\
%
 A representation theoretical background of the formula can be found in 
the following paper. \\
%
I.V.\ Cherednik, 
 Quantum groups as hidden symmetries of classic representation theory, 
in Proceedings of the XVII International Conference on Differential Geometric
 Methods in Theoretical Physics, Chester,
ed. A. I. Solomon (World Scientific, Singapore, 1989), 47-54.

 
\bibitem{Q-systems}
A.N.\ Kirillov and N.Yu.\ Reshetikhin,
 \textit{ ``Representations of Yangians and multiplicities of the inclusion of the irreducible
 components of the tensor product of representations of simple Lie algebras,''}
 J. Soviet Math.  {\bf 52}, 3156 (1990);  
 \\
%\cite{Kuniba:2001mj}
%\bibitem{Kuniba:2001mj}
  A.~Kuniba, T.~Nakanishi and Z.~Tsuboi,
  \textit{ ``The Canonical solutions of the Q-systems and the Kirillov-Reshetikhin
  conjecture,''}
  Commun.\ Math.\ Phys.\  {\bf 227} (2002) 155-190 [arXiv:math/0105145]; 
%
\\
P.D. Francesco and R. Kedem,
\textit{ ``Q-systems as cluster algebras II: Cartan matrix of finite type and the polynomial property,''}
Lett.Math.Phys.\ {\bf 89} (2009) 183-216 [arXiv:0803.0362 [math.RT]].


\bibitem{Gromov09}
  N.~Gromov,
  \textit{``Y-system and Quasi-Classical Strings,''}
  JHEP 1001 (2010) 112 
  [arXiv:0910.3608 [hep-th]].
 
\bibitem{Bax72}
R.J.\ Baxter, 
Partition function of the eight-vertex lattice model, 
 Ann. Phys. 70 (1972) 193-228.
 
\bibitem{Zamolodchikov90}
Al. B. Zamolodchikov, 
Thermodynamic Bethe ansatz in relativistic models: Scaling 3-state potts and Lee-Yang models,
Nucl. Phys. B342 (1990) 695-720. 
 
\bibitem{BLZ97}
%
V.V.\ Bazhanov, S.L.\ Lukyanov, A.B.\ Zamolodchikov, 
Integrable Structure of Conformal Field Theory III. The Yang-Baxter Relation, 
Commun.Math.Phys. 200 (1999) 297-324 
[arXiv:hep-th/9805008]. 

\bibitem{CKW09}
S.-J. Cheng, J.-H. Kwon, W. Wang, 
Kostant homology formulas for oscillator modules of Lie superalgebras, 
Adv. Math. 224 (2010) 1548-1588
[arXiv:0901.0247 [math.RT]]

\bibitem{Kwon06}
J.-H. Kwon,
\textit{ ``Rational semistandard tableaux and character formula for the Lie superalgebra $\hat{\frak{gl}}_{\infty|\infty}$,''}
 Adv. Math. {\bf 217} (2008) 713-739
 [arXiv:math/0605005 [math.RT]].
 
\bibitem{Conrey05}
J.B. Conrey, D.W. Farmer and M.R. Zirnbauer,
\textit{ ``Howe pairs, supersymmetry, and ratios of random characteristic polynomials for the unitary groups $U_{N}$,''}
arXiv:math-ph/0511024.

\bibitem{Volin10-2}
D.\ Volin, 
String hypothesis for $gl(n|m)$ spin chains: a particle/hole democracy
arXiv:1012.3454v1 [hep-th]. 

\bibitem{BKSZ05}
N.\ Beisert, V.A.\ Kazakov, K.\ Sakai, K.\ Zarembo, 
Complete Spectrum of Long Operators in N=4 SYM at One Loop, 
JHEP 0507 (2005) 030 [arXiv:hep-th/0503200].

%%%%%%
\bibitem{MV03}
E.M.\ Moens, J.\ Van der Jeugt:
 A determinant formula for supersymmetric Schur polynomials,  
J. Algebraic Combinatorics 17 (2003) 283-307. 
%%%%

\bibitem{Zabrodin07}
A.\ Zabrodin, 
Backlund transformations for the difference Hirota equation and the supersymmetric Bethe Ansatz, Theor. Math. Phys. 155 (2008) 567-584 
[arXiv:0705.4006[hep-th]].

\bibitem{FSS89}
L.\ Frappat, A.\ Sciarrino, and P.\ Sorba,
Structure of Basic Lie Superalgebras
and of their Affine Extensions, 
Commun.\ Math.\ Phys.\ 121 (1989) 457-500.

\bibitem{Pronko-Stroganov00}
G.P.\ Pronko, Yu.G.\ Stroganov,   
Families of solutions of the nested Bethe Ansatz for the $A_2$ spin chain, 
J. Phys. A: Math. Gen. 33 (2000) 8267-8273 
[arXiv:hep-th/9902085].


\bibitem{BHK02}
V.V.\ Bazhanov, A.N.\ Hibberd, S.M.\ Khoroshkin, 
 Integrable structure of ${W}_3$ Conformal Field Theory, Quantum
  {B}oussinesq Theory and Boundary Affine {T}oda Theory, 
Nucl. Phys. B622 (2002) 475--547 [arXiv:hep-th/0105177].


\bibitem{BDKM06} 
A.V.\ Belitsky, S.E.\ Derkachov, G.P.\ Korchemsky, A.N.\ Manashov:
 Baxter Q-operator for graded $SL(2|1)$ spin chain,
 J.Stat.Mech. 0701 (2007) P005 [arXiv:hep-th/0610332].

\bibitem{Kojima08}
T.\ Kojima,
\textit{``The Baxter's Q-operator for the W-algebra \(W_N\),''}
J.Phys.A: Math. Theor. 41 (2008) 355206
[arXiv:0803.3505 [nlin.SI]].

%%%%%%
\bibitem{BT08}
V.V.\ Bazhanov, Z.\ Tsuboi: 
Baxter's Q-operators for supersymmetric spin chains, 
Nucl. Phys. B 805 [FS] (2008) 451-516 [arXiv:0805.4274 [hep-th]]. 
%
\footnote{
In section 2.4 of this paper, 
solutions of the Yang-Baxter relation (L-operators) for the Q-operators (for trigonometric models) 
were presented.  
The Q-operators were given as the (super)trace of these L-operators over some 
oscillator representations. 
%These solutions were obtained based on a similar idea for the solutions of 
% the Yang-Baxter relation for the Q-operators for \(U_{q}(\widehat{sl}(3))\) 
%by V.V.\ Bazhanov and S.M.\ Khoroshkin (2001), 
%where some contractions on the Cartan subalgebra part (diagonal elements) of the L-operators were used. 
%There are \(2^{K+M}\) ways of these contractions, which correspond to \(2^{K+M}\) different Q-operators. 
%
A preliminary idea on the derivation of the L-operators  
was presented first in many conferences 
%\footnote{
(these include the following two: 
``Workshop and Summer School: 
From Statistical Mechanics to Conformal and Quantum Field Theory'', 
the university of Melbourne, January, 2007; 
%[http://www.smft2007.ms.unimelb.edu.au/program/LectureSeries.html];  
%
La 79eme Rencontre entre physiciens theoriciens et mathematiciens 
``Supersymmetry and Integrability'', IRMA Strasbourg, June, 2007  %[http://www-irma.u-strasbg.fr/article383.html].
%}
) in 2007. 
Recently, a method closely related to this (for rational models) has been developed rapidly \cite{BFLMS10}. 
%Only the final results on  \(U_{q}(\widehat{sl}(2|1))\) case 
%were published in the above paper in relation to the universal 
%R-matrix without detailed derivation. 
%
}


\bibitem{DM10}
S.E.\ Derkachov, A.N.\ Manashov,
\textit{``Noncompact $sl(N)$ spin chains: BGG-resolution, Q-operators and alternating sum representation 
for finite dimensional transfer matrices,''}
Lett.Math.Phys.97 (2011) 185-202 [arXiv:1008.4734 [nlin.SI]].
%%%%%%%%%%%%%%%%%


\bibitem{FR99}
E.\ Frenkel, N.\ Reshetikhin, 
 The $q$-characters of representations of quantum affine algebras and
 deformations of  W-algebras, Contemporary Math. 248 (1999) 163-205
 [arXiv:math/9810055[math.QA]];\\
%
E.\ Frenkel, E.\ Mukhin, 
Combinatorics of $q$-characters of finite-dimensional representations 
of quantum affine algebras, Commun. Math. Phys. 216 (2001) 23-57
[arXiv:math/9911112[math.QA]]. 

\bibitem{Knight95}
H.\ Knight, 
Spectra of Tensor Products of Finite Dimensional Representations of Yangians, 
J. Algebra 174 (1995) 187-196.


\bibitem{GS03}
F.\ G\"ohmann, A.\ Seel: 
A note on the Bethe Ansatz solution of the supersymmetric t-J model, 
Czech.J.Phys. 53 (2003) 1041-1046 
[arXiv:cond-mat/0309138].
%

\bibitem{DDMST06}
P.\ Dorey, C.\ Dunning, D.\ Masoero, J.\ Suzuki, R.\ Tateo, 
Pseudo-differential equations, and the Bethe Ansatz for the classical
Lie algebras,  
Nucl. Phys. B 772 (2007) 249-289 [arXiv:hep-th/0612298];\\
%
%\bibitem{GV07}
N.\ Gromov, P.\ Vieira, 
Complete 1-loop test of AdS/CFT, 
JHEP 04 (2008) 046, [arXiv:0709.3487 [hep-th]].

\bibitem{OR10}
D.\ Orlando, S.\ Reffert, 
Relating Gauge Theories via Gauge/Bethe Correspondence, 
JHEP 1010 (2010) 071
[arXiv:1005.4445 [hep-th]].

\bibitem{Woynarovich83}
F.\ Woynarovich, 
Low-energy excited states in a Hubbard chain with on-site attraction, 
J.Phys.C: Solid State Phys. 16 (1983) 6593-6604; \\
%
%\bibitem{BCFH92}
P.A.\ Bares, I.M.P.\ Carmelo, J.\ Ferrer, P.\ Horsch,  
Charge-spin recombination in the one-dimensional supersymmetric t-J model,  
  Phys. Rev. B46 (1992) 14624-14654; \\
%
%\bibitem{EKS92}
F.H.L.\ Essler, V.E.\ Korepin, K.\ Schoutens, 
Exact Solution of an Electronic Model of Superconductivity, 
Int. J. Mod. Phys. B 8 (1994) 3205-3242 [arXiv:cond-mat/9211001]. 
%%%%%%%%%%%%%%%%%%%
\bibitem{Z97}
R.\ B.\ Zhang, 
Symmetrizable quantum affine superalgebras 
and their representations, 
J.\ Math.\ Phys.\ 38 (1997) 535-543.  


%
\bibitem{KS94}
A.\ Kuniba, J.\ Suzuki, 
Functional Relations and Analytic Bethe Ansatz for Twisted Quantum Affine Algebras, 
J. Phys. A: Math. Gen. 28 (1995) 711-722 [arXiv:hep-th/9408135].
%
\bibitem{T02}
Z.\ Tsuboi, 
Difference $L$ operators and a Casorati determinant solution to the $T$-system 
for twisted quantum affine algebras, 
J. Phys. A: Math. Gen. 35 (2002) 4363-4373 [arXiv:0911.5368 [math-ph]].
%%

\bibitem{T99} 
Z.\ Tsuboi, 
Analytic Bethe ansatz and functional relations related to 
tensor-like representations of type-II Lie superalgebras $B(r|s)$ and $D(r|s)$, 
J. Phys. A 32: Math. Gen. (1999) 7175-7206 [arXiv:0911.5393 [math-ph]]. 

\bibitem{BLMS10}
V.\ V.\ Bazhanov, T.\ Lukowski, C.\ Meneghelli, M.\ Staudacher, 
A Shortcut to the Q-Operator, 
J.Stat.Mech.1011: (2010) P11002 [arXiv:1005.3261 [hep-th]]. 

\bibitem{JKS98} 
G.\ J\"uttner, A.\ Kl\"umper, J.\ Suzuki, 
{}From fusion hierarchy to excited state TBA, 
Nucl. Phys. B512 (1998) 581-600 [arXiv:hep-th/9707074]. 

\bibitem{Saleur99} 
H.\ Saleur, 
The continuum limit of $sl(N/K)$ integrable super spin chains, 
Nucl.Phys. B578 (2000) 552-576, [arXiv:solv-int/9905007].

\bibitem{Takahashi00} 
M. Takahashi, 
Simplification of thermodynamic Bethe-ansatz equations, 
in Physics and Combinatorics, 
eds. A. N. Kirillov and N. Liskova, (2001) 299-304 
(World Scientific, Singapore); 
[arXiv:cond-mat/0010486]. 

\bibitem{Tsuboi06}
Z.\ Tsuboi, 
Nonlinear integral equations and high temperature expansion for 
the $U_{q}(\hat{sl}(r+1|s+1))$ Perk-Schultz Model, 
Nucl. Phys. B737 (2006) 261-290 [arXiv:cond-mat/0510458]; 
Z.\ Tsuboi, M.\ Takahashi, 
Nonlinear Integral Equations for Thermodynamics of the $U_{q}(\hat{sl}(r+1))$ Perk-Schultz Model, 
J.Phys.Soc.Jan. 74 (2005) 898-904 [arXiv:cond-mat/0412698].

\bibitem{DD92}
C.\ Destri, H.J.\ de Vega, 
New thermodynamic Bethe ansatz equations without strings, 
Phys. Rev. Lett. 69 (1992) 2313-2317;\\
%
 A.\ Kl\"umper, 
Thermodynamics of the anisotropic spin-1/2 Heisenberg chain and related quantum chains, 
Z. Phys. B91 (1993) 507-519;\\
%
A.\ Kl\"umper, 
Free energy and correlation lengths of quantum chains related to restricted 
solid-on-solid lattice models, 
Ann. Physik 504 (1992) 540-553.

\bibitem{KL10}
V.\ Kazakov, S.\ Leurent, 
Finite Size Spectrum of $SU(N)$ Principal Chiral Field from Discrete Hirota Dynamics, 
arXiv:1007.1770 [hep-th].

\bibitem{nlie}
P.\ Zinn-Justin, 
Non-Linear Integral Equations for complex Affine Toda associated to simply laced Lie algebras, 
 J. Phys. A: Math. Gen. 31 (1998) 6747-6770 [arXiv:hep-th/9712222]; \\
%\bibitem{DDT00}
P.\ Dorey, C.\ Dunning and R.\ Tateo, 
 Differential equations for general $SU(n)$ Bethe ansatz systems, 
 J. Phys. A: Math. Gen. 33 (2000) 8427-8441 [arXiv:hep-th/0008039].

\bibitem{Suzuki99}
J.\ Suzuki, 
Spinons in magnetic chains of arbitrary spins at 
finite temperatures, 
J. Phys. A: Math. Gen. 32 (1999) 2341-2359 [arXiv:cond-mat/9807076]. \\
%
There is an attempt to generalize this for the T-hook: \\
R.\ Suzuki, Hybrid NLIE for the Mirror $AdS_5 \times S^{5}$,
J.Phys.A44: (2011) 235401 [arXiv:1101.5165 [hep-th]].

\bibitem{DK06}
J.\ Damerau, A.\ Kl\"{u}mper, 
Nonlinear integral equations for the thermodynamics of the $sl(4)$-symmetric Uimin-Sutherland model,
J.Stat.Mech.0612 (2006) P12014 [arXiv:cond-mat/0610559]. 

\bibitem{T98-2}
Z.\ Tsuboi, Analytic Bethe Ansatz related to a  
 one-parameter family of finite-dimensional  
 representations of the Lie superalgebra 
 $sl(r+1|s+1)$, 
J.Phys.A: Math. Gen. 31 (1998) 5485-5498 [arXiv:0911.5389 [math-ph]]; 
%
Z.\ Tsuboi, 
Analytic Bethe ansatz related to the Lie superalgebra $C(s)$, 
Physica A 267 (1999)173-208 [arXiv:0911.5390 [math-ph]]; 
%
Z.\ Tsuboi, A.\ Kuniba, 
Solutions of a discretized Toda field equation for $D_{r}$ from Analytic Bethe Ansatz, 
J.Phys.A29: (1996) 7785-7796 [arXiv:hep-th/9608002].


\bibitem{KV07}
  V.~Kazakov and P.~Vieira,
  ``From Characters to Quantum (Super)Spin Chains via Fusion,''
  JHEP 0810 (2008) 050
  [arXiv:0711.2470 [hep-th]].
  
\bibitem{KLT10}
V.\ Kazakov, S.\ Leurent, Z.\ Tsuboi, 
Baxter's Q-operators and operatorial Backlund flow for quantum (super)-spin chains, 
 Commun.\ Math.\ Phys. 311 (2012) 787-814 [arXiv:1010.4022 [math-ph]].

\bibitem{Benichou10}
R.\ Benichou, 
Fusion of line operators in conformal sigma-models on supergroups, and the Hirota equation, 
JHEP 1101 (2011) 066 [arXiv:1011.3158 [hep-th]].

\bibitem{osc-app}
%\bibitem{Kulish:2005qc}
  P.~P.~Kulish, A.~M.~Zeitlin,
  ``Superconformal field theory and SUSY N=1 KDV hierarchy II: The Q-operator,''
  Nucl.\ Phys.\  {\bf B709 } (2005)  578 
  [hep-th/0501019];
\\
H.\ Boos, M.\ Jimbo, T.\ Miwa, F.\ Smirnov, Y.\ Takeyama, 
 Hidden Grassmann structure in the XXZ model, 
Commun. Math. Phys. 272 (2007) 263-281 [arXiv:hep-th/0606280]; 
%
\\
%
H.\ Boos, F.\ G\"{o}hmann, A.\ Kl\"{u}mper, K.S.\ Nirov, A.V.\ Razumov, 
Exercises with the universal R-matrix 
J. Phys. A: Math. Theor. 43 (2010) 415208 [arXiv:1004.5342 [math-ph]].


\bibitem{BFLMS10}
V. Bazhanov, R. Frassek, T. Lukowski, C. Meneghelli, M. Staudacher, 
Baxter Q-Operators and Representations of Yangians, 
Nucl.Phys. B850 (2011) 148-174 [arXiv:1010.3699 [math-ph]]; 
%
\\
R. Frassek, T. Lukowski, C. Meneghelli, M. Staudacher, 
Oscillator Construction of $su(n|m)$ Q-Operators, 
Nucl. Phys. B850 (2011) 175-198 
[arXiv:1012.6021 [math-ph]]. 

\bibitem{Talalaev04}
D. Talalaev, 
The Quantum Gaudin System, 
Funct. Anal. Its Appl., 40 (2006) 73-77
[arXiv:hep-th/0404153].

\bibitem{KLV10}
V.\ Kazakov, S.\ Leurent, D.\ Volin, 
private communication, November 2010. 
%%%%%%%%%%%%%%%%%%%%%%%%%%



%%%%%


\end{thebibliography}
\end{document}